\def\mearth{{\rm\,M_\oplus}}

\def\deg{^\circ}
\def\micron{\mu {\rm m}}

\documentclass[traditabstract]{aa}

\usepackage{graphicx}
\usepackage{txfonts}

\usepackage{natbib}
\usepackage{aas_macros}

\begin{document}

\titlerunning{Debris disks as signposts of terrestrial planet formation}
\authorrunning{Raymond et al.}

\title{Debris disks as signposts of terrestrial planet formation}

\author{Sean N. Raymond\inst{1,2}, 
Philip J. Armitage\inst{3,4},
Amaya Moro-Mart{\' i}n\inst{5,6},
Mark Booth\inst{7,8},
Mark C. Wyatt\inst{7},
John. C. Armstrong\inst{9},
Avi M. Mandell\inst{10},
Franck Selsis\inst{1,2}, \&
Andrew A. West\inst{11,12}
}

\institute{
Universit{\'e} de Bordeaux, Observatoire Aquitain des Sciences de l'Univers, 2 rue de l'Observatoire, BP 89, F-33271 Floirac Cedex, France\\ \email{rayray.sean@gmail.com}
\and
CNRS, UMR 5804, Laboratoire d'Astrophysique de Bordeaux, 2 rue de l'Observatoire, BP 89, F-33271 Floirac Cedex, France
\and
JILA, University of Colorado, Boulder CO 80309, USA
\and
Department of Astrophysical and Planetary Sciences, University of Colorado, Boulder CO 80309, USA
\and
Department of Astrophysics, Center for Astrobiology, Ctra. de Ajalvir, km 4, Torrej{\' o}n de Ardoz, 28850, Madrid, Spain
\and
Department of Astrophysical Sciences, Princeton University, Peyton
Hall, Ivy Lane, Princeton, NJ 08544, USA
\and
Institute of Astronomy, Cambridge University, Madingley Road, Cambridge, UK
\and
University of Victoria, 3800 Finnerty Road, Victoria, BC, V8P 1A1 Canada
\and
Department of Physics, Weber State University, Ogden, UT, USA
\and
NASA Goddard Space Flight Center, Code 693, Greenbelt, MD 20771, USA
\and
Department of Astronomy, Boston University, 725 Commonwealth Ave, Boston, MA, 02215 USA
\and
Visiting Investigator, Department of Terrestrial Magnetism, Carnegie Institute of Washington, 5241 Broad Branch Road, NW, Washington, DC 20015, USA
}

\date{}

\abstract{
There exists strong circumstantial evidence from their eccentric orbits that most of the known extra-solar planetary systems are the survivors of violent dynamical instabilities. 
Here we explore the effect of giant planet instabilities on the formation and survival of terrestrial planets.  
We numerically simulate the evolution of planetary systems around Sun-like stars that include three components: (i) an inner disk of planetesimals and planetary embryos, (ii) three giant planets at Jupiter-Saturn distances, and (iii) an outer disk of planetesimals comparable to estimates of the primitive Kuiper belt.  
We calculate the dust production and spectral energy distribution of each system by assuming that each planetesimal particle represents an ensemble of smaller bodies in collisional equilibrium.  
Our main result is a strong correlation between the evolution of the inner and outer parts of planetary systems, i.e. between the presence of terrestrial planets and debris disks.  Strong giant planet instabilities -- that produce very eccentric surviving planets -- destroy all rocky material in the system, including fully-formed terrestrial planets if the instabilities occur late, and also destroy the icy planetesimal population.  Stable or weakly unstable systems allow terrestrial planets to accrete in their inner regions and significant dust to be produced in their outer regions, detectable at mid-infrared wavelengths as debris disks.  
Stars older than $\sim 100$ Myr with bright cold dust emission (in particular at $\lambda \sim$ $70 \micron$) signpost dynamically calm environments that were conducive to efficient terrestrial accretion.  Such emission is present around $\sim$16\% of billion-year old Solar-type stars.  

Our simulations yield numerous secondary results: 1) The typical eccentricities of as-yet undetected terrestrial planets are $\sim$0.1 but there exists a novel class of terrestrial planet system whose single planet undergoes large amplitude oscillations in orbital eccentricity and inclination; 2) By scaling our systems to match the observed semimajor axis distribution of giant exoplanets, we predict that terrestrial exoplanets in the same systems should be a few times more abundant at $\sim 0.5$ AU than giant or terrestrial exoplanets at 1 AU; 3) The Solar System appears to be unusual in terms of its combination of a rich terrestrial planet system and a low dust content.  This may be explained by the weak, outward-directed instability that is thought to have caused the late heavy bombardment.
}{}{}{}{}

\maketitle

\section{Introduction}

Circumstellar disks of gas and dust are expected to produce three broad classes of planets in radially-segregated
zones~\citep{kokubo02}.  The inner disk forms terrestrial (rocky) planets because it contains too little solid mass to rapidly accrete
giant planet cores, which are thought to form preferentially beyond the snow line where the surface density in solids is higher because
of ice condensation and the isolation mass is larger~\citep{lissauer87}.  Terrestrial planets form in 10-100 million years (Myr) via
collisional agglomeration of Moon- to Mars-sized planetary embryos and a swarm of 1-10$^3$~km sized
planetesimals~\citep{chambers01,kenyon06,raymond06,raymond09c,obrien06}.  From roughly a few to a few tens of AU, giant planet cores
grow and accrete gaseous envelopes if the conditions are right~\citep{pollack96,alibert05}.  Despite their large masses, gas giants
must form within the few million year lifetime of gaseous disks~\citep{haisch01,pascucci06,kennedy09} and be present during the late phases of
terrestrial planet growth.  Resonant interactions, arising not just from giant planets but also from the changing surface 
density of the gas disk itself \citep{nagasawa05}, thus likely play a role in terrestrial planet formation.
Finally, in the outer regions of planetary systems the growth time scale exceeds the lifetime of the gas
disk, and the end point of accretion is a belt of Pluto-sized (and smaller) bodies~\citep{kenyon98}.  

Our understanding of planetary formation in these different zones is constrained by a variety of observations.  The initial conditions of inner disks can be probed by observations of hot dust.  Such observations have shown evidence for grain growth in individual protoplanetary disks~\citep{bouwman08}, and larger samples have shown that hot dust disappears from these disks within a few Myr and that cooler dust takes longer to disappear~\citep{haisch01,mamajek04,meyer08}.  Hundreds of close-in planets -- the outcome of planet formation -- have been detected by radial velocity and transit observations, with masses down to a few $\mearth$~\citep[e.g.,][]{butler06,udry07,leger09,batalha11}.  The frequency of planets on short-period orbits has been found to be a function of the planet mass; less massive planets are significantly more common than high-mass ``hot Jupiters''~\citep{mayor09,howard10}.  The frequency of 3-10 $\mearth$ planets has been measured at 12\%, and by extrapolation the frequency of 0.5-2 $\mearth$ is 23\%~\citep[for orbital periods less than 50 days;][]{howard10}.

Despite the existence of hot Jupiters around $\sim$ 1\% of solar-type stars~\citep{howard10}, the vast majority of giant planets is found beyond 1 AU~\citep{butler06,udry07b}.  The absolute frequency of giant planets is poorly constrained; estimates range from 10\%~\citep{cumming08} to more than 50\%~\citep{gould10}.  These planets are characterized by their broad eccentricity distribution that includes several planets with $e \ge 0.9$~\cite{butler06}.  This distribution is quantitatively reproduced if giant planets form in systems with multiple planets, and dynamical instabilities occurred in 70-100\% of all observed systems~\citep{chatterjee08,juric08,raymond10}.  The onset of instability may be caused by the changing planet-planet stability criterion as the gas disk dissipates~\citep{iwasaki01}, resonant migration~\citep{adams03}, or chaotic dynamics~\citep{chambers96}.  Whatever the trigger, the instability leads to a phase of planet-planet scattering and in most cases to the eventual removal of one or more planets from the system by collision or hyperbolic ejection~\citep{rasio96,weidenschilling96}.  It is the {\it surviving} planets that match the observed distribution.  The properties of the outer Solar System are consistent with the giant planets having formed in a similarly unstable configuration~\citep{thommes99}, though the low eccentricities of Jupiter and Saturn ($e_{\rm J}\approx e_{\rm S} \approx 0.05$), and the late timing of the Late Heavy Bombardment~\citep{strom05}, hint that the instability that occurred in our Solar System was weak and occurred too late to affect terrestrial accretion~\citep{morby10}.

Sub-millimeter observations of dust disks around young stars provide information on the initial conditions in outer planet-forming
disks, and by extrapolation to entire disks.  These observations suggest that the typical protoplanetary disk has a mass of 0.001-0.1
Solar masses~\citep{andrews05,andrews07a,eisner08} and a radial surface density profile of roughly
$r^{-(0.5-1)}$~\citep{mundy00,andrews07b,isella10}.  There appears to be a roughly linear trend between the dust mass and the stellar mass such
that the typical protoplanetary disk contains a few percent of the stellar mass~\citep{andrews07a}, which has important implications
for the planet frequency as a function of stellar mass~\citep{ida05,raymond07,greaves10}.

Debris disks -- warm or cold dust observed around older stars, typically at infrared wavelengths ($\lambda \sim 10-100 \mu m$) -- probe outer disks after planet formation has completed.  Debris disks provide evidence for the existence of leftover planetesimals because the lifetime of observed dust particles under the effects of collisions and radiation forces is far shorter than the typical stellar age, implying a replenishment via collisional grinding of larger bodies.  {\em Spitzer} observations show that about 15\% of solar-type stars younger than 300 Myr have significant dust excesses at $24 \micron$ but that this fraction decreases to about 3\% for stars older than 1 Gyr~\citep{meyer08,carpenter09}.  At $70 \micron$ the fraction of stars showing significant excesses is roughly constant in time (at $\sim 16\%$) but the upper envelope of the distribution decreases with the stellar age~\citep{trilling08,gaspar09,carpenter09}.  A stars have a significantly higher fraction of dust excesses at both 24 and $70 \micron$ and the excesses themselves are brighter than for FGK stars but the dust lifetime is shorter~\citep{su06}.  Very few debris disks are known around M dwarfs and the dust brightness is necessarily far lower than for higher-mass stars~\citep{gautier07}.  

Here we perform a large ensemble of N-body simulations to model the interactions between the different radial components of planetary systems: formation and survival of terrestrial planets, dynamical evolution and scattering of giant planets, and dust production from collisional grinding.  By matching the orbital distribution of the giant planets, we infer the characteristics of as-yet undetected terrestrial planets in those same systems.  We post-process the simulations to compute the spectral energy distribution of dust by treating planetesimal particles as aggregates in collisional equilibrium~\citep{dohnanyi69} to calculate the incident and re-emitted flux~\citep{booth09}.  We then correlate outcomes in the different radial zones and link to two key observational quantities: the orbital properties of giant planets and debris disks.  

The paper is structured as follows.  In section 2 we explain our choice of initial conditions, the integration method, and our debris disk calculations.  In section 3 we demonstrate the detailed evolution of a single simulation in terms of its dynamics, the formation of terrestrial planets, and the dust evolution.  In section 4 we present results from our fiducial set of 152 simulations and explore the correlations between terrestrial planet formation efficiency, giant planet characteristics and debris disks. In Section 5 we discuss the implications of our results: we scale our simulations to the known eccentricity-semimajor axis distribution of giant planets, discuss the observed debris disks in known exoplanet systems, and evaluate the Solar System in the context of our results.   We conclude in Section 6.

In a companion paper (Raymond et al 2011; referred to in the text as Paper~2) we test the effects of several system parameters on this process: the giant planet masses and mass distribution, the width of the outer planetesimal disk, the mass distribution of the outer planetesimal disk, and the presence of disk gas at the time of giant planet instabilities.  We also calibrate our results to the statistics of known debris disks to produce an estimate of the fraction of solar-type stars that host terrestrial planets ($\eta_{Earth}$ in the Drake equation).   

\section{Methods}
We use numerical simulations to study the formation of terrestrial planets from massive embryos, the dynamical 
evolution of fully-formed gas and ice giants, and the long-term evolution of debris disks, starting from an 
assumed set of initial conditions that are specified at the epoch when the protoplanetary gas disk dissipates. 
The goal is to study how these processes, occurring in spatially distinct regions of the planetary system, are 
coupled, and to quantify the expected dispersion in the final outcome that arises from the chaotic nature of the 
evolution. In what follows, we describe in detail the specific initial conditions that we adopt, together with the 
numerical methods. It is worth emphasizing at the outset that our initial conditions are not unique. Observationally, 
little is known about the radial distribution of planetesimal formation within protoplanetary disks, and as a result 
the single most important initial condition for subsequent planet formation is not empirically well-constrained. Our 
fundamental assumption is that planetesimals form across a wide range of disk radii (extending out, in particular, 
to encompass a cold outer debris disk), with a smooth radial distribution. We further assume that cores are able 
to form quickly enough to yield planets at least as massive as ice giants in typical disks\footnote{Recent calculations 
of planet formation via core accretion support such an assumption~\citep{movshovitz10}, though it should be noted that 
these are subject to substantial uncertainties because the physics of Type~I migration remains imperfectly understood~\citep{lubow10}.}. 
Given these assumptions, it is likely that the system-to-system variation in the masses of giant planets -- arising 
ultimately from dispersion in the masses and radii of the protoplanetary disks -- exceeds that of the terrestrial 
planets or that of the debris disk, because small variations in the formation time scale of cores result in large 
changes to the final envelope mass.

Our models are an extension of what might be described as the ``classical" scenario for planet formation. 
Substantially different scenarios are also possible. In particular, planetesimal formation is poorly understood 
\citep[for a review, see][]{chiang10}, and may be coupled to the level of intrinsic turbulence within the 
gaseous disk, which probably varies with radius \citep{gammie96,armitage11}.  
If planetesimal formation efficiency varies dramatically with radius, the initial conditions for subsequent planet 
formation would be radically different from what we have assumed, leading to qualitatively different conclusions. 
For example, it could be true that the zones of terrestrial and giant planet formation are typically dynamically 
well-separated, contrary to what is implied by our initial conditions. In such a model, the coupling between the 
giant and terrestrial planets would be much weaker than occurs given our assumptions, such that only exceptionally 
violent giant planet scattering -- such as occurs when instabilities drive $e \rightarrow 1$ -- would affect the 
terrestrial zone. Similar caveats apply to the outer debris disk region.

\subsection{Initial conditions} 
\begin{figure} %[p]
\center\includegraphics[width=0.48\textwidth]{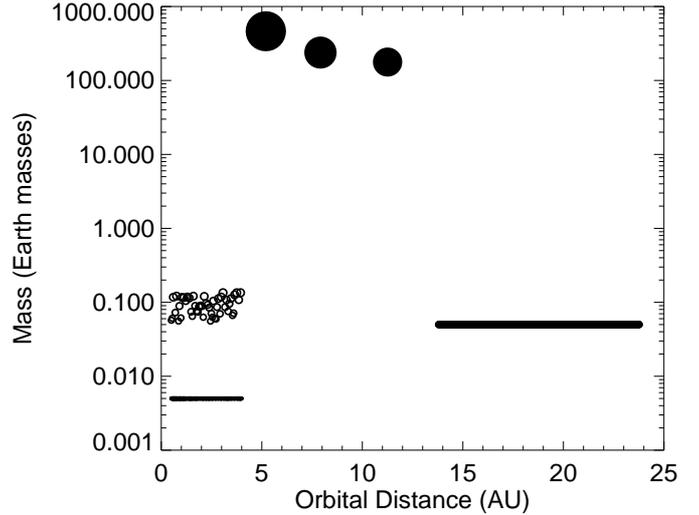}
\caption{An example of the initial conditions assumed to exist shortly after the epoch of gas disk 
dispersal. The terrestrial planet formation zone contains planetary embryos and lower mass 
planetesimals, while the outer planetesimal belt contains a population of relatively low-mass 
(compared to the giant planets) bodies that are represented numerically by $N$ equal mass particles. The 
intermediate zone contains 3 fully-formed giant planets whose spacing is such that they 
are marginally unstable against dynamical instability.}
\label{fig:init}
\end{figure}

The initial conditions for planet formation are expected to vary with stellar mass, due both to the 
relatively well-understood variation of the location of the snow line with stellar type \citep{kennedy08}, and as a 
result of any systematic variations in the star-to-disk mass ratio \citep{alexander06}. We consider 
Solar-mass stars, and start our simulations 
from highly idealized initial conditions that represent the
predicted state of a planetary system at the time of the dissipation of the gaseous protoplanetary disk.  There are three radial zones:
an inner zone containing $9 M_\oplus$ in planetary embryos and planetesimals from 0.5 to 4 AU, three giant planets on marginally stable
orbits from Jupiter's orbital distance of 5.2~AU out to $\sim$10~AU (depending on the masses), and an outer 10~AU-wide disk of
planetesimals containing 50~$M_\oplus$.  In the majority of simulations the giant planets underwent a violent phase of scattering but
in a significant fraction they did not.  Figure~\ref{fig:init} shows an example set of initial conditions.  

\subsubsection{Terrestrial planet-forming region} The terrestrial planet forming region extends from 0.5 to 4 AU.   As in the bulk of
terrestrial planet formation simulations the inner boundary was chosen as a compromise between simulation run time and capturing the
dynamics of the inner disk~\citep[e.g.][]{chambers01}.  The outer boundary corresponds to the approximate stability boundary with the
innermost giant planet (at 5.2 AU), depending on that planet's mass, such that objects beyond this limit would be immediately
destabilized~\citep{marchal82,gladman93}. Within this zone, we adopt initial conditions that are motivated by ``chaotic 
growth" models, of the type summarized by \cite{goldreich04} and calculated in detail by \cite{kenyon06}. Specifically,  
the terrestrial zone contains a total of 9~$M_\oplus$ in 49 planetary embryos and 500
planetesimals, with equal mass in each component. The embryo mass of $0.09 \ M_\oplus$ is a factor of a few to ten larger than that used in recent simulations of late-stage terrestrial accretion~\citep{chambers01,raymond06,raymond09c,morishima10}, but is within a factor of $\sim$2 of that obtained from semi-analytic models of oligarchic growth~\citep{chambers06} over 
the radial range between 1~AU and 3~AU at 3~Myr.  Of course, the outcome of late-stage accretion has been shown to be insensitive to the embryo mass distribution but rather to depend mainly on the large-scale mass distribution of the disk~\citep{kokubo06}.  
The mass is distributed following a radial surface density profile $\Sigma \propto
r^{-1}$.  This profile is consistent with an inward extrapolation of that measured for DG~Tau by \cite{isella10},  who infer a
slope for the disk surface density via interferometric observations of thermal dust emission at  mm wavelengths. We do not consider the
possibility of variations in the disk mass, due either to dispersion in the surface density  of the protoplanetary disk at early times,
or arising from dynamical effects such as the passage of massive planets  migrating inward due to Type~II migration
\citep{fogg07,raymond06,mandell07}.  Nor do we consider the effect of the changing gravitational potential of the dissipating gas disk, which in some cases could influence terrestrial accretion~\citep{nagasawa05,morishima10}.
Modeling these effects would require starting our calculations at an early epoch, when the gas
disk was still present, and would introduce additional  physical uncertainties.  Our disk is comparable in mass to the ``minimum-mass
solar nebula'' (MMSN) model~\citep{weidenschilling77}, with 2.6 [5.1] $M_\oplus$ inside 1.5 [2.5] AU, though we do not 
adopt the MMSN surface density slope.

The embryo mass increases only slightly with orbital distance, in agreement with models for embryo growth~\citep{chambers06}.  In practice, embryos are spaced by $\Delta$ mutual Hill radii $R_{H,m}$: \begin{equation}
R_{H,m} = \frac{1}{2}  (a_1 + a_2)  \left( \frac{m_1+m_2}{3 M_\star} \right)^{1/3}, 
\end{equation}
where $a_1$ and $a_2$ are orbital distances of two adjacent embryos, $m_1$ and $m_2$ are their masses, and $M_\star$ is the stellar mass.  We chose $\Delta$ to lie between 8-16 but decreasing systematically with orbital distance to avoid large variations in embryo mass in different regions of the disk.  In practice, we used $\Delta = 8 + \delta/a$, where $\delta$ is randomly chosen between zero and 8 and $a$ is the orbital distance in AU.  The mass resolution of  the terrestrial component of our model is within a factor of 2-4 of the best current simulations of terrestrial planet formation~\citep{raymond09c,morishima10}.  Embryos were given randomly-chosen initial eccentricities of up to 0.02 and initial inclinations of up to 0.1$^\circ$.  Planetesimals were given initial eccentricities of up to 0.02 and initial inclinations of up to 1$^\circ$.

\subsubsection{Giant Planets} Our giant planet distribution is motivated by observations of extrasolar planetary systems, which show
that the giant planet population detected at $a < 5$~AU exhibits a broad eccentricity distribution~\citep{butler06}. 
The eccentricity distibution is not strongly affected by selection effects \citep{cumming08}, which do however 
dominate the radial distribution (which is essentially unknown beyond about 5~AU).
There is no unique
theoretical interpretation of this result, but the most common explanation is that the eccentricity
represents the endpoint of a dynamical scattering phase. Models built upon this assumption~\citep{juric08,chatterjee08} can reproduce
quantitatively the observed $f(e)$ for extrasolar planets. When the effects of outer planetesimal disks are included, they are also
consistent with near-circular orbits being typical for low mass giant planet systems at larger orbital radii~\citep[][as is found in
the Solar System]{raymond10}.

Here, we assume that marginally unstable initial conditions are 
also the rule for orbital separations modestly larger than those needed to explain the 
current observations of massive extrasolar planets.  This may not be true, but it 
represents the simplest extrapolation of current observational results. 
We model systems of three giant planets, with 
the innermost giant planet given an orbital semimajor axis of 5.2~AU, the same as Jupiter's.  This is an
arbitrary choice which allows for easy comparison with the Solar System (assuming that Jupiter formed at 5.2 AU).  Two
additional planets are spaced outward with randomly-chosen separations of (in the fiducial runs) $4-5 R_{H,m}$.   This
separation was chosen to select for systems that will likely become unstable on a timescale of 100,000 years or longer~\citep{marzari02,chatterjee08}.  This value of $10^5$ years is the expected timescale for the final dissipation of the gaseous protoplanetary disk~\citep{wolk96,ercolano11}, although substantial structural changes to the disk occur over longer timescales~\citep{currie09}.  Because damping within the disk tends to stabilize planetary systems~\citep{iwasaki02}, we expect the natural instability timescale to be on the order of the gas dissipation timescale.

The outcome of scattering is dependent on both the initial mass distribution~\citep[which is 
related to the well-constrained observed mass distribution][]{butler06}, and on the 
distribution of {\em mass ratios} between planets in individual systems~\citep{ford03,raymond10}. For the results shown in the paper we focus on a fiducial set of runs (called {\tt mixed} in paper 2), in which the planet masses are chosen to follow the observed distribution of extra-solar planets, 
\begin{equation}
\frac{{\rm d}N}{{\rm d}M} \propto M^{-1.1},
\end{equation}
where masses are chosen between one Saturn mass and 3~Jupiter masses. This range is chosen so as to include the 
majority of well-observed exoplanet systems (more massive planets are relatively uncommon, while selection effects 
become stronger for very low mass giants), but there is nothing particularly special about our choice. Similar models, 
that fit the observed constraints equally well, could almost certainly be constructed assuming different mass ranges, 
although it might be necessary to adopt different numbers of planets or to assume correlations between their 
masses \citep{raymond10}. In our runs, the masses of individual planets are chosen independently.  

The radii of giant planets affect their early dynamics, by altering the ratio of physical collisions 
to scattering events. We adopt a constant bulk density, $\rho = 1.3 \ {\rm g \ cm}^{-3}$, independent of mass, 
that matches that of the (present-day) Jupiter. Since young giant planets will assuredly have larger radii, our 
assumption underestimates the true number of physical collisions. We do not, however, consider this to be a 
major source of error. Planets in our mass range, formed via core accretion, are only modestly inflated at ages as 
short as a Myr \citep{marley07}, and collisions are suppressed at $a > 5 \ {\rm AU}$ because of the relatively modest orbital 
velocities at these radii \citep{ford01}. 

As noted above, the initial conditions that we adopt for the giant planets are motivated  by strictly empirical conditions: the
observed mass function of extrasolar planets, and  the requirement (within the context of a scattering model) that most multiple
planet  systems are eventually dynamically unstable. However, they are also broadly consistent with first principles
models of giant planet formation \citep{bromley11}, which suggest that radial migration and dynamical instability ought to be 
common, even prior to the dispersal of the gas disk. The specific separation of planets that we have assumed matches that 
expected if multiple giant planets interact with the gas disk such that they end up trapped in mean-motion
resonances~\citep{snellgrove01,lee02,kley04}, {\em provided} that the majority of these resonances are broken before or shortly after
the dispersal of the gas disk. Under some conditions fluctuating gravitational perturbations from disk turbulence may be able to break
resonances~\citep{adams08,lecoanet09}, though the ubiquity of this process remains unclear since the true strength of turbulence within
disks cannot yet be reliably determined.

In paper 2 we test the effect of the mass distribution of giant planets, including systems with much lower-mass giant planets that can undergo Nice model-like instabilities and systems with equal-mass giant planets that produce the strongest instabilities for a given planet mass~\citep{raymond10}.

\subsubsection{Outer Planetesimal Belt}
In all cases, the outer planetesimal disk contains 50 $M_\oplus$ spread over a 10 AU-wide annulus.  The inner edge of the annulus is chosen to be 4 linear Hill radii exterior to the outermost giant planet, such that the innermost planetesimals are ``Hill stable" at the start of the simulation.  This disk's mass and mass distribution is comparable to that invoked by recent models of the evolution of the Solar System's giant planets (the ``Nice model'') to explain the late heavy bombardment~\citep{tsiganis05,gomes05}.  Note that, for numerical reasons and also to account for the much larger total mass, the mass of a ``cometary'' planetesimal particle in the outer system is roughly an order of magnitude larger than an ``asteroidal'' planetesimal in the inner system, although in both cases each planetesimal particle is assumed to represent an ensemble of smaller bodies (see below).

In paper 2 we test the effect of varying certain properties of the outer planetesimal disk mass, including the width of the outer planetesimal disk and the presence of $\sim$ Earth-mass ``seeds'' in the disk.

\subsection{Integration Method}
Our simulations were run using the hybrid integrator in the {\tt Mercury} integration package~\citep{chambers99}, which combines the
speed of a symplectic mapping scheme~\citep{wisdom91} while particles are well-separated with the Bulirsch-Stoer method during closer
encounters.  We used a timestep of 6 days for each of our simulations.  Particles were removed from a simulation when they
attained perihelion distances of less than 0.2~AU, at which point they were assumed to collide with the star, or if they 
reached aphelion at more than 100~AU, when they were assumed to be ejected.
Collisions were treated an inelastic mergers.  We performed extensive numerical tests
to validate our choice of timestep as well as the threshold integration error in orbital energy above which simulations were rejected as unreliable.  These tests are described in Appendix A.

\subsection{Debris Disk Modeling}
To calculate the dust flux in a planetary system from a simulation we follow the procedure outlined in section 2 of~\cite{booth09} with only a few small modifications.  This approach makes the assumption that planetesimal particles act as tracers of a collisional populations of small bodies and adopts a simple model for the evolution of the population.  The properties of this collisional population are drawn from previous studies that fit models of the collisional evolution of planetesimal belts to the statistics for the evolution of debris disks~\citep{dominik03,krivov05,krivov06,wyatt07b,lohne08,wyatt08,krivov10,kains11}.  Although the parameters in this simple model have a precise physical meaning for the model as it is presented below, in practice they represent ``effective'' parameters that determine the mass evolution and are calibrated to match observations.  For example, although we adopt the classic single power law size distribution for a collisional cascade~\citep{dohnanyi69,williams94}, the true size distribution of a collisional population is certainly more complicated~\citep{obrien05,kobayashi10}.  We also assume a constant value for $Q_D^\star$, the impact energy required to catastrophically disrupt an object of size $D$ that is not realistic because $Q_D^\star$ is undoubtedly a function of $D$~\citep[e.g.,][]{benz99}. However, this type of simple model does a reasonably good job at matching more detailed calculations of debris disk evolution that include a size dependent $Q_D^\star$ and (consequently) a multiphase size distribution~\citep[see Fig. 11 of][]{lohne08,kenyon08,kenyon10,krivov10}. Furthermore the parameters of the model have been tuned to match the observed debris disk evolution~\citep{wyatt07b,kains11}.

We divide our planetesimal populations into two components: asteroidal and cometary: asteroids are simply planetesimals interior to the giant planets' initial orbits and comets are exterior.  We assume that each planetesimal population is in a collision-dominated regime from the start of the simulation such that its differential size distribution is represented by $n(D) \propto D^{2-3q_d}$, where D is the object diameter and $q_d = 11/6$ for an infinite collisional cascade~\citep{dohnanyi69,williams94}.  This distribution spans from the smallest assumed particles $D_{bl} = 2.2 \ {\mu}{\rm m}$ up to the largest $D_c = 2000$ km.  Smaller particles than $D_{bl}$ are assumed to be blown out of
the system by radiation pressure on short enough timescales to not contribute to the size distribution~\citep[see, e.g.,][]{wyatt05b}.  The largest bodies in the distribution were chosen to match the largest known objects in the Kuiper belt, assuming that the primitive Kuiper belt represents a proxy for the accretion that would have occurred in planetesimal disks of this mass.  If $D_c$ were significantly smaller then the collisional timescales would be shorter and the collisional evolution would proceed more quickly.  

The surface area of a population of bodies with a collisional size distribution can be easily calculated. Using our chosen values
for $D_{bl}$, $D_c$ and $q_d$, and assuming a physical density for all objects of 1 ${\rm g}\, {\rm cm}^{-3}$, the total cross-sectional area
in a given radial bin is related to the total mass in that bin by:
\begin{equation}
\frac{\sigma(R)}{M(R)} = 0.19 {\rm AU}^2 M_\oplus^{-1},
\end{equation}
\noindent where $\sigma(R)$ is in AU$^2$ and $M(R)$ is in $M_\oplus$.  

For each population of bodies (asteroidal and cometary), we calculate the collisional timescale $t_c$, which is a function of the
particle size $D$ as well as the orbital properties of the population~\citep{wyatt99,wyatt07a}.  This represents the mean timescale
between collisions that are violent enough to destroy bodies of a given size at a mean orbital distance $R_m$ with an annular width
$dr$.  
\begin{equation}
t_c(D) = \left(\frac{R_m^{2.5} \, dr}{M_\star^{0.5} \sigma_{tot}}\right) \left(\frac{2\left[1+1.25
(e/I)^2\right]^{-1/2}}{f_{cc}(D)}\right) {\rm yr},
\end{equation}
where 
\begin{equation}
\sigma_{tot} = \frac{\sigma(R)}{M(R)} M_{tot}. 
\end{equation}
Here $t_c$ is in years, $M_\star$ is the stellar mass in solar masses and $\sigma_{tot}$ is the total surface area in AU$^2$.  The orbital characteristics of the planetesimal population are represented by their mean eccentricities $e$ and inclinations $I$ in radians.  The factor $f_{cc}(D)$ represents the fraction of the total size distribution that could cause a catastrophic collision with a particle of size D, and for our assumptions can be expressed as
\begin{eqnarray}
f_{cc}(D) &=& \frac{3 q_d -5}{D_{bl}^{5-3q_d}}
\left[
\frac{D_{cc}(D)^{5-3q_d} -D_c^{5-3q_d}}{3q_d - 5} \right. \nonumber \\
 &+& 2D \frac{D_{cc}(D)^{4-3q_d} -D_c^{4-3q_d}}{3q_d - 4} \nonumber \\ &+&
\left. D^2 \frac{D_{cc}(D)^{3-3q_d} -D_c^{3-3q_d}}{3q_d - 3}
\right],
\end{eqnarray}
\noindent where $D_{cc}(D) = X_c D$ where $X_c D > D_{bl}$ and $D_{cc} (D) = D_{bl}$ otherwise, and 
\begin{equation}
X_c = 1.3 \times 10^{-3} \left(\frac{Q_D^\star R_m M_\star^{-1}}{1.25e^2 + I^2}\right)^{1/3}.
\end{equation}

\noindent $Q_D^\star$ is the impact energy required to catastrophically disrupt and disperse a body of size $D$.  Here we assume
that $Q_D^\star = 200 {\rm \, J \, kg}^{-1}$ and that it does not vary with $D$.  As discussed above, this assumption is not realistic in terms of the collisional dynamics~\citep{benz99}, but for our purposes $Q_D^\star$ represents an effective strength that determines the dust production and mass loss rate from the planetesimal belt, and using a constant value allows for a good fit to the statistics of debris disks around A stars~\citep{wyatt07b}.  In addition, it may not be reasonable to assume the same $Q_D^\star$ for asteroidal and cometary planetesimals given that their compositions are likely to be different.  However, the typical collisional timescales for the largest (2000 km) bodies are very short, just a few$\times 10^4$ years, for the asteroidal planetesimal population (compared with $1-3 \times 10^8$ years for the cometary planetesimal population).  Thus, asteroidal dust is severely depleted within the first 0.1-1 Myr of each system's evolution and our choice of $Q_D^\star$ has virtually no effect on the results.  

For a given simulation timestep, we calculate the spectral energy distribution (SED) of the system dust as follows (again, as in Booth et al. 2009 with only small modifications):
\begin{itemize}
\item Calculate $t_c$ for the largest bodies in both asteroidal and cometary planetesimal populations.  We then artificially decrease the mass of the planetesimals in each population by a factor of $\left[1+t/tc(D_c)\right]^{-1}$ to account for collisional mass loss.  This represents a very slow decrease for the comets but a significant mass loss among the asteroids.\footnote{Our procedure of decreasing the particle mass according to the collisional timescale is not completely self-consistent as our planetesimal mass did not change during the simulation itself.  Given the long collisional timescales in the outer disk, this assumption has little to no effect on the outer planetary systems.  This assumption also has little effect on cases in which the giant planets were unstable quickly.  However, for stable simulations or simulations with delayed instabilities, this means that, in the inner disk, our dynamics is not completely true to our calculated dust flux.  In other words, the amount of damping provided to growing embryos by planetesimals (via viscous stirring and dynamical friction) should realistically have been lower if we had accounted for collisional evolution of planetesimals. However, given that we compare our simulations with observed debris disks that are generally far older than the 10-100 Myr timescale for terrestrial planet formation, this does not affect our results.  For a careful treatment of dust production during terrestrial planet formation, see~\cite{kenyon04}}.
\item Divide the radial domain into $N_{bin}$ radial bins spaced logarithmically between 0.2 and 100 AU, which are the inner and outer boundaries of our simulation.  We tested a range of $N_{bin}$ values and found good convergence with $N_{bin} \ge 30-50$ and so we use $N_{bin}=100$ to be conservative.  
\item Calculate the total planetesimal mass in each radial bin for asteroids and comets by sampling the orbit of each planetesimal $N_{samp}$ times along its orbit at intervals that are equally spaced in time (i.e., in mean anomaly).  We calculate $N_{samp} = 1+e/e_{limit}$, where $e_{limit}$ represents the minimum statistical eccentricity needed to cross between radial bins: $e_{limit} \approx 0.06$ in our case but we use half of that value.  Thus, we sample circular orbits sparsely but more eccentric orbits up to 30 times per orbit to allow the dust to be spread over a range of orbital distances.  
\item For each radial bin and for asteroids and comets, calculate the blackbody temperature
\footnote{Small grains with sizes of $2.2 \micron$ do not actually radiate as blackbodies.  Rather, the effective temperature of small grains is likely to be higher than the blackbody temperature. The consequence of this in the context of our model is discussed extensively in~\cite{bonsor10} and~\cite{kains11}. To summarize, the actual flux may be slightly higher or lower than that predicted with our assumptions, since on the one hand the higher temperature means that we underestimate the dust flux for a given dust mass, however this is compensated to some extent by the fact that we also overestimate the evolutionary timescale (since the evolutionary timescale for a disk of observed temperature is set observationally).}
\begin{equation}
T_{bb} = 278.3 \left(\frac{L_\star}{L_\odot}\right)^{1/4}\left(\frac{R}{\rm 1 AU}\right)^{-1/2},
\end{equation}
\noindent where $L_\star$ is the stellar luminosity, R is the radial distance, and $T_{bb}$ is in Kelvin.  Then, assuming that all objects radiate as blackbodies, we calculate the flux density (in Janskys) $F_\nu$ seen by an observer at distance $d$:
\begin{eqnarray}
F_\nu = \sum_R 2.35 \times 10^{-11} \sigma(R) \times \nonumber \\ 
B_\nu(\lambda, T_{bb}(R)) \left(\frac{d}{\rm 1 pc}\right)^{-2} X_\lambda^{-1},
\end{eqnarray}
\noindent where $B_\nu$ is the Planck function in units of $\rm Jy \, sr^{-1}$ and $X_\lambda$ is a factor derived by Wyatt et al. (2007b) to account for a decrease in emission beyond $\sim 200 {\mu}{\rm m}$ needed to match observed sub-mm measurements of debris disks: $X_\lambda = 1$ for $\lambda < 210 {\mu}{\rm m}$ and $X_\lambda = \lambda/210$ for longer wavelengths. 
\item Sum the contributions from each radial bin over the whole SED, and then sum the asteroidal and cometary contributions.  We consider the wavelength range from 1 to 1000 ${\mu}{\rm m}$.  
\end{itemize}

Using this method, we calculate the SED of each simulation from the orbits of all planetesimal particles at each simulation timestep.  To make the SED useful for comparison with observations, the most convenient quantity is the dust flux relative to the stellar flux.  The stellar flux $F_{\nu \star}$ in Jy can be calculated as
\begin{equation}
F_{\nu \star} = 1.77 B_\nu (\lambda, T_\star) \left(\frac{L_\star }{L_\odot}\right)T_\star^{-4} \left(\frac{d}{\rm 1 pc}\right)^{-2}
\end{equation}
\noindent where $B_\nu$ is again the Planck function in units of $\rm Jy \, sr^{-1}$ and $T_\star$ is the stellar effective
temperature, which we take to be 5777 K for all cases.  Armed with the stellar flux as well as the dust flux, we can isolate the
dust-to-stellar flux ratio at any wavelength of interest such as those observed with {\em Spitzer} or {\em Herschel}.  

Our simulations ran for 100-200 Myr, but we want to compare fluxes with stars at a range of ages.  Thus, we need to extrapolate the
expected dust fluxes into the future.  We do this by making the assumption that there is no more significant dynamical evolution of
the system, i.e. that the orbital characteristics of the planetesimals are not going to change in the future.  Clearly, this
assumption does not account for punctual events like the late heavy bombardment~\citep{gomes05} or very late dynamical
instabilities.  For a given time $t$ after the end of the simulation, we decrease the planetesimal mass by a simple factor related
to the collisional timescale at the end of the simulation, i.e., as $\left(1+t/t_c\right)^{-1}$.  In terms of the global analysis of our sample of simulations,
we retain snapshots of each simulation at wavelengths of 1, 3, 5, 15, 20, 25, 50, 70, 100, 160, 250, 350, 500 and 850 ${\mu}{\rm
m}$ at time intervals of 1, 3, 10, 30, 100, 300, 1000, and 3000 Myr.  

\begin{figure*} %[p]
\center\includegraphics[width=0.8\textwidth]{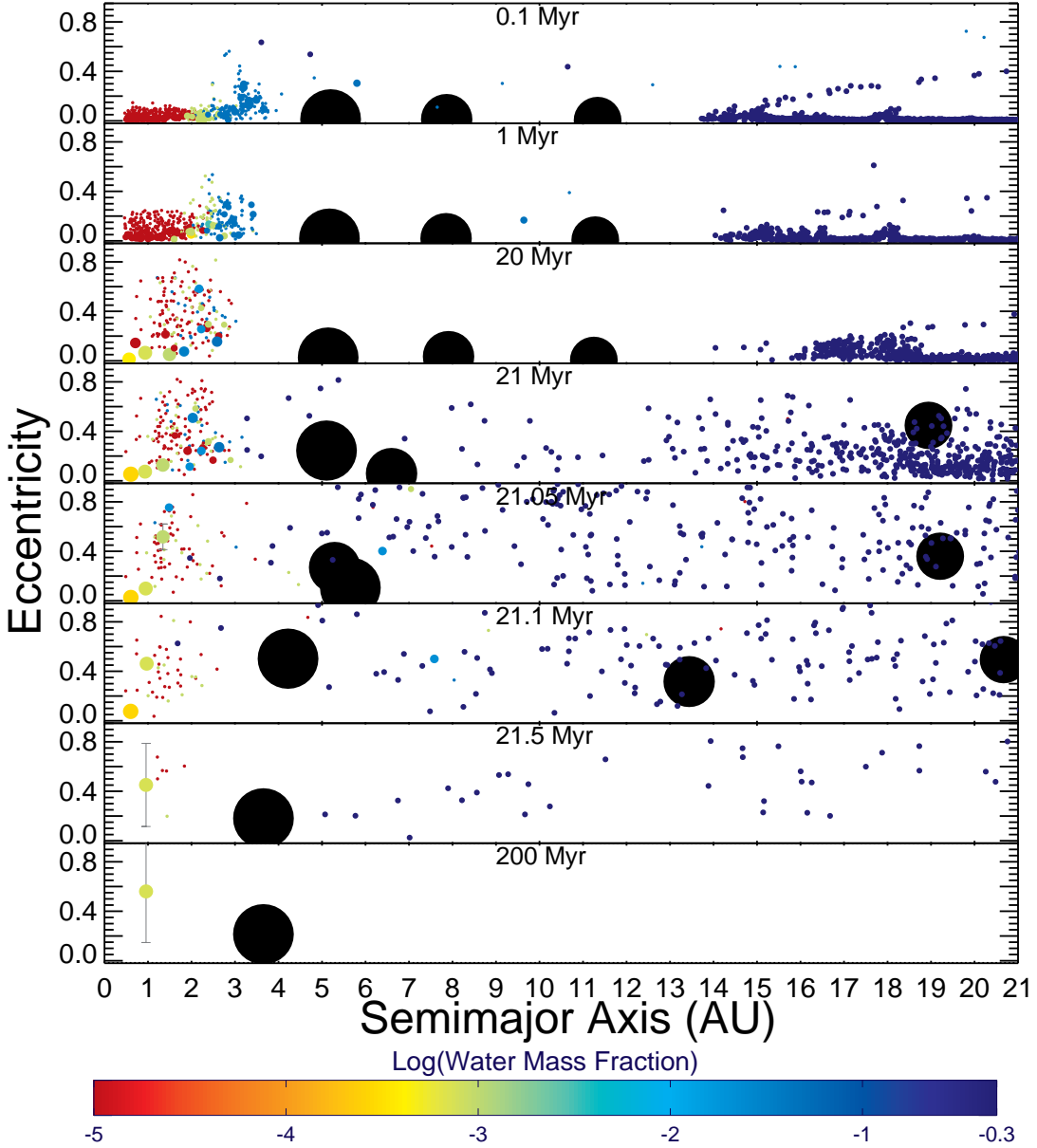}
\caption{Evolution of a system with a relatively late (21~Myr) instability among the giant planets. Each panel shows a snapshot in time of orbital eccentricity vs. semi-major axis for all particles; vertical bars denote $\sin(i)$ for terrestrial bodies with $M_p > 0.2 \ M_\oplus$ and $i > 10^\circ$. The particle size is proportional to the mass$^{1/3}$, but giant planets (large black circles) are not on this scale.  Colors denote water content, assuming a Solar System-like initial distribution~\citep{raymond04}. The surviving terrestrial planet has a mass of 0.72~$M_\oplus$, a stable orbit within the habitable zone (semimajor axis of 0.96 AU), and a high eccentricity and inclination (and large oscillations in these quantities).  A movie of this simulation is available at http://www.obs.u-bordeaux1.fr/e3arths/raymond/scatterSED.mpg.
}
\label{fig:aeit}
\end{figure*}

Our dust flux calculations compare reasonably well with observations and other models.  We perform this comparison using systems that are dynamically calm (referred to as ``stable systems'' later in the paper), in which the giant planets remain on stable orbits throughout the simulation such that the outer planetesimal disk remains largely intact.  In these stable systems, our dust flux calculations yield typical values for the dust-to-stellar-flux ratio $F/F_{star}$ of 0.1-0.5 at $25 \micron$ and 10-35 at $70 \micron$ after 1 Gyr of simulated dynamical and calculated collisional evolution.  These values are broadly consistent with the fluxes detected around solar-type stars with observed excesses at 24 and $70 \micron$~\citep{habing01,beichman06,moor06,trilling08,hillenbrand08,carpenter09,gaspar09}, although our stable simulations yield very few systems with $F/F_{star} (70 \micron) \approx 1-10$, probably because of the relatively large masses in our outer planetesimal disks (note that our unstable systems can yield those flux levels).  Compared with the more sophisticated models of dust production during planetary accretion of~\cite{kenyon08,kenyon10}, our calculated fluxes are larger by a factor of a few, notably at $70 \micron$.  Our fluxes are only a factor of 2-3 larger than those calculated by~\cite{kennedy10}.

\section{An example simulation}
In this section we explore the dynamics and dust properties of a single simulation.  In following sections we will consider the outcome of an ensemble of many simulations, so this is an opportunity to inspect the detailed evolution of a particularly interesting system.  The initial conditions for the chosen simulation are shown in Fig.~\ref{fig:init}: the giant planet masses were 1.46 $M_J$ (at 5.2 AU), 0.75 $M_J$ (7.9 AU) and 0.55  $M_J$ (11.3 AU).

Figure~\ref{fig:aeit} shows the dynamical evolution of the system, which became unstable after 21 million years.  Before the instability, the system evolved in the expected quiescent fashion.  In the inner disk, embryos accreted planetesimals and other embryos from the inside-out.  Embryos' eccentricities and inclinations were kept small by the planetesimals via dynamical friction and viscous stirring.  After 21 million years there were several almost fully-grown terrestrial planets, including three planets more massive than 0.6 $\mearth$ at 0.61, 0.93 and 1.34 AU and a dozen other embryos of 0.06-0.3 $\mearth$ extending out to 2-3 AU.  During this phase of quiescent accretion, the giant planets' orbits remained almost perfectly circular, with eccentricities of less than 0.05 for each planet (less than 0.03 for the innermost gas giant).  Planetesimal scattering caused a slow drift in the giant planets' semimajor axes, of 0.06 AU for the innermost (and most massive) giant planet, and less than 0.01 AU for the other two giants.  During this time, the outer planetesimal disk was slowly sculpted by the giant planets.  The inner edge of the planetesimal disk was eroded, and strong resonances with the outermost giant planet (low-order mean motion resonances) were slowly cleared out.  At the time of the instability, roughly three quarters of the initial outer planetesimal disk was intact.

The instability began with a rapid eccentricity increase in the eccentricity of the two outer giant planets and was followed by a close encounter between those two.  That set off a series of close encounters between all three giant planets that lasted 84,000 years, involved 35 planet-planet scattering events, and culminated with the ejection of the outermost giant planet.  The behavior of the simulation was characteristic in terms of dynamical instabilities in that the least massive giant planet was ejected and the surviving planets were segregated by mass, with the most massive one closest to the star (at 3.65 AU) and the less massive one farther out~\citep[at 36.6 AU, e.g.][]{chatterjee08,raymond10}. 

\begin{figure} %[p]
%\centerline{
\center\includegraphics[width=0.4\textwidth]{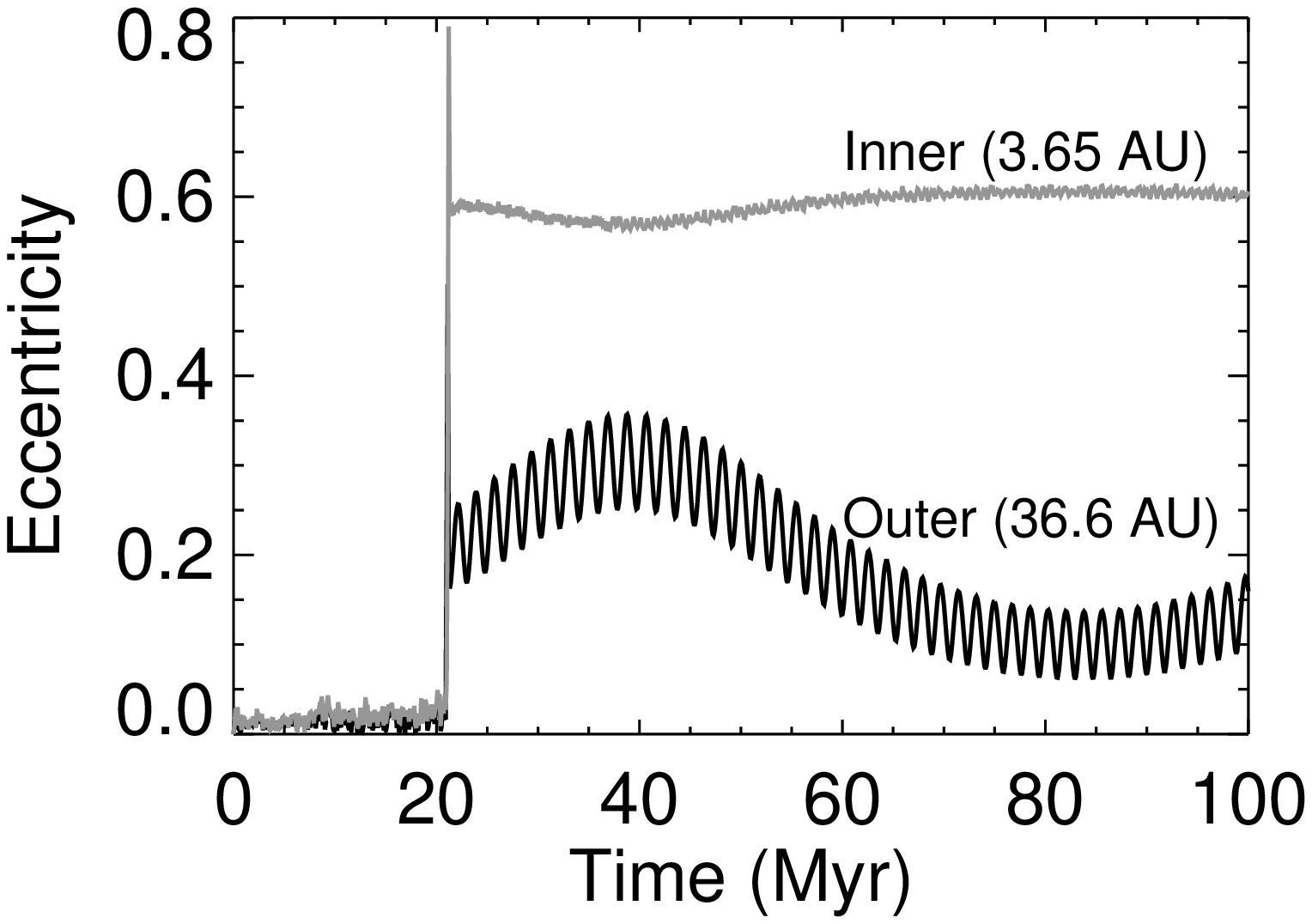}
\center\includegraphics[width=0.4\textwidth]{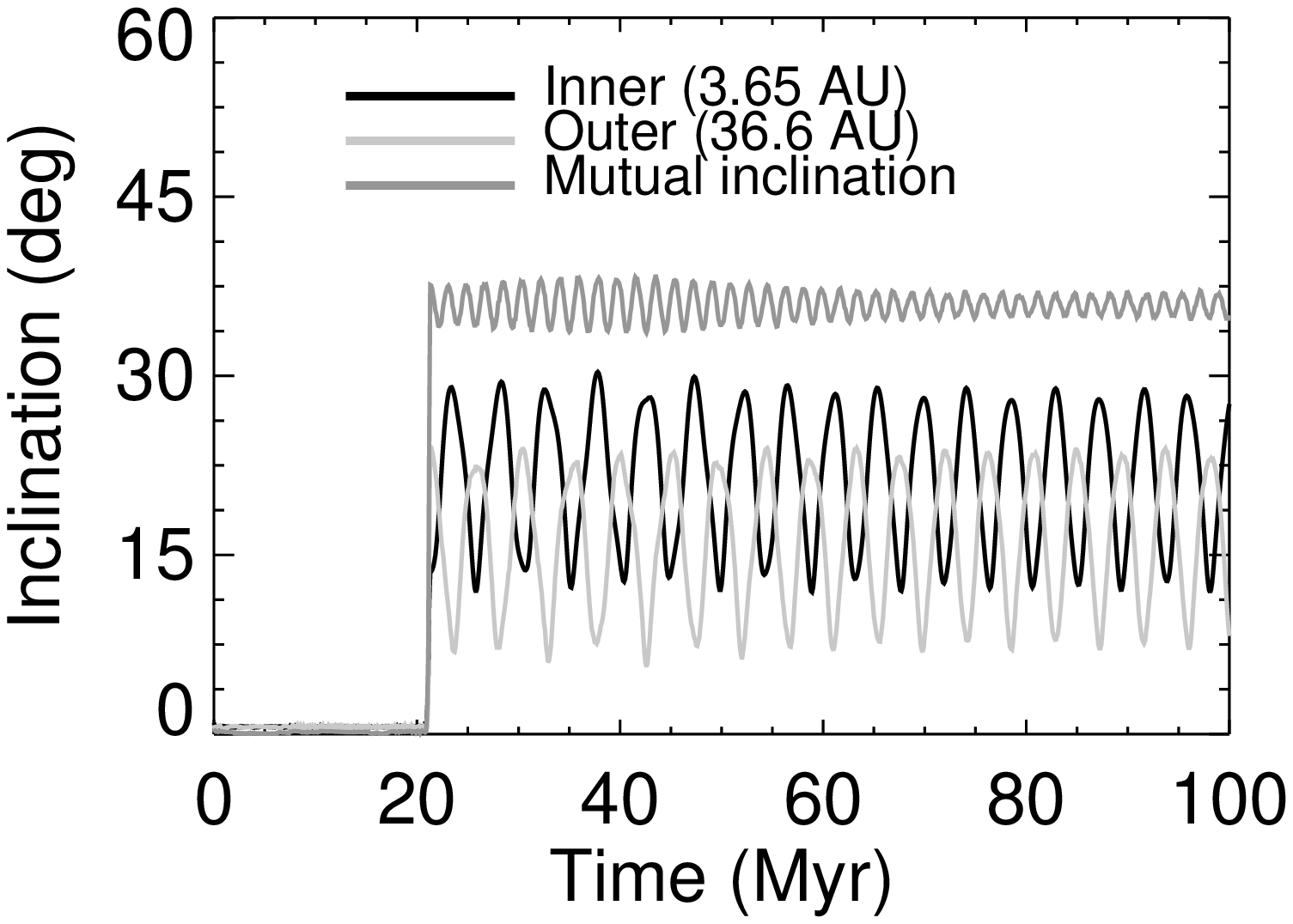}
%}
\caption{The eccentricity (top panel) and inclination (bottom panel) evolution of the two surviving giant planets in the example simulation.  The three curves in the right panel show the inclination of each planet relative to the starting plane of the system -- presumably the stellar equatorial plane -- in black and light grey and their mutual inclination, which is larger than each of their individual inclinations, in dark grey.}
\label{fig:giants}
\end{figure}

During and immediately after the giant planet instability, the inner disk was almost entirely thrown into the star.  This happened by a rapid increase in embryos' and planetesimals' eccentricities mainly by secular forcing from the scattered giant planets.\footnote{In the simulation, any body that came within 0.2 AU of the star is considered to have collided with it.  It is conceivable that in some cases, star-grazing embryos could have their orbits shrunk and re-circularized by tidal effects, although the efficiency of this process is uncertain and probably quite small~\citep{raymond08a}.}  Among the 3.9 $\mearth$ of terrestrial material that was destroyed were 13 embryos including a 0.95 $\mearth$ at 0.6 AU, a 0.62 $\mearth$ planet in the habitable zone at 1.34 AU, and five other embryos larger than 0.15 $\mearth$.  All but one of the embryos -- the sole survivor -- collided with the star within a few hundred thousand years of the start of the instability.  The rocky planetesimals were all destroyed within 1 Myr.  The outer planetesimal disk was completely destabilized by the instability, as the two lower-mass giant planets scattered each other to tens of AU.  All planetesimals were either  ejected from the system (about 85\%) or collided with the star (15\%).  The vast majority were destroyed within 1 Myr of the instability -- all but three within 5 Myr -- but the last planetesimal took almost 25 million years to be ejected.  

\begin{figure} %[p]
%\centerline{
\center\includegraphics[width=0.4\textwidth]{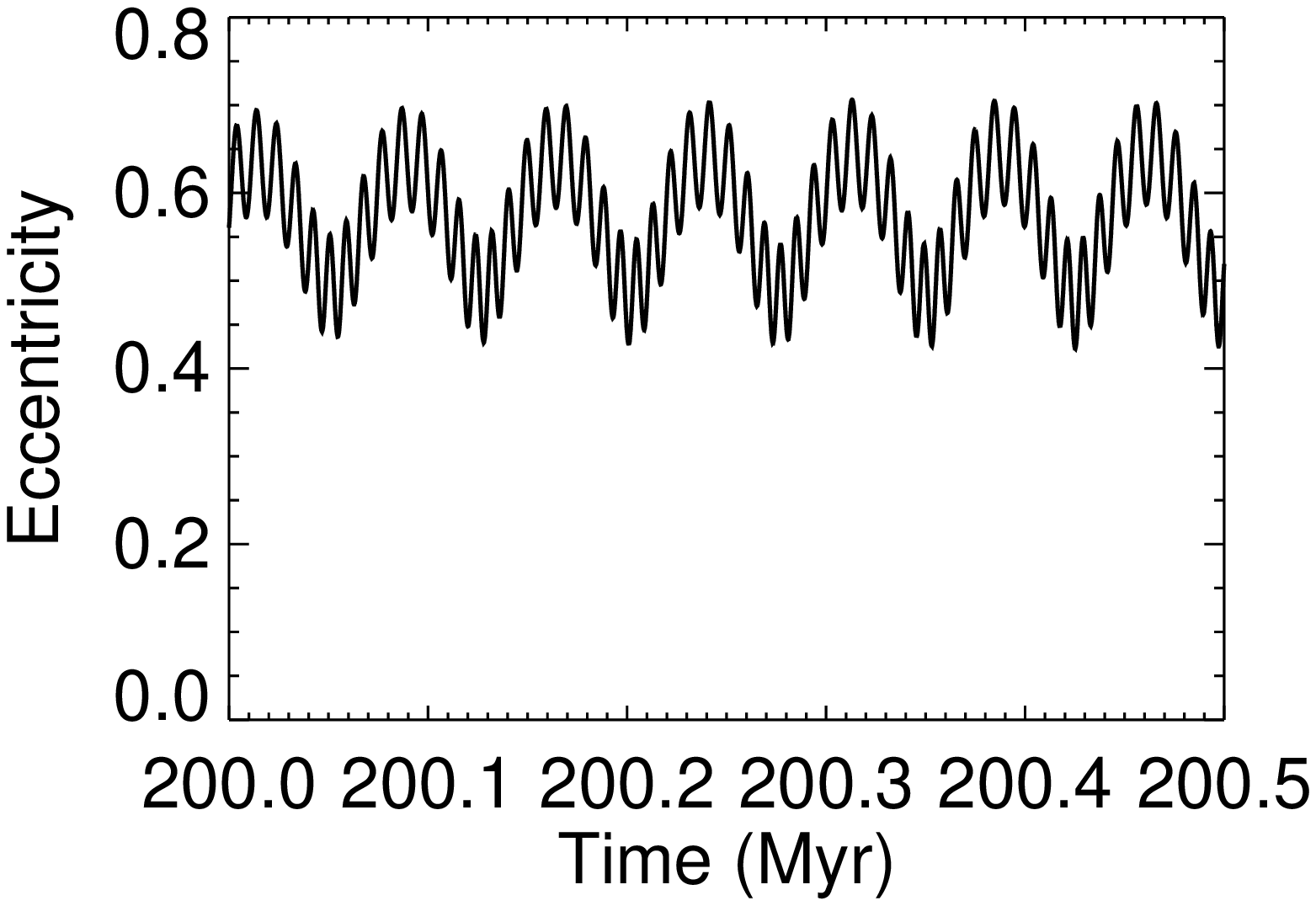}
\center\includegraphics[width=0.4\textwidth]{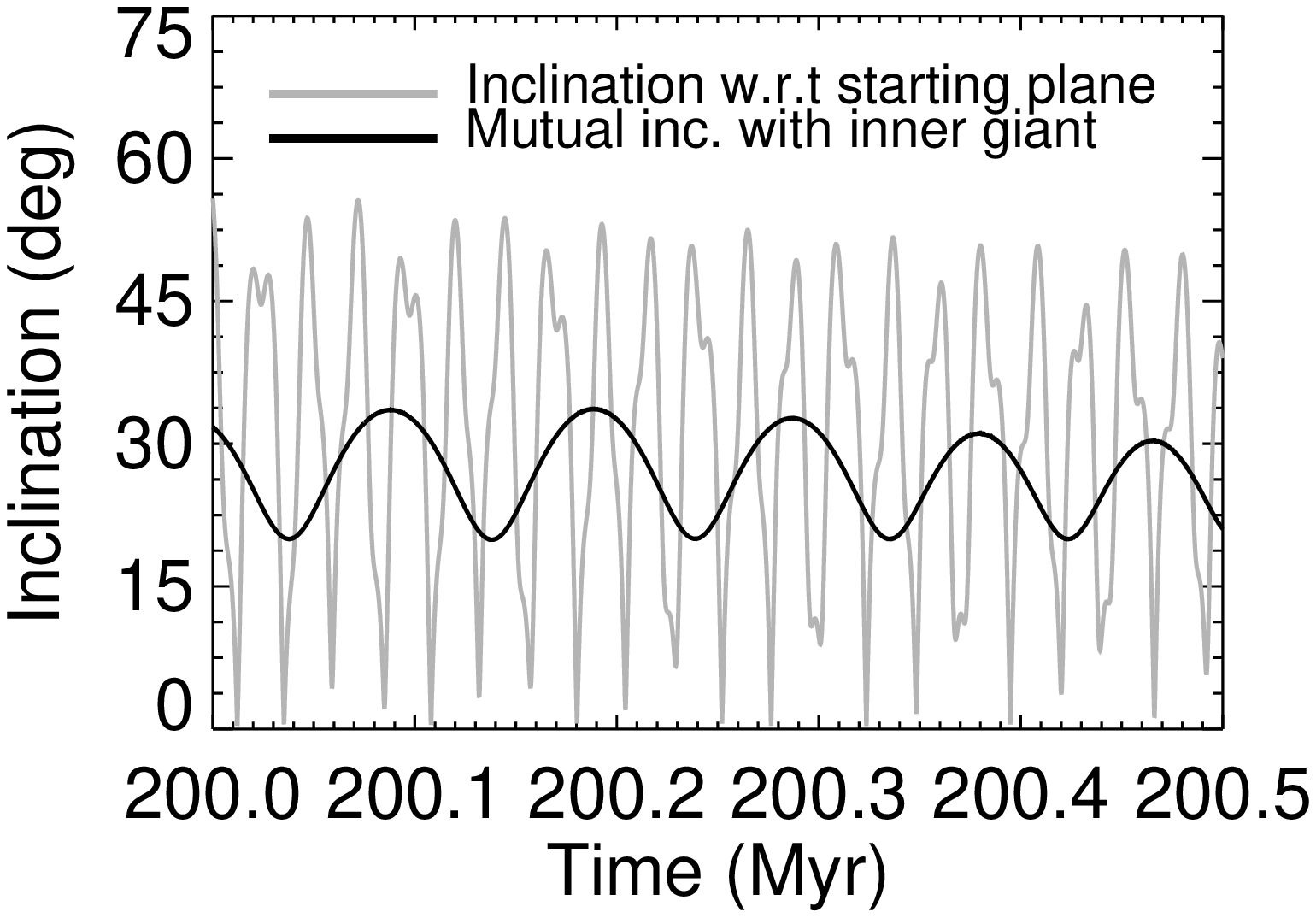}
%}
\caption{The eccentricity (top panel) and inclination (bottom panel) of the surviving terrestrial planet in the example simulation (mass of 0.72$\mearth$, semimajor axis of 0.93 AU).  The two curves in the right panel show the inclination relative to the starting plane of the system in grey and the mutual inclination with respect to the innermost giant planet in black.}
\label{fig:suic38}
\end{figure}

Three planets -- two gas giants and one terrestrial planet -- survived to the end of the simulation, all on excited, eccentric and inclined orbits.  The long-term dynamics of the two surviving giant planets is shown in Figure~\ref{fig:giants}. The eccentricities of the surviving giant planets are considerable as are their orbital inclinations with respect to the starting plane.  The two giant planets have a mutual orbital inclination of $\sim36^\circ$.  Although this is less than the formal limit of 39.2$^\circ$ required for the Kozai effect to take place~\citep{kozai62,takeda05}, out of phase, Kozai-like oscillations are evident in the planets' eccentricities and inclinations.  

The terrestrial planet's final mass is $0.72 \mearth$ and its semimajor axis of 0.935 AU places it in the circumstellar habitable zone~\citep{kasting93,selsis07}.  However, Figure~\ref{fig:suic38} shows that the planet's orbit is strongly perturbed.  Its eccentricity oscillates between 0.4 and 0.7 and its inclination between almost zero and more than $60^\circ$ on a $\sim$10,000 year timescale (although there other longer secular frequencies in the oscillations).  On Myr timescale the orbit has a minimum and maximum eccentricity of 0.39 and 0.73 and a minimum and maximum inclination of 0.03$^\circ$ and 63.3$^\circ$.  Given its large eccentricity, one would expect this planet's climate to be highly variable during the year~\citep{williams02,dressing10}.  In addition, the changes in both its eccentricity and inclination -- equivalent to changes in obliquity for a fixed spin axis -- would cause variations on the secular timescales~\citep{spiegel10}.  Given that the orbit-averaged stellar flux increases with the orbital eccentricity (as $\left(1-e^2\right)^{-1/2}$) and that the planet's closest approach to the star is only 0.25 AU, this planet may not be habitable during its high-eccentricity phases.  However, more detailed modeling of such planets is beyond the scope of the paper and the reader is referred to recent climate modeling papers that include the effects of varying the eccentricity and obliquity~\citep{williams02,williams03,spiegel09,spiegel10,dressing10}.

\begin{figure} %[p]
\center\includegraphics[width=0.48\textwidth]{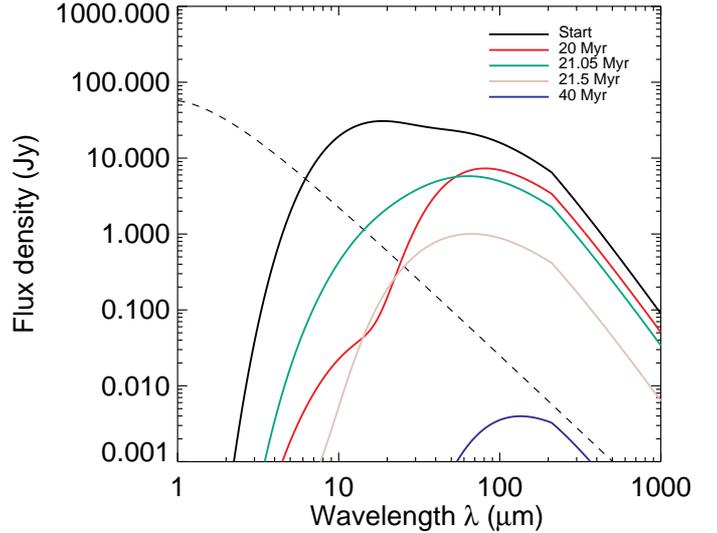}
\caption{Evolution of the spectral energy distribution (SED) of the simulation from Fig.~\ref{fig:aeit}.  Each curve shows the SED during a given simulation snapshot.  The instability occurred at 21 Myr, and the SED evolved dramatically in the immediate aftermath, as icy planetesimals were scattered onto high-eccentricity orbits -- thereby producing transient hot dust -- and then ejected from the system.  The dashed line represents the stellar photosphere.  The system is viewed at a distance of 10 pc.}
\label{fig:SED}
\end{figure}

The evolution of the dust brightness follows from the dynamical and collisional evolution of the planetesimals in the system. Figure~\ref{fig:SED} shows the resulting spectral energy distribution at different snapshots in time.  The collisional timescale for the largest planetesimals (2000 km) in the terrestrial planet forming region is only $t_{coll} \sim 10^4$ years, so the planetesimals in the inner disk are ground into hot dust within just a few million years.  Thus, the total dust brightness has dropped sharply at all wavelengths by the 20 million year snapshot, most dramatically at wavelengths shorter than roughly 50 $\micron$.  For the rest of the simulation, the primary source of dust is the outer planetesimal belt for which $t_{coll} \gtrsim 10^8$ years because the inner disk planetesimals have been ground away.\footnote{We do not consider regeneration of small bodies via giant impacts but we note that the debris from giant collisions could cause short-lived spikes in the dust brightness, in particular at mid- to near-infrared wavelengths~\citep{stern94,grogan01,kenyon05,lisse09}.  These peaks in brightness from collisions can only occur within a few AU because a very massive collision is needed to cause substantial brightening.  }

\begin{figure*}%[p]
\centerline{
\includegraphics[width=0.3\textwidth]{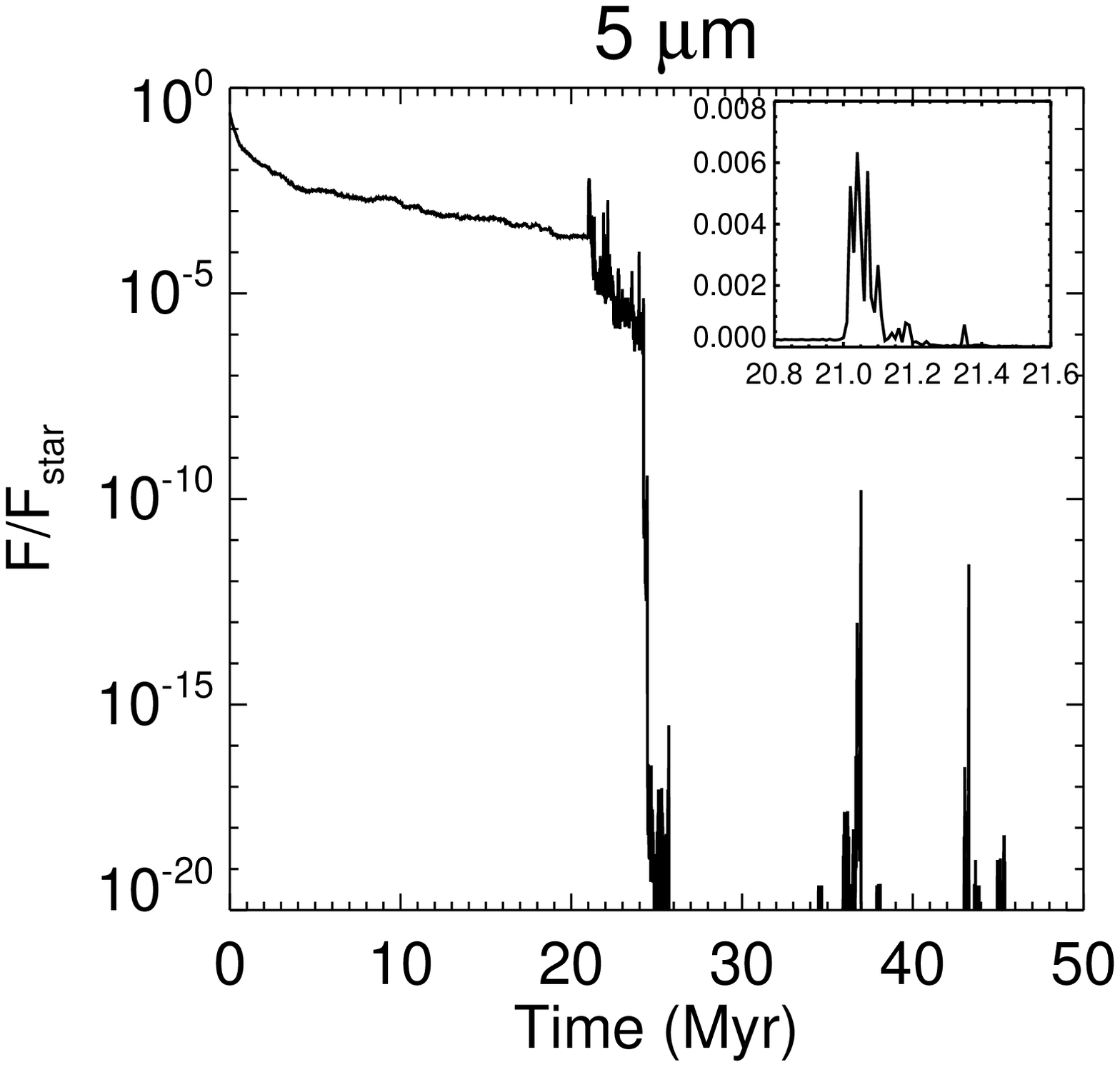}
\includegraphics[width=0.3\textwidth]{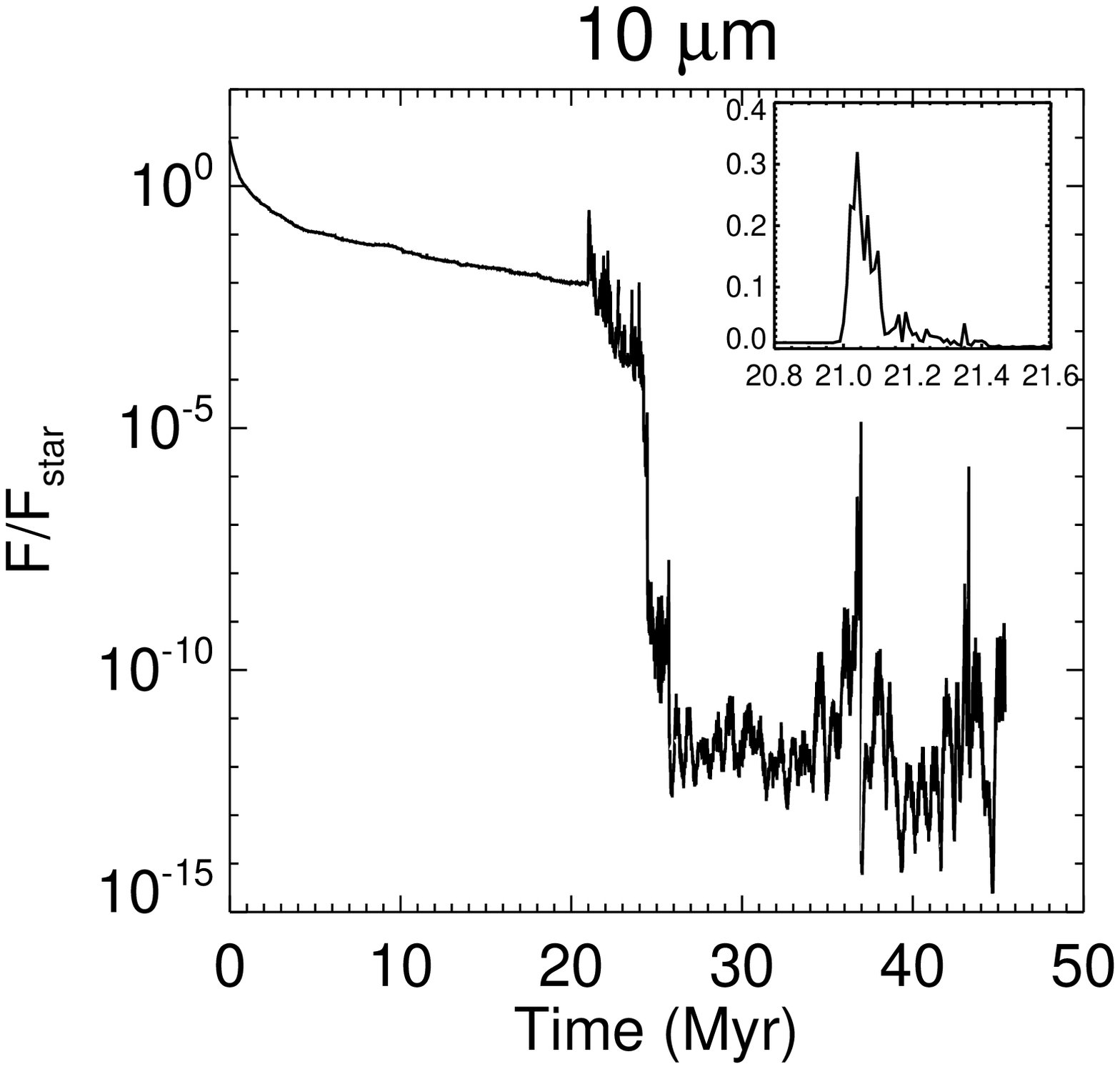}
\includegraphics[width=0.3\textwidth]{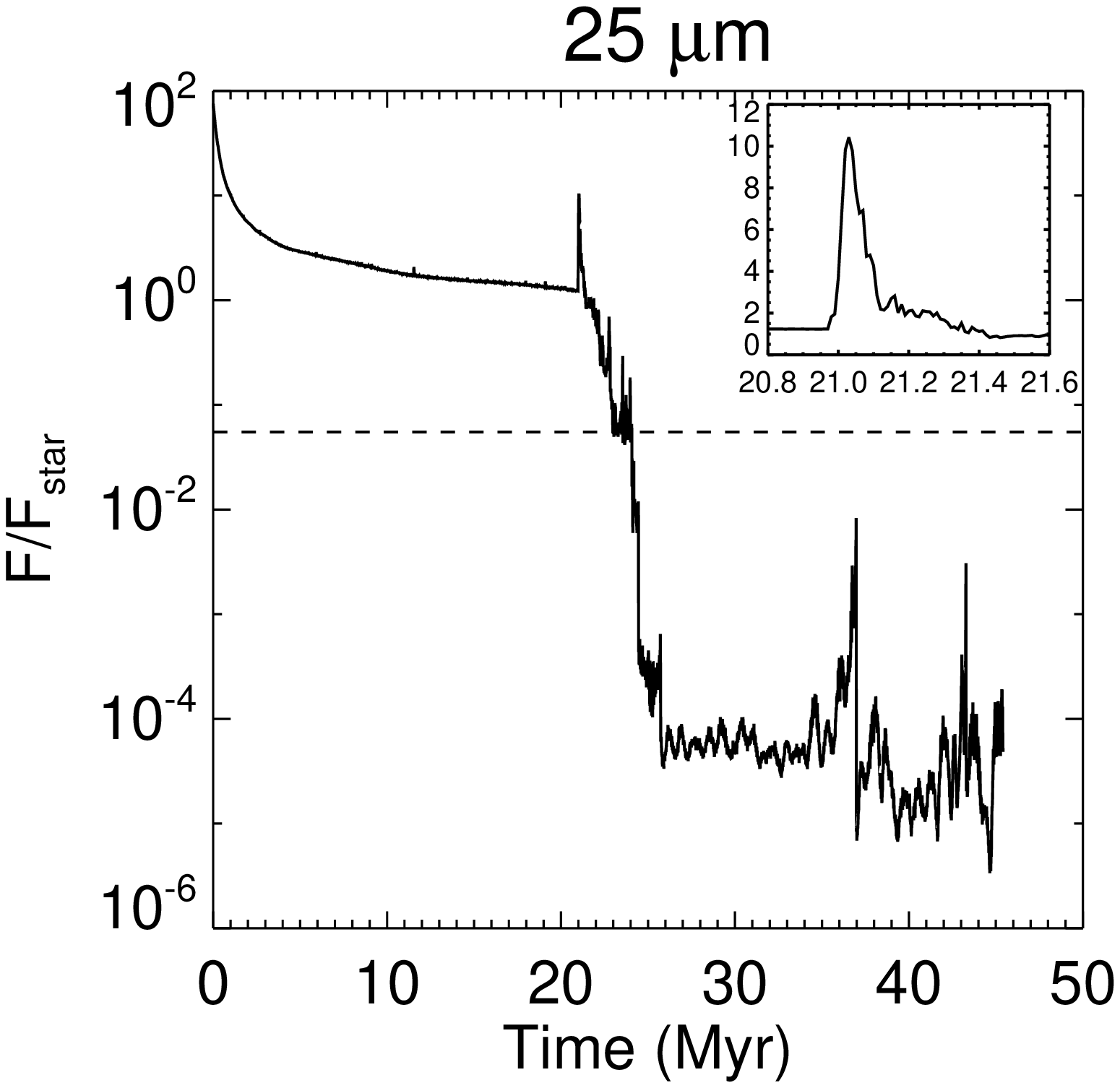}
}
\centerline{
\includegraphics[width=0.3\textwidth]{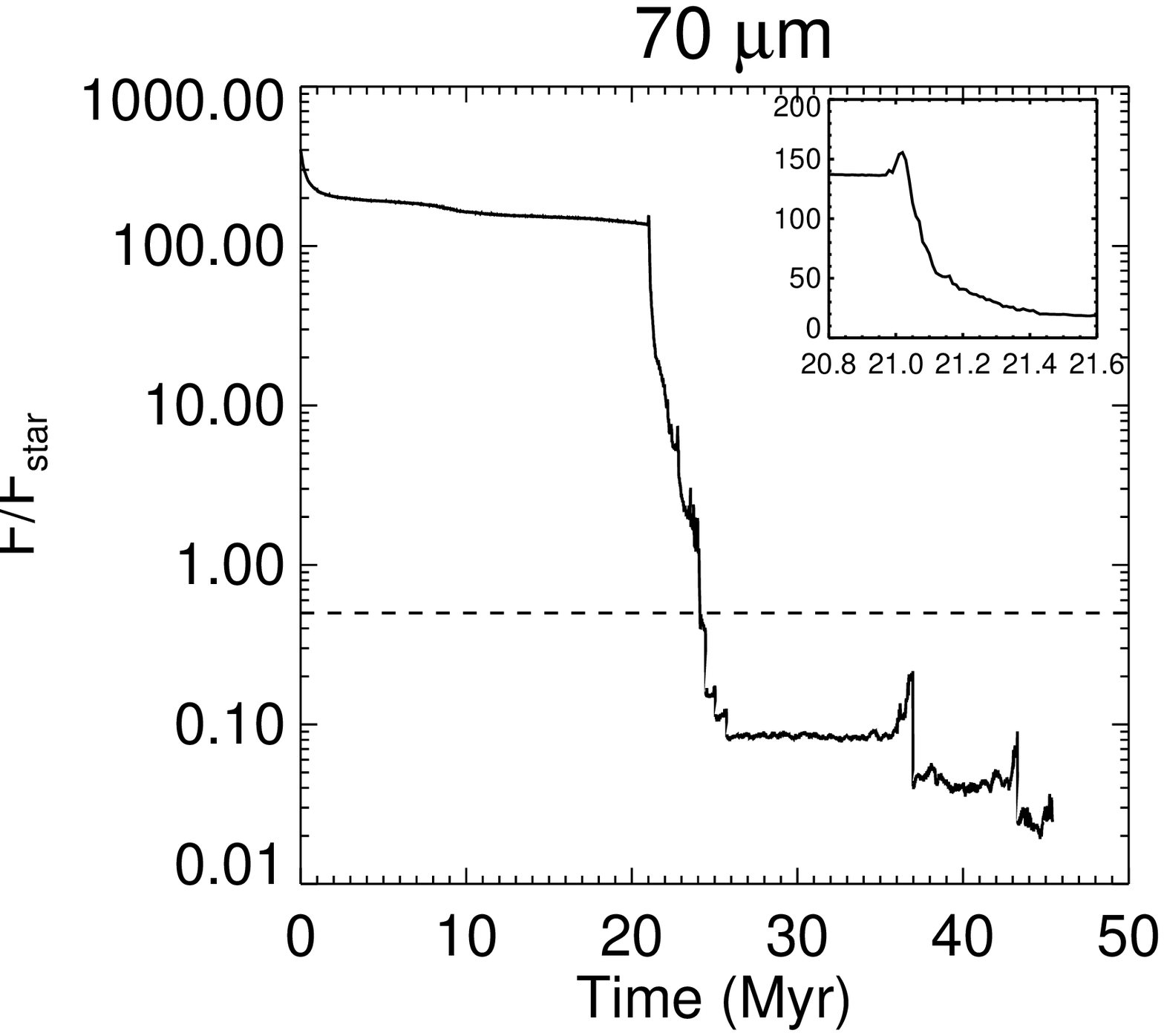}
\includegraphics[width=0.3\textwidth]{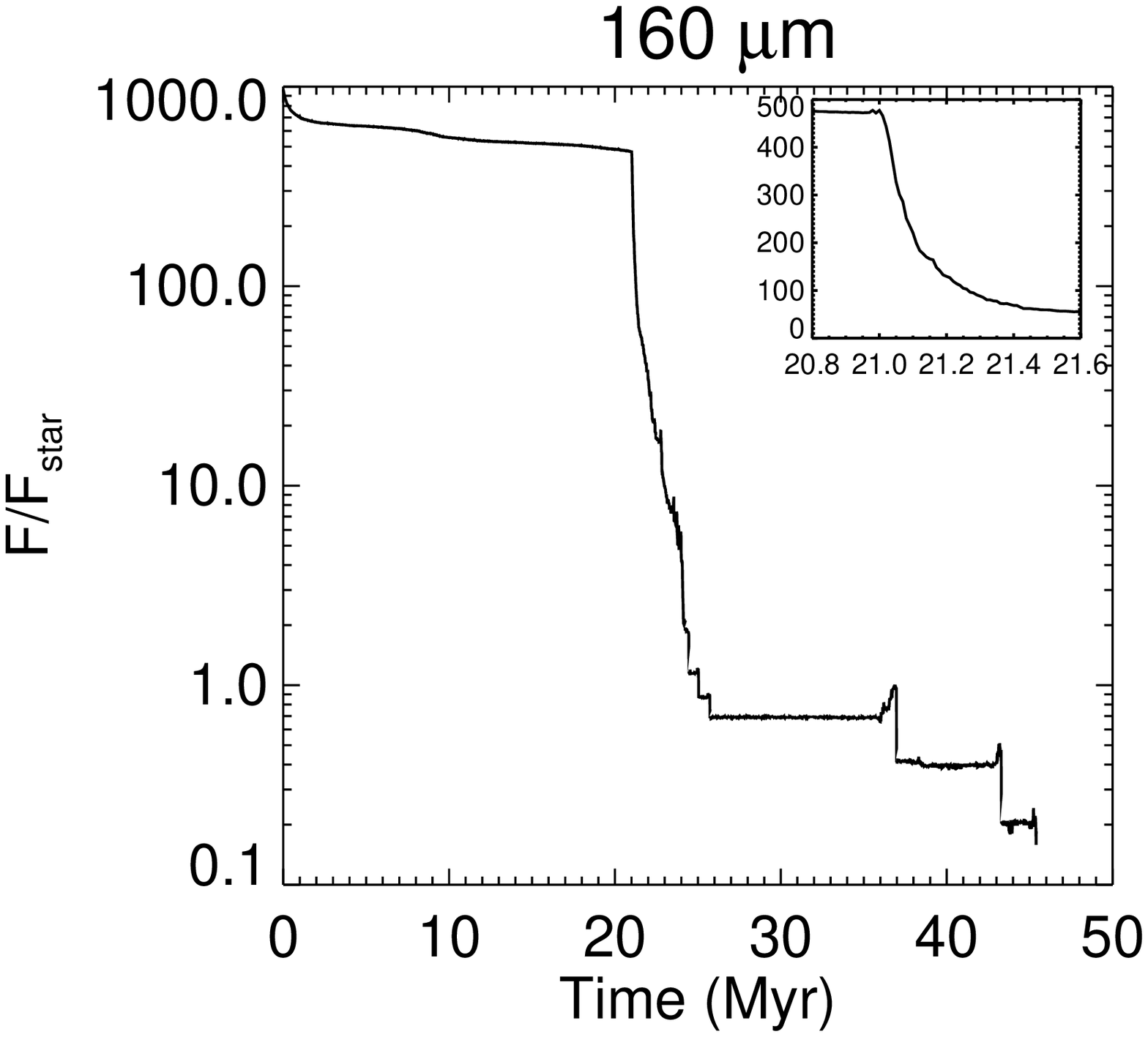}
\includegraphics[width=0.3\textwidth]{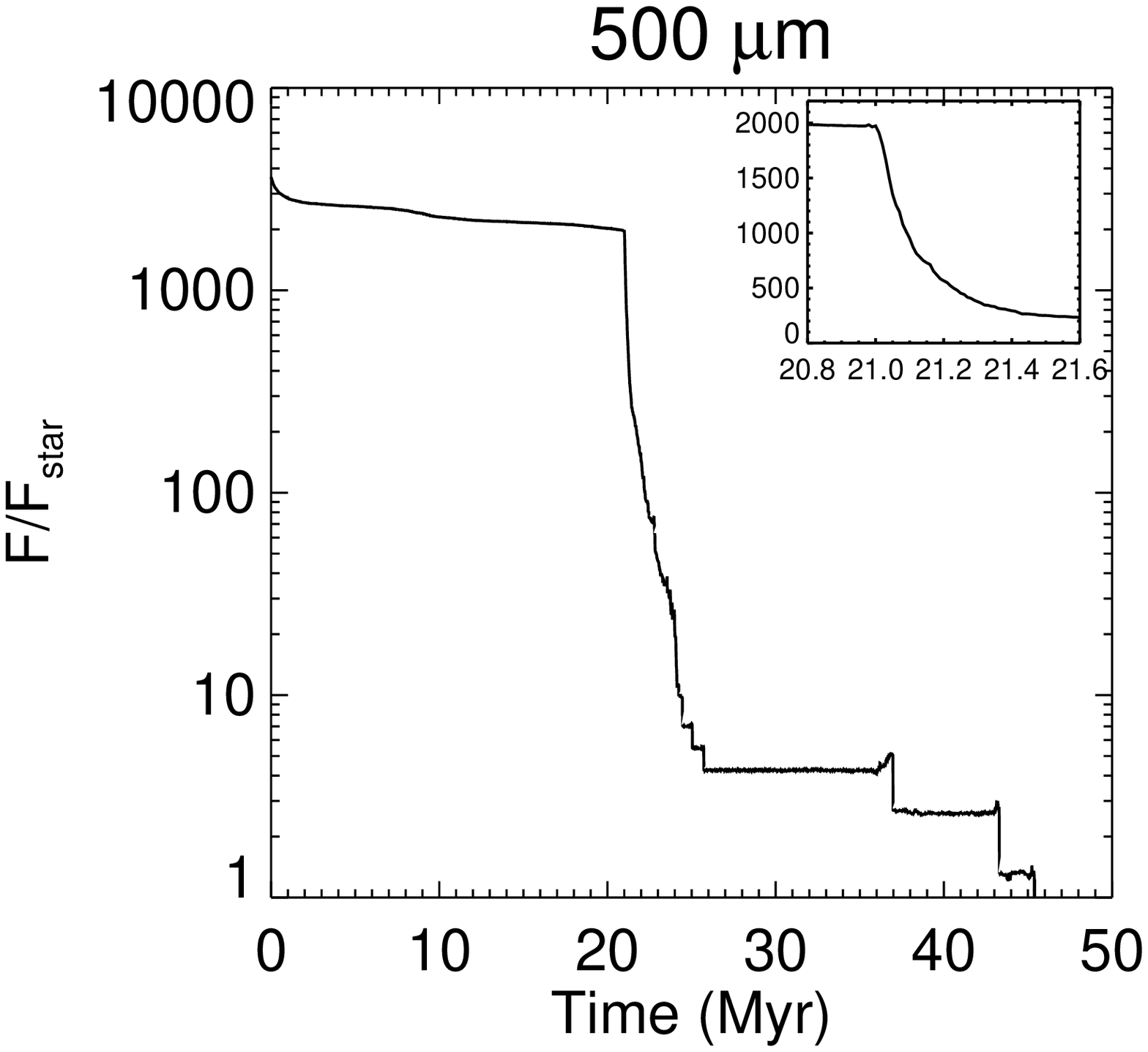}
}
\caption{The ratio of the dust-to-stellar flux as a function of time for six different wavelengths from 5 to 500 microns, including a zoom during the instability -- note that each main plot is on a log scale and each zoom-in is on a linear scale.  The rough observational limits of the {\it MIPS} instrument on NASA's {\it Spitzer Space Telescope} are shown for 25 and 70 $\micron$ with the dashed line~\cite{trilling08}.  All planetesimal particles were destroyed as of 45 Myr via either collision or ejection so there is no more dust in the system after that point.  }
\label{fig:flux-t}
\end{figure*}
When the instability occurs, the spectral energy distribution changes dramatically and quickly (Fig.~\ref{fig:SED}).  Comets from the outer planetesimal disk are scattered onto highly eccentric orbits.  Those that enter the inner Solar System can cause bombardments on the terrestrial planets akin to or often far more intense than the Solar System's late heavy bombardment~\citep{gomes05,strom05}, although given the much earlier timing of most instabilities, these bombardments could act to seed the terrestrial planet-forming region rather than impact fully-formed planets.  The decrease in the comets' perihelia introduces a large amount of warm dust into the system which lasts for the duration of the bombardment and leads to a spike in brightness at near- to mid-infrared wavelengths, shown in Figure~\ref{fig:flux-t} for $\lambda = 5, 10, 25, 70, 160$ and $500 \micron$~\citep[see also][]{booth09,nesvorny10}.  At shorter wavelengths the flux is more jagged than at longer wavelengths because the short wavelength flux is entirely due to hot dust that is produced sporadically by individual planetesimals entering the inner planetary system.  In contrast, the long wavelength flux combines the flux from a much larger range of dust temperatures and therefore includes a much larger number of particles.  The spike in flux associated with the instability is strong for wavelengths shorter than $\sim50 \micron$, but at longer wavelengths the spike is weak or absent.  As objects are dynamically removed from the system the system's brightness drops precipitously and out of the detectable range in the few million years following the instability.  We note that a more realistic model of dust production by new comets suggests that the mid-infrared peak during the bombardment in our model may be underestimated by a factor of a few~\citep{nesvorny10}.

\section{Results for the ensemble of simulations}

In this section we study the outcome of our ensemble of simulations (called {\tt mixed} in paper 2).  This set is of particular interest because the surviving giant planets match the observed eccentricity distribution without any fine-tuning~\citep{raymond09b,raymond10}.  We explore the formation process of terrestrial planets, the orbital characteristics of terrestrial planets that formed, giant planet dynamics, dust production from planetesimals, and correlations between these.  We also compare the simulations with the observed extra-solar giant planets and debris disks.  In paper~2 we explore the effect of a variety of parameters on the process.

\subsection{Giant Planet Instabilities}

Of the 152 simulations that ran for at least 100 Myr and met our energy conservation cutoff (see Appendix A), the giant planets were unstable in 96 systems (63\%).  The instability times ranged from 247 years to 180 Myr after the start of the simulation with a median of 91,600 years.  This is encouraging because it is close to the value expected based on our initial giant planet separations of $4-5 R_{H,m}$~\citep{marzari02,chatterjee08}. This means that our assumption of no disk gas at the start of the instability is marginally acceptable, because the typical timescale for the gas dissipation is thought to be $\sim 10^5$ years~\citep{wolk96,ercolano11}.  However, in many cases instabilities are likely to have occurred in the presence of a residual gas disk.  To test this assumption, we ran an additional set of simulations that include a simple prescription for the effects of gas damping that are presented in paper 2 (and show no significant changes from the gas-free simulations).  We note that later instabilities can certainly occur (e.g., the late heavy bombardment) but we expect the number of instabilities to decrease in time, and the required computing time made it impractical to search for instabilities beyond 100-200 Myr.

Although 75\% of instabilities occur within 1 Myr, there is a tail that extends to longer timescales.  One in six instabilities (16 out of 96) occurred after 10 Myr, one in ten (10/96) after 30 Myr, and one in fifty (2/96) after 100 Myr.  Given that our simulations only lasted 100-200 Myr, we undersampled the fraction of unstable systems in the 100-200 Myr timespan and we expect that even later instabilities should certainly occur in a fraction of systems due to long-term chaotic diffusion.  The timing of the instability is important in terms of the sizes of objects in the inner disk; instabilities later than the typical terrestrial planet formation time of 10-100 Myr may destroy fully-formed Earth-sized planets -- sometimes with appreciable water contents -- rather than embryos and planetesimals.

\begin{figure}%[p]
\center\includegraphics[width=0.4\textwidth]{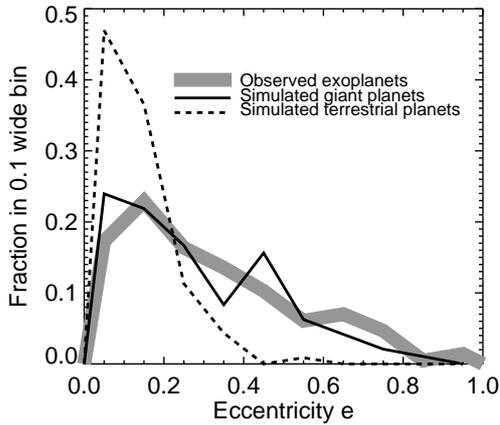}
\caption{Eccentricity distributions of the surviving giant (solid black line) and terrestrial (dashed black line) planets in the {\em unstable} systems.  The grey curve represents the known extra-solar planets beyond 0.2 AU -- this helps to exclude planets whose orbits have been tidally circularized as well as low-mass planets. }
\label{fig:hist_etpgp}
\end{figure}

As shown in Figure~\ref{fig:hist_etpgp}, the surviving giants in the {\em unstable} simulations provide a quantitative match to the observed extra-solar giant planets~\citep[$p = 0.49$ from a Kolmogorov-Smirnov test, consistent with previous work with very similar initial conditions but much larger samples;][]{raymond08b,raymond09a,raymond09b,raymond10}.  Thus, one might imagine that the unstable systems represent the appropriate subsample of simulations that we should use to represent the known exoplanet systems, and that our stable simulations are unrealistic in that they somehow lack a trigger to make them unstable~\citep[e.g.,][]{malmberg10}.  However, as we discuss in section 5.1, the observed exoplanets do appear to require a contribution from dynamically stable systems.  We note that additional constraints exist in the exoplanet sample (e.g., the mass-eccentricity correlation).  Combinations of different sets of simulations can match all of the observed characteristics of the exoplanet distribution, and we perform this exercise in paper 2~\citep[see also section 5 in][]{raymond10}.  

A prediction of the planet-planet scattering model is that most planetary systems should contain multiple giant planets, i.e., additional giant planets should exist exterior to most of the known ones~\citep[e.g.,][]{rasio96,marzari02}.  Likewise, the vast majority of unstable simulations in our sample (85/96 = 89\%) contain multiple giant planets; the remaining 11\% contain just a single surviving giant planet.  The outer giant planet is typically 5-10 AU more distant than the inner giant planet (with a tail to $> 30$ AU), so long-duration observations are needed to follow up the known giant exoplanets.  The distribution of separations between planets in scattered systems may actually provide constraints on their initial mass distribution because the surviving planets tend to be more widely-separated if they are equal-mass than if their masses differ markedly~\citep{raymond09a}.

\subsection{Terrestrial Planet Formation}
The number and spacing of terrestrial planets that form is governed by the eccentricities of the planetary embryos from which they form~\citep{levison03}.  These eccentricities are a result of gravitational forcing both from nearby embryos and the giant planets.  There are two differences in terrestrial planet formation between stable and unstable systems: 1) perturbations during the instability can play a major role in shaping the embryo distribution; and 2) the dynamical state of the giant planets after an instability is generally more excited than before the instability.   In some unlucky cases the surviving giant planets provide an accretion-friendly environment that is empty because the instability has already removed all rocky bodies.  Past simulations of terrestrial planet formation with giant planets on excited orbits have all neglected the planet-planet scattering phase during which the giant planets actually acquired their eccentricities~\citep{chambers02,levison03,raymond04,raymond06a}, which clearly plays a very important role in the dynamics. 

Giant planet perturbations span a continuous range but the outcome is quantized into a discrete number of terrestrial planets during the accretion process.  If perturbations are weak -- if the giant planets collide rather than scattering (or are dynamically stable or low-mass) -- then embryos' eccentricities remain small and feeding zones narrow and several terrestrial planets form.  For stronger giant planet perturbations, feeding zones widen and fewer terrestrial planets form, although the total mass in planets tends to decrease because stronger perturbations imply that the giant planets were scattered closer to the terrestrial planet region so more embryos end up on unstable orbits.  In systems where embryos' radial excursions are comparable to the radial extent of the surviving disk only one planet forms, usually on an excited orbit.  In the simulation from Fig.~\ref{fig:aeit}, the lone terrestrial planet did not accrete from a disk of excited embryos but rather was the only planet to {\em survive} the instability.  Perturbations during, not after, the instability determined the outcome.  

\begin{figure}%[p]
\center\includegraphics[width=0.4\textwidth]{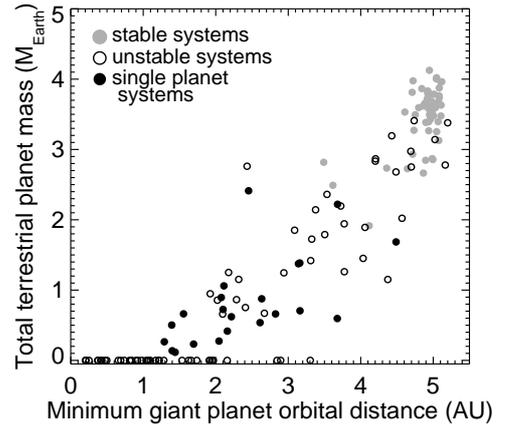}
\caption{The total mass in surviving terrestrial planets as a function of the minimum perihelion distance of a giant planet during the simulation.  Black dots represent systems in which the giant planets were dynamically unstable and grey dots are systems that were stable.  Filled black dots are systems in which a single terrestrial planet formed.  }
\label{fig:mterr-minperi}
\end{figure}

The strength of the giant planet perturbations is directly related to the smallest giant planet perihelion distance $q_{GP,min}$.  Figure~\ref{fig:mterr-minperi} shows that the efficiency of terrestrial planet formation is directly related to $q_{GP,min}$.  This is not surprising: planets with smaller $q_{GP,min}$ have higher eccentricities (since all giant planets start with the same range of semimajor axes) and have therefore undergone more violent scattering events.    Every system with a giant planet scattered to $q_{GP,min} < 1.3$ AU destroyed all terrestrial material in the system, as did some simulations out to 3 AU.  This is consistent with the results of~\cite{veras06}, who found a similar scaling between $q_{GP,min}$ and the survival of test particles in the inner disk.  As expected, systems that form a single terrestrial planet are those with giant planet perturbations that are almost strong enough to completely empty the terrestrial material from the system.  In these cases the embryos' eccentricities are large and, as we will see below, the single planet that survives maintains an excited orbit.

\subsection{Characteristics of surviving terrestrial planets}
\begin{figure}%[p]
\center\includegraphics[width=0.4\textwidth]{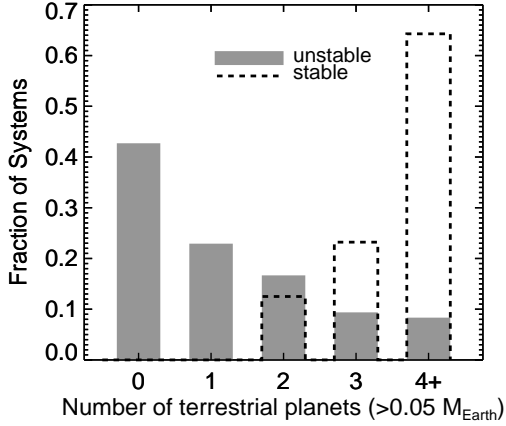}
\caption{Distribution of the number of surviving terrestrial planets (defined as $M_p > 0.05 \ M_\oplus$) in our fiducial {\tt mixed} set of simulations. The grey shaded histogram shows the unstable simulations and the dashed line shows the stable simulations.  
}
\label{fig:hist-nterr}
\end{figure}

All terrestrial material was destroyed in 41 out of 96 unstable simulations (43\%; Figure~\ref{fig:hist-nterr}).   Likewise, 22 unstable systems (23\%) formed a single terrestrial planet, 16 formed two terrestrial planets (17\%), and and the remaining 17 systems (18\%) formed three or more terrestrial planets.  Among the 56 stable simulations, only 7 (12.5\%) formed two planets, and the remaining 87.5\% of simulations formed three or more planets.  

As a population, the surviving terrestrial planets have smaller eccentricities than the giant planets (Figure~\ref{fig:hist_etpgp}); the median eccentricity is 0.10 for terrestrials and 0.21 for giants.  The terrestrial planets that formed in systems with unstable giant planets have only slightly more eccentric and inclined orbits than the terrestrial planets that formed in stable systems.  The median $e$ and $i$ were 0.10 and 5.1$\deg$ for the unstable systems and 0.08 and 2.5$\deg$ for the stable systems, respectively.  The single-terrestrial planet systems were the most excited, with median $e$ of 0.14 and $i$ of 8.7$\deg$.  If we neglect the single-terrestrial planet systems, the eccentricities and inclinations of the terrestrial planets that formed in stable and unstable systems is very close.\footnote{We expect that numerical resolution should play a role here.  Given that they contain only 500 planetesimal particles, our simulations cannot fully resolve dynamical friction at late times~\citep[e.g.][]{obrien06,raymond06b}.  We expect that eccentricities and inclinations are somewhat overestimated in the stable systems, but since the instabilities remove most of the planetesimals, not in unstable systems.}

\begin{figure}%[p]
\center\includegraphics[width=0.4\textwidth]{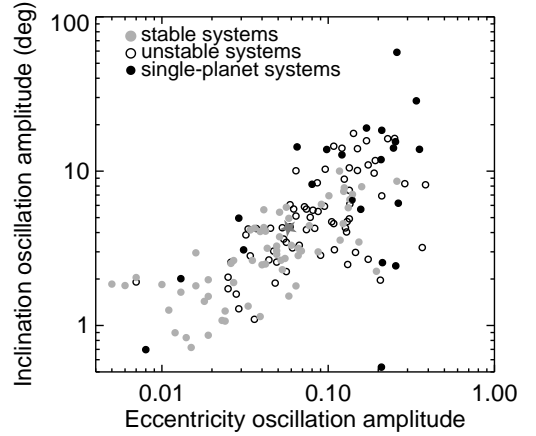}
\caption{The peak-to-peak oscillation amplitudes in eccentricity $e$ and inclination $i$ for the surviving terrestrial planets in simulations with stable (grey) and unstable (black) giant planets. The single-terrestrial planet systems are shown with filled black circles, and Earth is the grey star.}
\label{fig:eiosc}
\end{figure}

Although their time-averaged orbits are similar, terrestrial planets in unstable systems undergo much stronger orbital oscillations (Figure~\ref{fig:eiosc}).  The median peak-to-peak eccentricity and inclination oscillation amplitudes are 0.11 and 5$\deg$ for the unstable systems and 0.043 and 2.8$\deg$ for the stable systems, respectively.  Again, the single-terrestrial planets are the most excited and undergo the largest oscillations, with median $e$ and $i$ amplitudes of 0.21 and 11.9$\deg$.  The climates of the single planets are likely to vary dramatically on the secular timescale of $10^3-10^6$ years~\citep{spiegel10}.

\subsection{Correlations with observable quantities: giant planets and debris disks}
\begin{figure}%[p]
\center\includegraphics[width=0.4\textwidth]{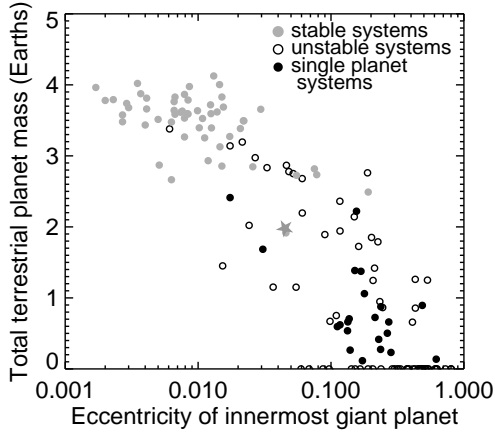}
\caption{The total mass in surviving terrestrial planets as a function of the eccentricity of the innermost surviving giant planet.  The Solar System is shown as the grey star.  }
\label{fig:eg-mterr}
\end{figure}

Observations of massive exoplanets can only very roughly diagnose the outcome of terrestrial accretion.  Figure~\ref{fig:eg-mterr} shows a strong negative correlation between the efficiency of terrestrial planet formation and the eccentricity of the innermost giant planet $e_g$.  For stable systems $e_g$ tends to be very small (median $e_g = 0.008$) while for the unstable systems $e_g$ spans a very wide range, from $<$0.01 to 0.8 (median $e_g = 0.21$).  The correlation between the total terrestrial planet mass and $e_g$ is expected but the range in terrestrial planet mass for different systems with a similar $e_g$ shows the importance of the orbital history. 

\begin{figure}%[p]
\center\includegraphics[width=0.4\textwidth]{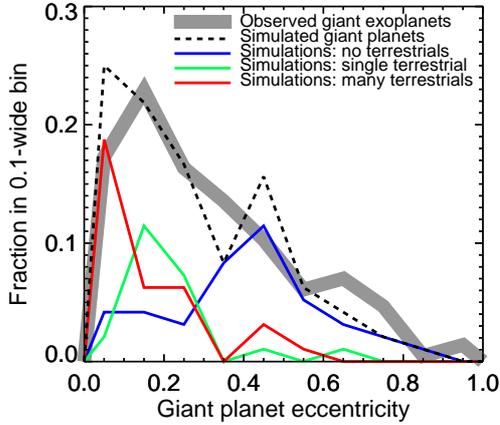}
\caption{Distribution of eccentricities of the innermost giant planet in simulations which formed zero (blue line), one (green), and two or more (red) terrestrial planets.  The sum of the three distributions (black dashed line) provides a quantitative match ($p$=0.49 from a Kolmogorov-Smirnov test) to the observed exoplanets (thick grey line).  }
\label{fig:hist_ecc_nterr}
\end{figure}

The orbits of surviving giant planets retain a memory of the strength of the instability, or lack thereof.  Figure~\ref{fig:hist_ecc_nterr} breaks down the giant planet eccentricity distribution by outcome in terms of the number of terrestrial planets that formed.  Multiple terrestrial planets form preferentially in systems with low giant planet eccentricities, because these represent very weak instabilities.  By the same argument, highly eccentric giant planets tend to destroy all terrestrial material in their systems.  The intermediate regime is represented by the single-terrestrial planet systems; for these systems $e_g$ is typically in the range 0.1-0.3 (median of 0.17).  These eccentricities are close to the median of the exoplanet distribution~\citep{butler06,udry07b} and there is considerable overlap from systems with zero or many terrestrial planets.  It is therefore very difficult to diagnose a single-planet system from a measurement of just the eccentricity of a giant exoplanet.  

\begin{figure}%[p]
\center\includegraphics[width=0.4\textwidth]{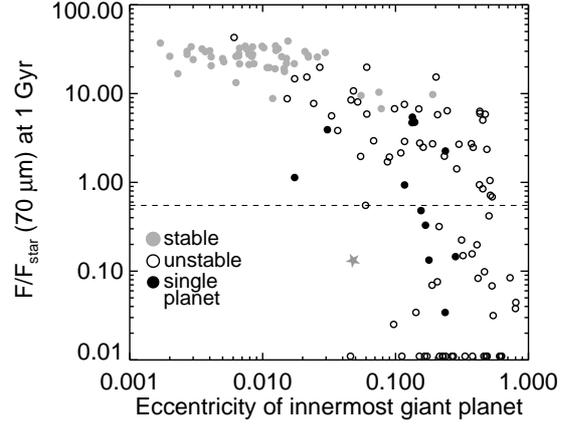}
\caption{The dust-to-stellar flux ratio at 70~$\mu$m after 1~Gyr of dynamical and collisional evolution, plotted as a function of the eccentricity of the innermost surviving giant planet. Unstable systems are plotted in black and stable systems in grey (systems with a single surviving planet are shown with solid circles).  The pileups very close to the $x$ axis represent systems with virtually zero 70~$\mu$m flux. The dashed line represents an approximate threshold above which excess emission was detectable using {\em Spitzer} data~\citep{trilling08}.  The star shows the estimated flux from the Kuiper Belt at 1~Gyr, as calculated previously~\citep{booth09} based on dynamical simulations of the outer Solar System~\citep{gomes05}. }
\label{fig:ff70-eg}
\end{figure}

The outer disk's evolution is also governed by giant planet perturbations.  Icy planetesimals -- whose collisional erosion creates debris disks -- survive in dynamically calm environments where the giant planets were either stable, low-mass, or underwent a relatively weak instability.  Indeed, Figure~\ref{fig:ff70-eg} shows a strong anti-correlation between the $70 \micron$ dust flux and the giant planet eccentricity.  This arises simply because eccentric giant planets have increased apocenter distances and impinge on the planetesimal disk, thereby dynamically removing planetesimals via ejection.  In addition, if planetesimals survive on highly-eccentric orbits their collisional lifetimes may be shortened~\citep{wyatt10}.  

\begin{figure}%[p]
\center\includegraphics[width=0.4\textwidth]{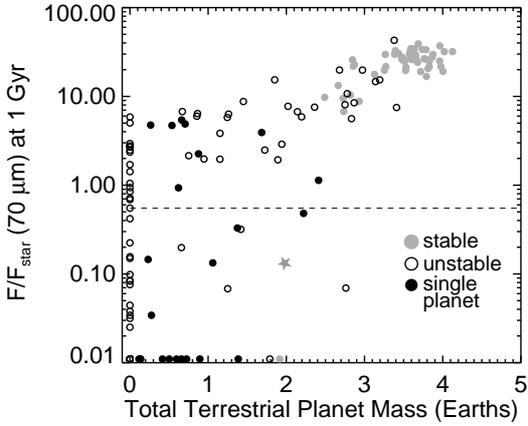}
\caption{The dust-to-stellar flux ratio at $70 \micron$ after 1~Gyr of dynamical and collisional evolution, plotted as a function of the total mass in terrestrial planets. Unstable systems are plotted in black and stable systems in grey.  The pileups close to the $x$ and $y$ axes represent systems with either no terrestrial planets or virtually zero $70 \micron$ flux. The dashed line represents an approximate threshold above which excess emission was detectable using {\em Spitzer} data~\citep{trilling08}.  The star shows the estimated flux from the Kuiper Belt at 1~Gyr, as calculated previously~\citep{booth09} using models based on dynamical simulations of the outer Solar System~\citep{gomes05}. The Solar System falls into an intermediate region of parameter space: the giant planets may have been unstable, but only weakly so (there is no clear evidence for ejections or planetary collisions), while the Kuiper Belt would have been bright for several hundred Myr prior to the Late Heavy Bombardment.}
\label{fig:ff70-mterr}
\end{figure}

Given that the efficiency of terrestrial planet formation anti-correlates with the giant planet eccentricity (Fig.~\ref{fig:eg-mterr}) and the dust flux also anti-correlates with the giant planet eccentricity, Figure~\ref{fig:ff70-mterr} shows that the efficiency of terrestrial planet formation correlates with the dust flux.  Stars with bright dust almost all contain terrestrial planets: the median system for which $F/F_{star} > 10$ contains $3.6 \mearth$ in terrestrial planets, and every single system contains at least $1.85 \mearth$ in terrestrial planets.  Systems that are extremely bright at long wavelengths should therefore be considered prime targets in the search for terrestrial planets.  

The debris disk-terrestrial planet correlation seen in Fig.~\ref{fig:ff70-mterr} exists because the inner and outer planetary system are subject to the same dynamical environment: the violent instabilities that abort terrestrial planet formation also tend to remove or erode their outer planetesimal disks.  As was the case for the giant planet eccentricities, single-terrestrial planet systems populate the intermediate area of the correlation and overlap with systems with no planets as well as those with several.  The terrestrial planet - debris disk correlation is not perfect as there exist ``false positives'' with bright dust emission and no terrestrial planets, corresponding to systems with asymmetric, inward-directed giant planet instabilities. Conversely, ``false negatives'' with terrestrial planets but little to no dust are systems that underwent asymmetric but outward-directed instabilities.  We discuss these caveats in detail in paper~2 when we compare our simulations with the observed debris disk statistics and use them to derive a crude estimate of $\eta_{Earth}$.  

\begin{figure*}%[p]
\centerline{
\includegraphics[width=0.3\textwidth]{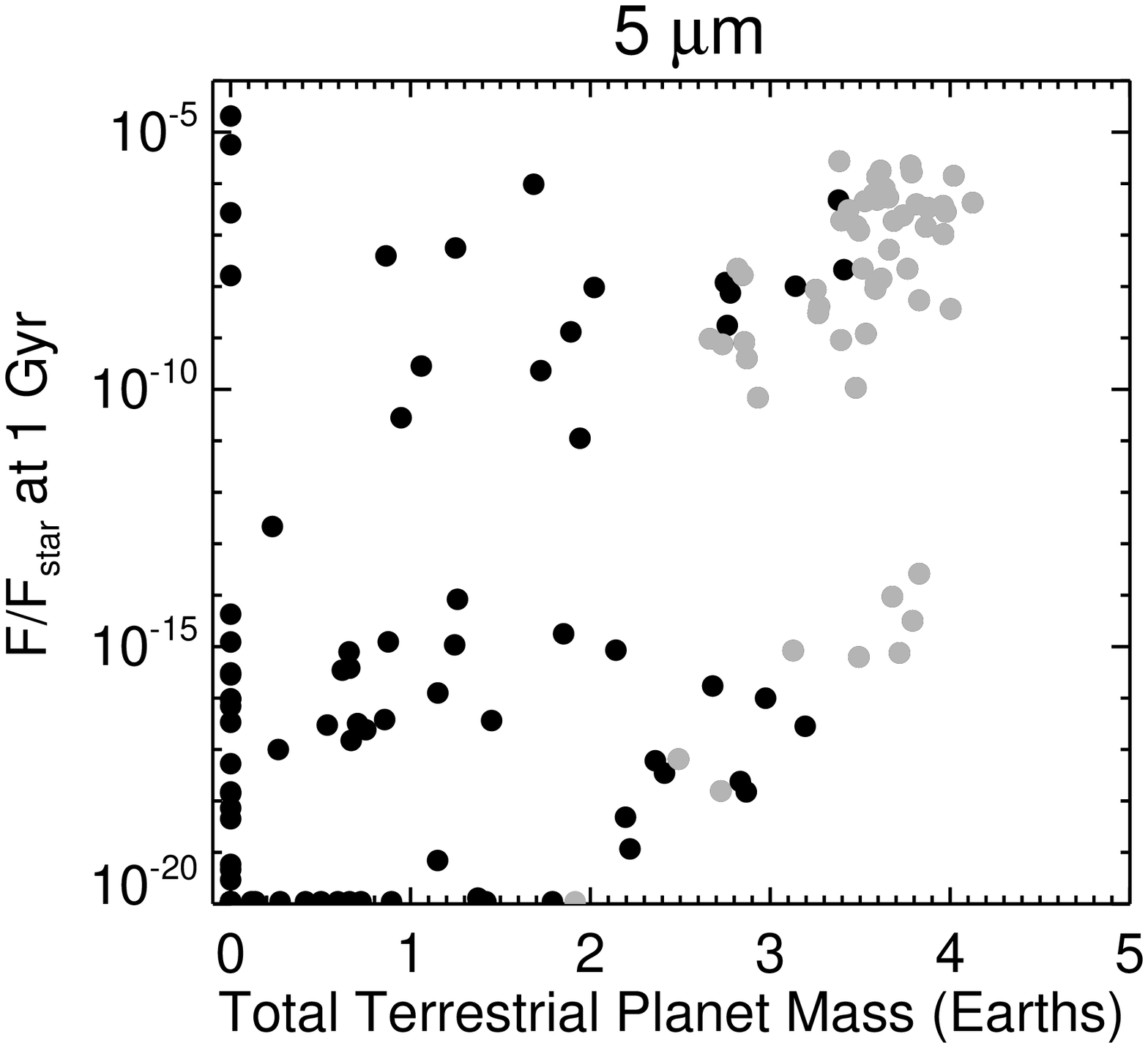}
\includegraphics[width=0.3\textwidth]{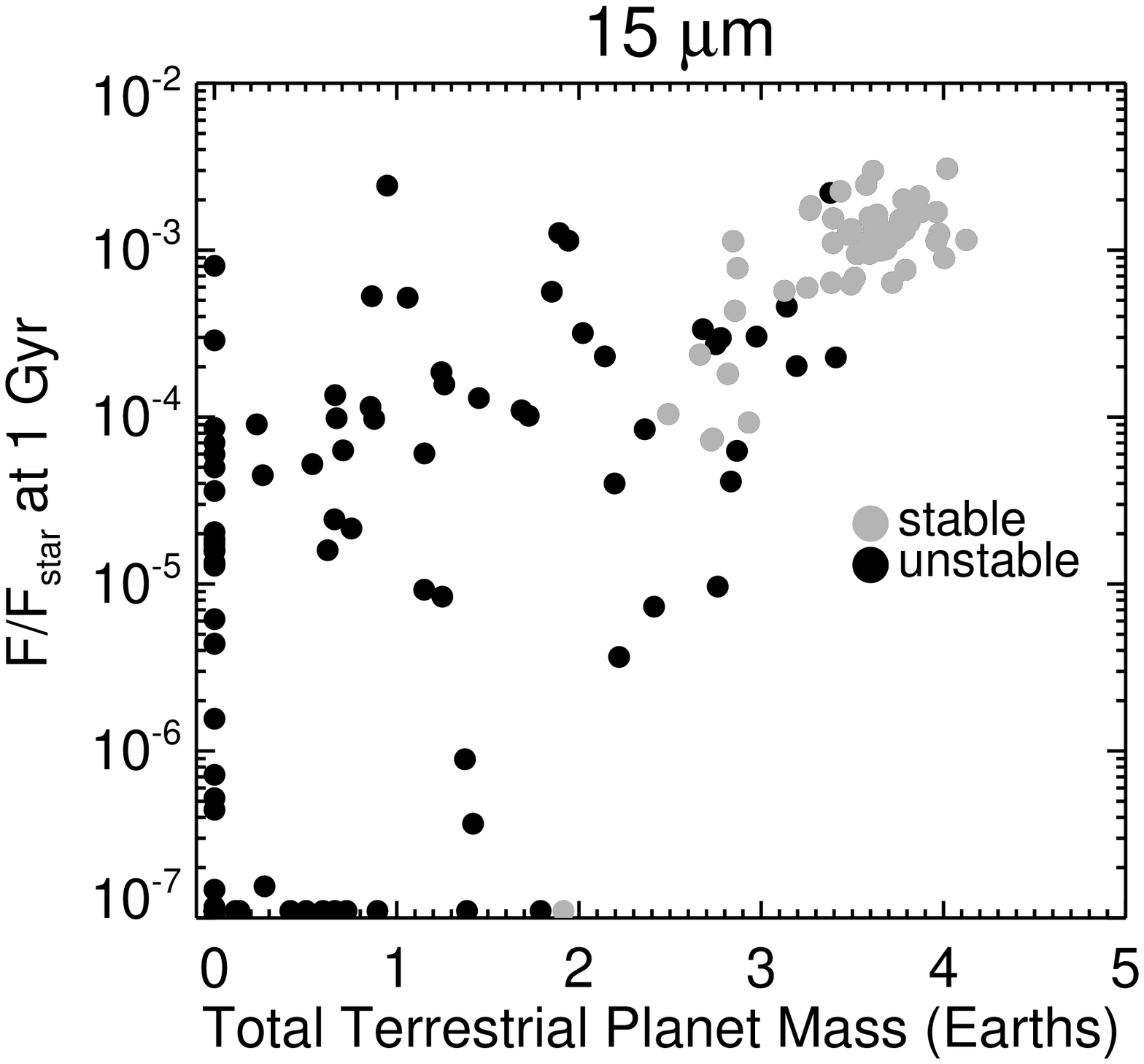}
\includegraphics[width=0.3\textwidth]{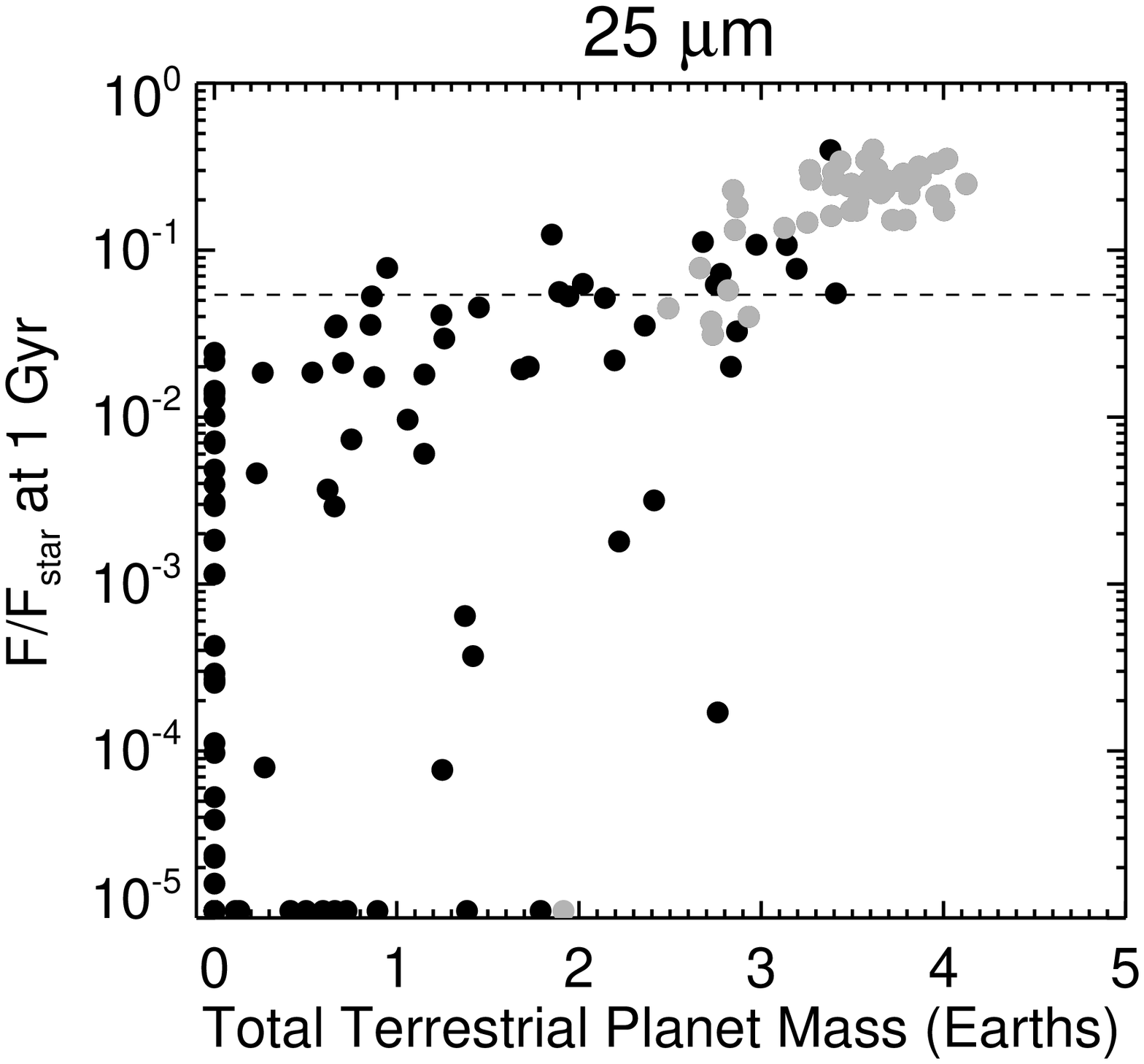}
}
\centerline{
\includegraphics[width=0.3\textwidth]{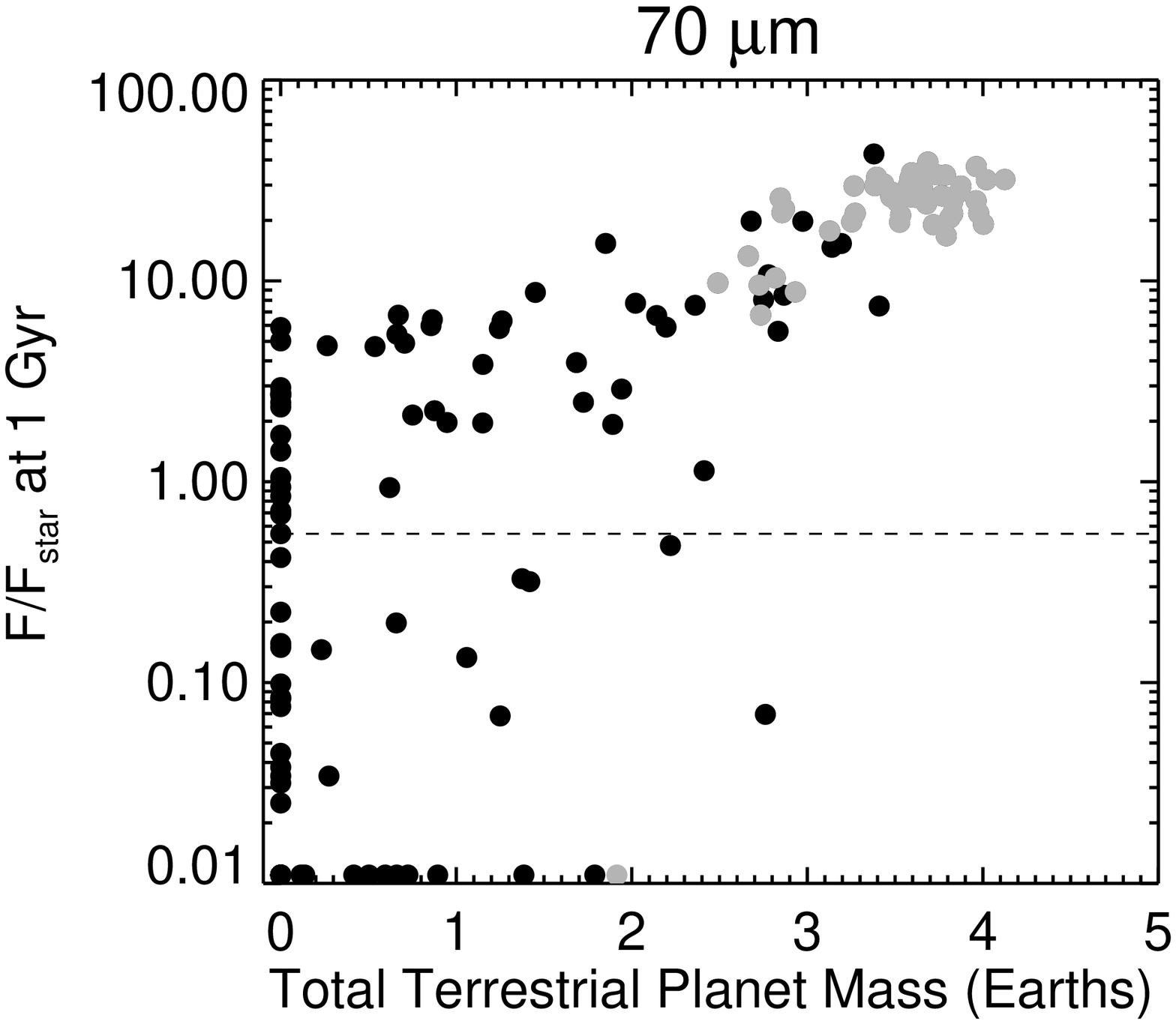}
\includegraphics[width=0.3\textwidth]{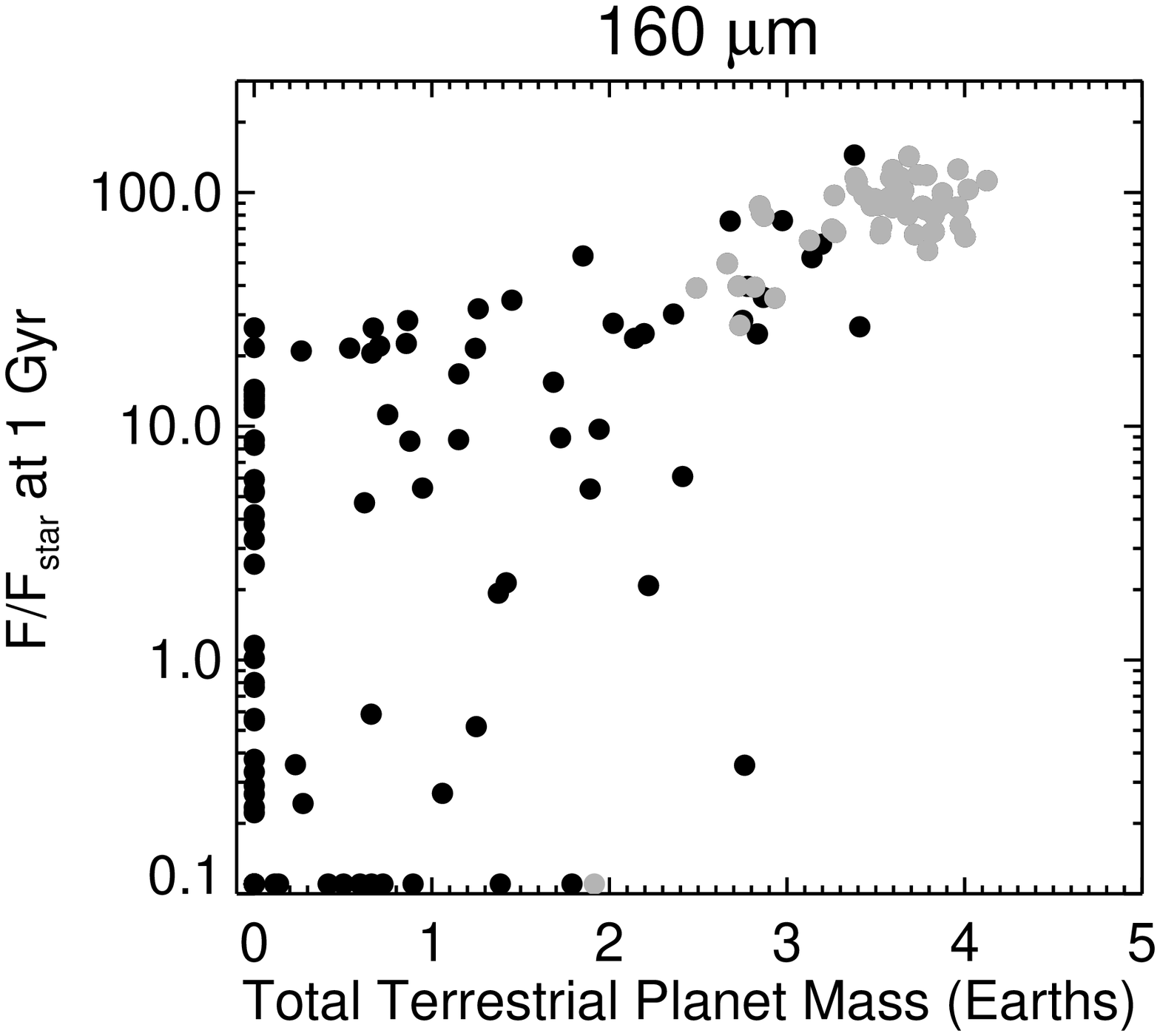}
\includegraphics[width=0.3\textwidth]{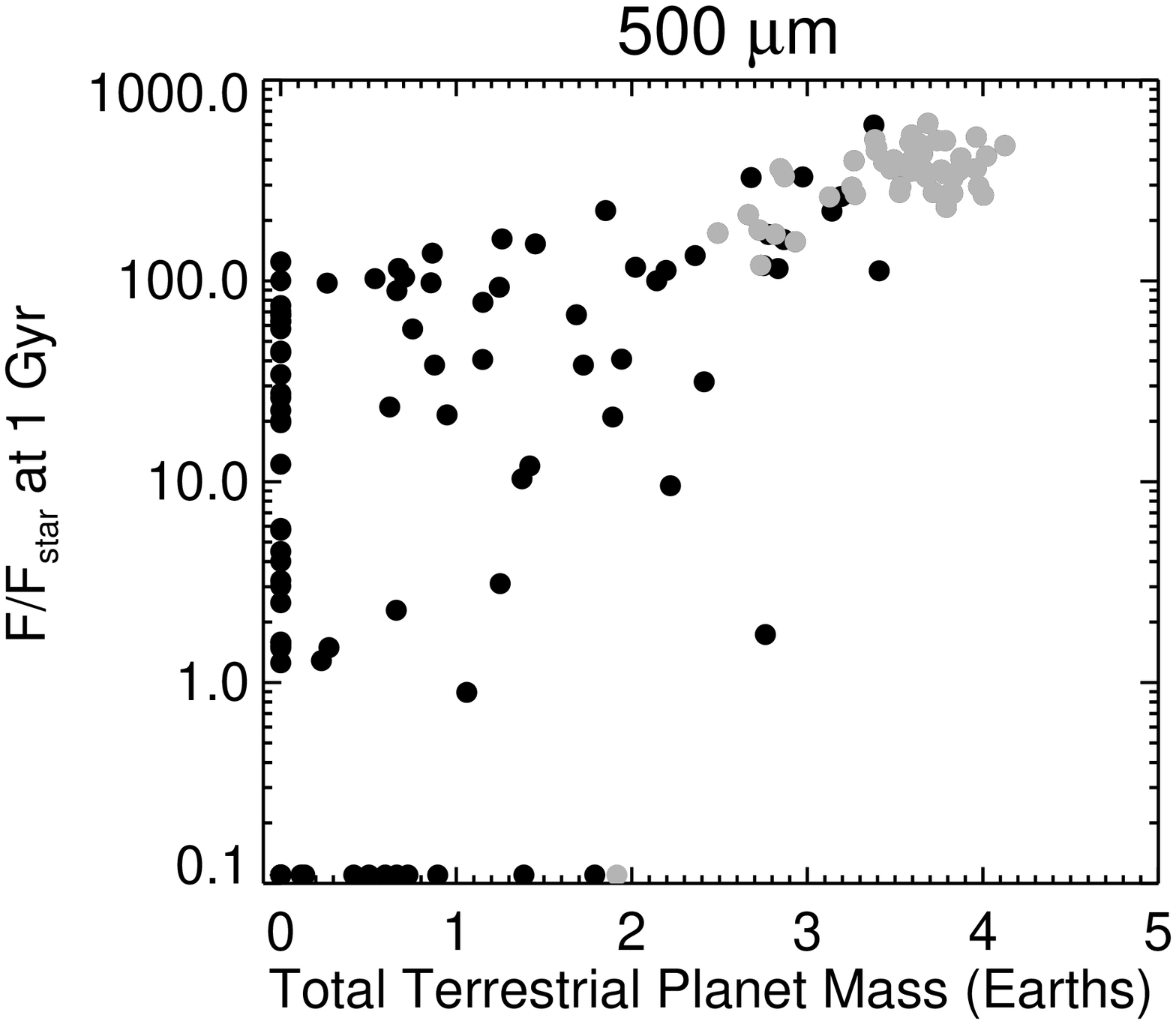}
}
\caption{The dust-to-stellar flux ratio $F/F_{star}$ after 1~Gyr of dynamical and collisional evolution as a function of the total mass in terrestrial planets, for six wavelengths between 5 and 500 microns.  The {\it Spitzer} observational limits are shown for 25 and 70 $\micron$ with the dashed line~\citep{trilling08}.  In each panel, grey circles represent stable simulations and black circles represent unstable simulations.  }
\label{fig:ff-mterr-wav}
\end{figure*}

The debris disk-terrestrial planet correlation is a function of wavelength.  Figure~\ref{fig:ff-mterr-wav} shows the correlation at 1 Gyr for six wavelengths between 5 and 500 $\micron$.  At $5 \micron$ there is no correlation but mainly a scatter plot.  The only hint of a correlation is a lower envelope of dust fluxes larger than $F/F_{star} \gtrsim 10^{-20}$ for systems with more than $\sim 2 \mearth$ in terrestrial planets.  This non-correlation comes from the fact that the short-wavelength flux can be dominated by either a small amount of hot dust or a large amount of colder dust.   
At $15 \micron$ the debris disk-terrestrial planet correlation clearly exists, although there are still some outliers with large fluxes and small terrestrial planet masses.  At longer wavelengths these outliers slowly disappear because the signal from small amounts of transient hot dust is overwhelmed by the large amount of cold dust produced in quiescent outer planetesimal disks.  The debris disk-terrestrial planet correlation is evident for all wavelengths longer than $\sim 25 \micron$.  

\begin{figure*}%[p]
\includegraphics[width=0.48\textwidth]{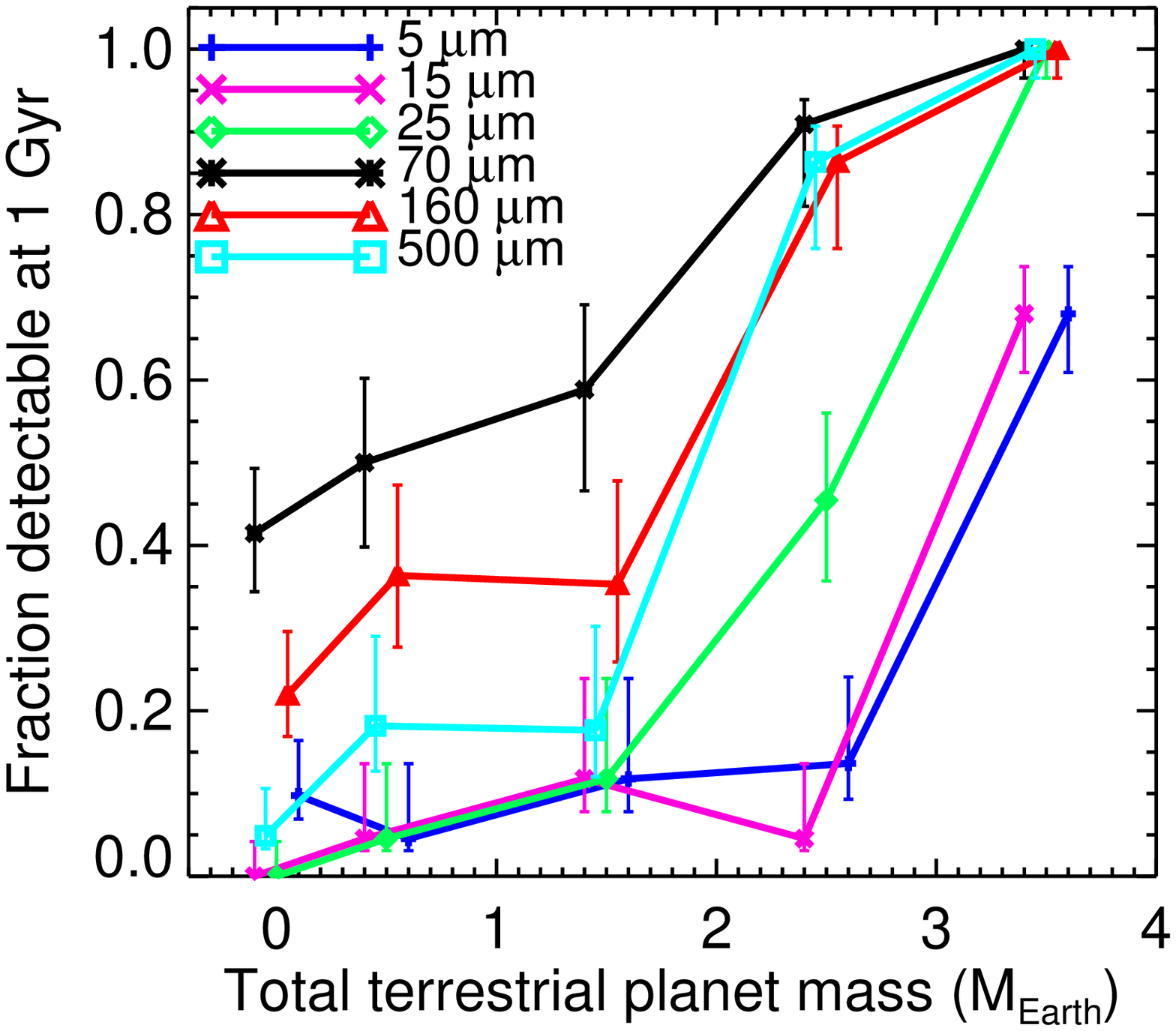}
\includegraphics[width=0.48\textwidth]{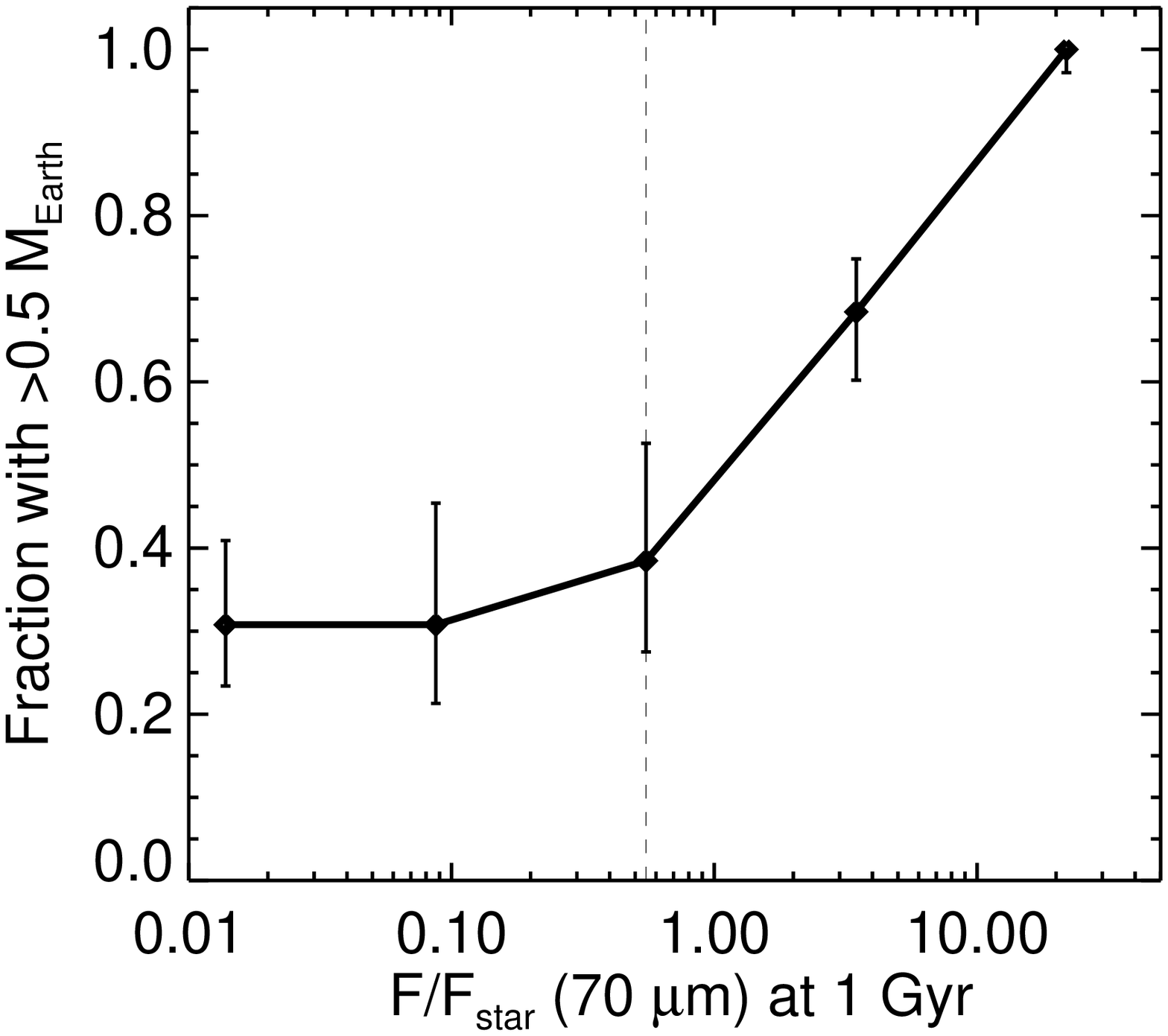}
\caption{{\bf Left:} The fraction of systems that would be detectable after 1 Gyr as a function of the total mass in surviving terrestrial planets for six different wavelengths between 5 and $500 \micron$.  Each curve represents a horizontal slice through Fig~\ref{fig:ff-mterr-wav} at a different vertical height.  The chosen ``detection limits'' in $F/F_{star}$ were: $10^{-8}$ at $5 \micron$, $10^{-3}$ at $15 \micron$, 0.054 at $25 \micron$, 0.55 at $70 \micron$, 10 at $160 \micron$, and 100 at $500 \micron$.  The error bars are 1-$\sigma$ and were calculated using binomial statistics~\citep{burgasser03}.  Note that the different curves are offset by up to $\pm 0.1 \mearth$ for clarity.  The bins themselves are at: zero, $0-1 \mearth$, $1-2 \mearth$, $2-3 \mearth$, and $> 3 \mearth$.  {\bf Right:} The fraction of systems with $0.5 \mearth$ or more in terrestrial planets as a function of $F/F_{star} (70 \micron)$ (1 Gyr).  Systems with $F/F_{star} < 10^{-2}$ are included in the bin at $F/F_{star} \approx 10^{-2}$.  The Spitzer detection limit is shown as the dashed line.  This represents a vertical slice through Fig.~\ref{fig:ff70-mterr}. }
\label{fig:slice-mterr}
\end{figure*}

Figure~\ref{fig:slice-mterr} (left panel) shows histograms that represent horizontal slices through Fig.~\ref{fig:ff-mterr-wav}.  For each wavelength we chose a ``detection limit'' and tabulated the fraction of systems that were deemed detectable as a function of the total terrestrial planet mass.   These detection limits were chosen for illustrative purposes and are in many cases unrealistic.   For example, the current detection threshold at $5 \micron$ is probably closer to $\sim 10^{-3}$, but none of our simulations would be detectable at that limit.  The reader is referred to other work discussing current dust detection limits at short~\citep{absil06,akeson09} and long~\citep{smith09} wavelengths~\citep[see also][]{wyatt08}.  The fraction of systems with no terrestrial planets that is detectable varies significantly between the different wavelengths and simply reflects the location of the detection limit with respect to the general trends in Fig.~\ref{fig:ff-mterr-wav}.  The detection limits at $25 \micron$ and $70 \micron$, which correspond to the actual {\it Spitzer} limits~\citep{trilling08}, are close to the extremes: at $25 \micron$ not a single terrestrial planet-free system is detectable while at $70 \micron$ more than 40\% of all systems are detectable.  

At each wavelength, the fraction of systems that is detectable increases with the terrestrial planet mass (Fig.~\ref{fig:slice-mterr}).  At all wavelengths the trend is relatively flat to a certain point where it increases significantly, by several standard deviations between adjacent bins.  At all wavelengths this transition occurs between either the $1-2 \mearth$ and $2-3 \mearth$ bins or the $2-3 \mearth$ and the $>3 \mearth$ bin.  Even at $5 \micron$, which showed no obvious debris disk-terrestrial planet correlation in Fig.~\ref{fig:ff-mterr-wav}, there is marked jump in the fraction of detectable objects in the last bin.  This jump is also seen at $15 \micron$, although these two shortest wavelengths are the only ones for which some systems with more than $3 \mearth$ in planets are not detectable.  Although our chosen detection limits at certain wavelengths may be overly optimistic or pessimistic (e.g., $F/F_{star} (5 \micron) \geq 10^{-8}$ is likely to be difficult to achieve in the near term), this shows that the debris disk-terrestrial planet correlation is present for all of these wavelengths.

The right panel of Fig.~\ref{fig:slice-mterr} shows the fraction of systems with significant mass in terrestrial planets as a function of $F/F_{star}$ at $70 \micron$.  This plot represents a {\em vertical} slice through Fig.~\ref{fig:ff70-mterr}.  At small dust fluxes, a minority of systems contain terrestrial planets -- these are referred to in section 5 as ``false negatives''.  The fraction of systems with terrestrial planets increases dramatically beyond the detection limit, and {\em every single one} of the 58 systems with $F/F_{star} > 10$ (including 7 unstable systems) contains at least $1.9 \mearth$ in terrestrial planets, with a median value of $3.5 \mearth$.  Of the 53 unstable systems above the detection threshold at $70 \micron$, 35 ($66\%^{+5.8\%}_{-6.9\%}$) contain terrestrial planets.  Thus, in our simulations debris disks appear to represent signposts for terrestrial planets with a confidence of at least 66\% at $70 \micron$.

\begin{figure*}%[p]
\includegraphics[width=0.48\textwidth]{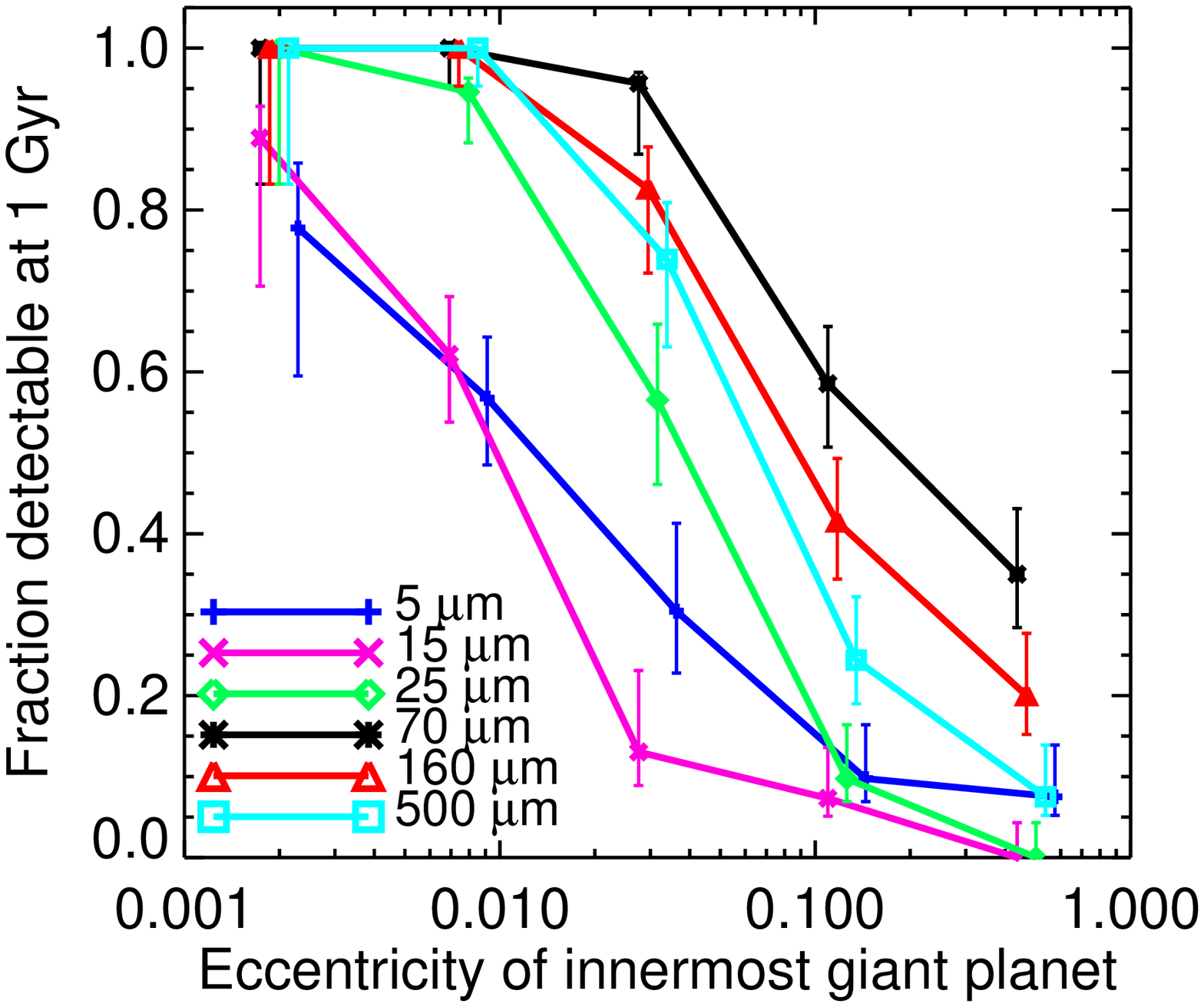}
\includegraphics[width=0.48\textwidth]{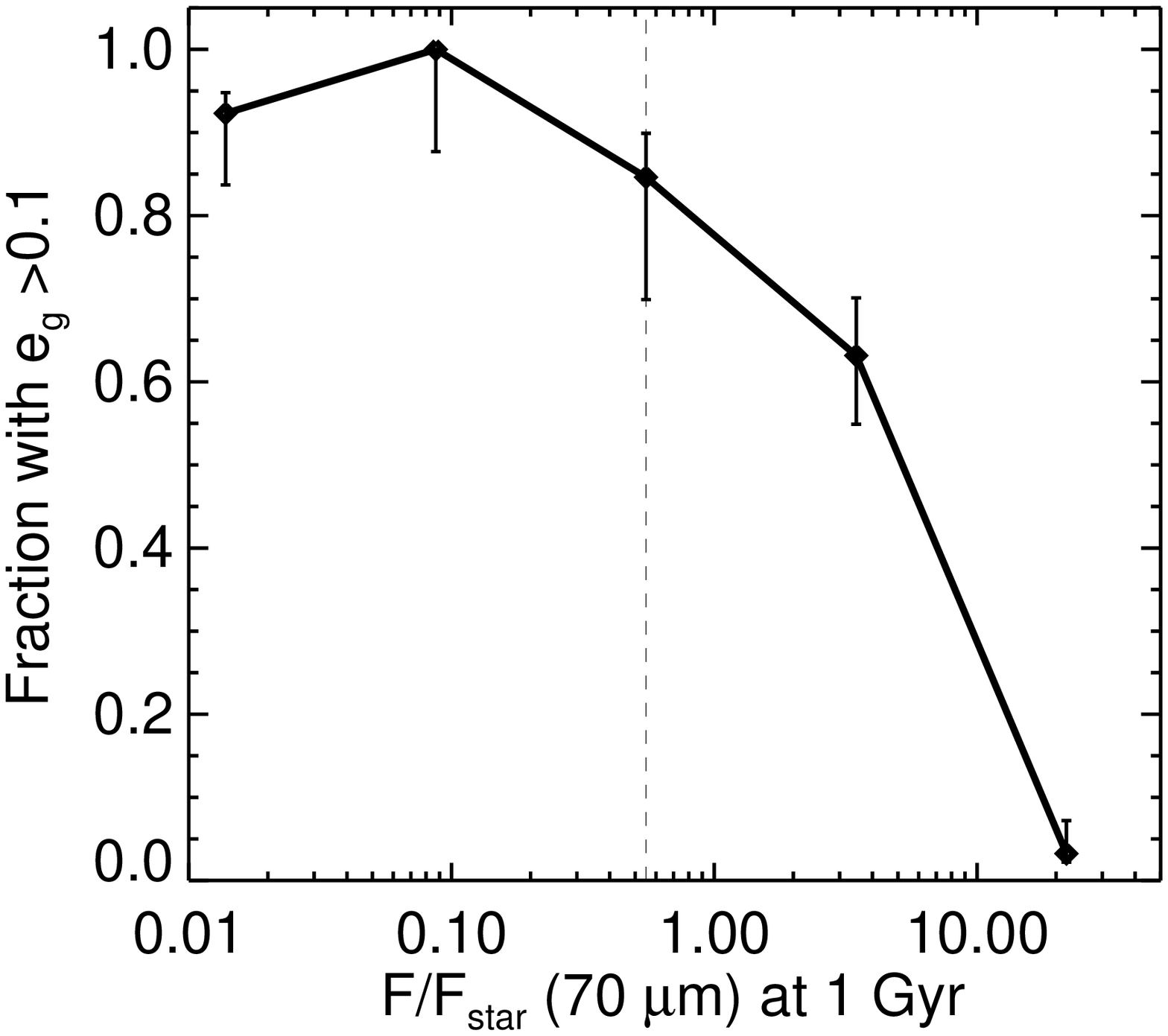}
\caption{{\bf Left:} The fraction of systems that would be detectable after 1 Gyr as a function of the eccentricity of the innermost surviving giant planet $e_g$ for six different wavelengths between 5 and $500 \micron$.  Each curve represents a horizontal slice through plots such as Fig~\ref{fig:ff70-eg} at a different vertical height.  The chosen ``detection limits'' in $F/F_{star}$ were the same as in Fig.~\ref{fig:slice-mterr}: $10^{-8}$ at $5 \micron$, $10^{-3}$ at $15 \micron$, 0.054 at $25 \micron$, 0.55 at $70 \micron$, 10 at $160 \micron$, and 100 at $500 \micron$.  The error bars are 1-$\sigma$ and were calculated using binomial statistics.  Note that the different curves are offset for clarity.  The bins themselves are logarithmically spaced between $10^{-3}$ and 1.  {\bf Right:} The fraction of systems with $e_g > 0.1$ as a function of $F/F_{star} (70 \micron)$ (1 Gyr).  Systems with $F/F_{star} < 10^{-2}$ are included in the bin at $F/F_{star} \approx 10^{-2}$.  The Spitzer detection limit is shown as the dashed line.  This represents a vertical slice through Fig.~\ref{fig:ff70-eg}. }
\label{fig:slice-eg}
\end{figure*}

The anti-correlation between debris disks and eccentric giant planets in our simulations also holds across many wavelengths.  Figure~\ref{fig:slice-eg} (left panel) shows the fraction of stars that is detectable as a function of the giant planet eccentricity for a range of different wavelengths, similar to Fig.~\ref{fig:slice-mterr}.  Again, the trends are stronger for wavelengths of $25 \micron$ and longer, but are still clear at shorter wavelengths (although the detection thresholds at these wavelengths are certainly much smaller than in reality).   Fig.~\ref{fig:slice-eg} (right panel) shows that the fraction of systems containing an eccentric giant planet is anti-correlated with the dust emission at $70 \micron$, simply because in our simulations the calmest giant planets destroy the fewest number of outer planetesimals and therefore produce the brightest debris disks.  

\begin{figure*} %[p]
%\centerline{
\includegraphics[width=0.3\textwidth]{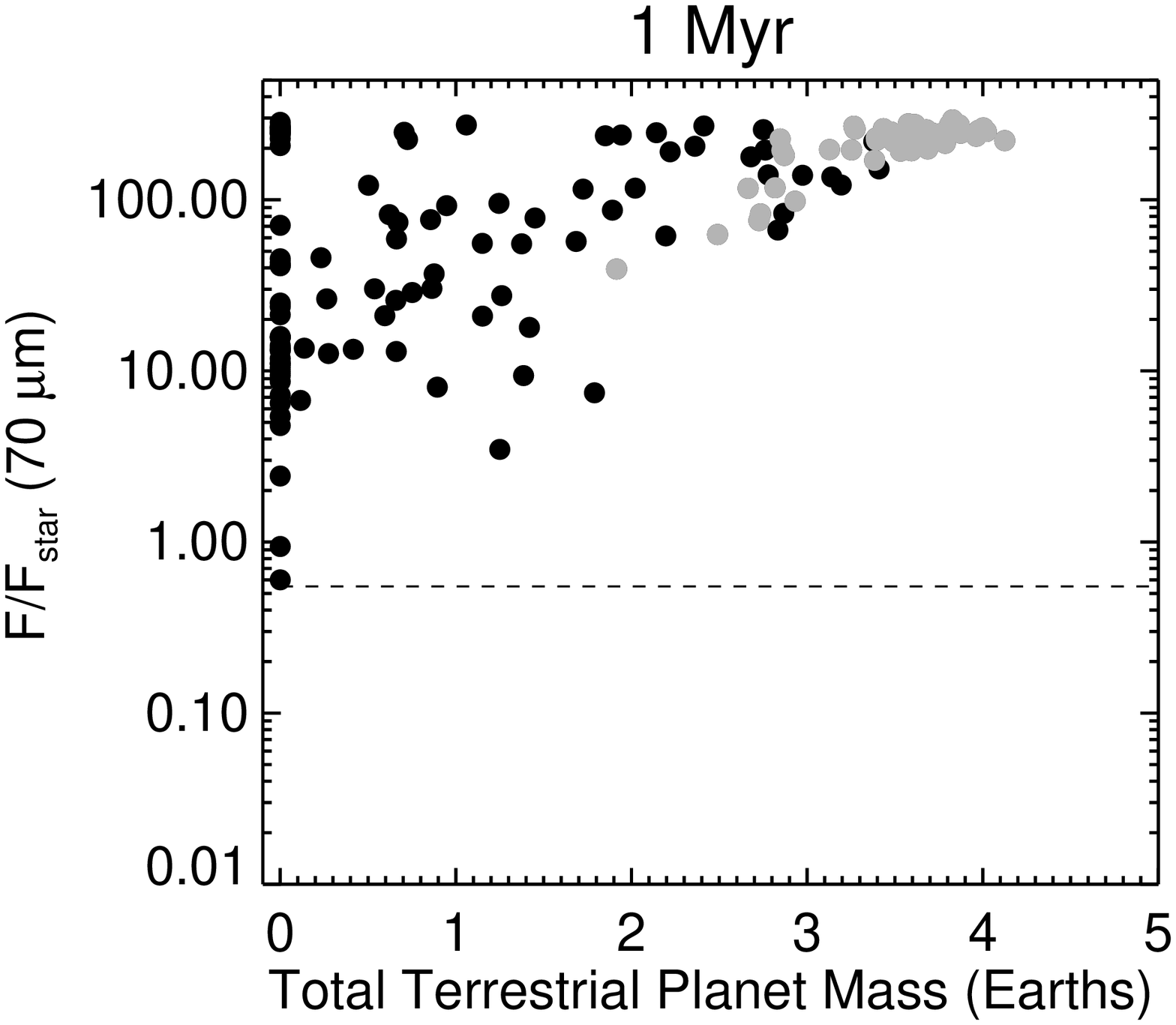}
\includegraphics[width=0.3\textwidth]{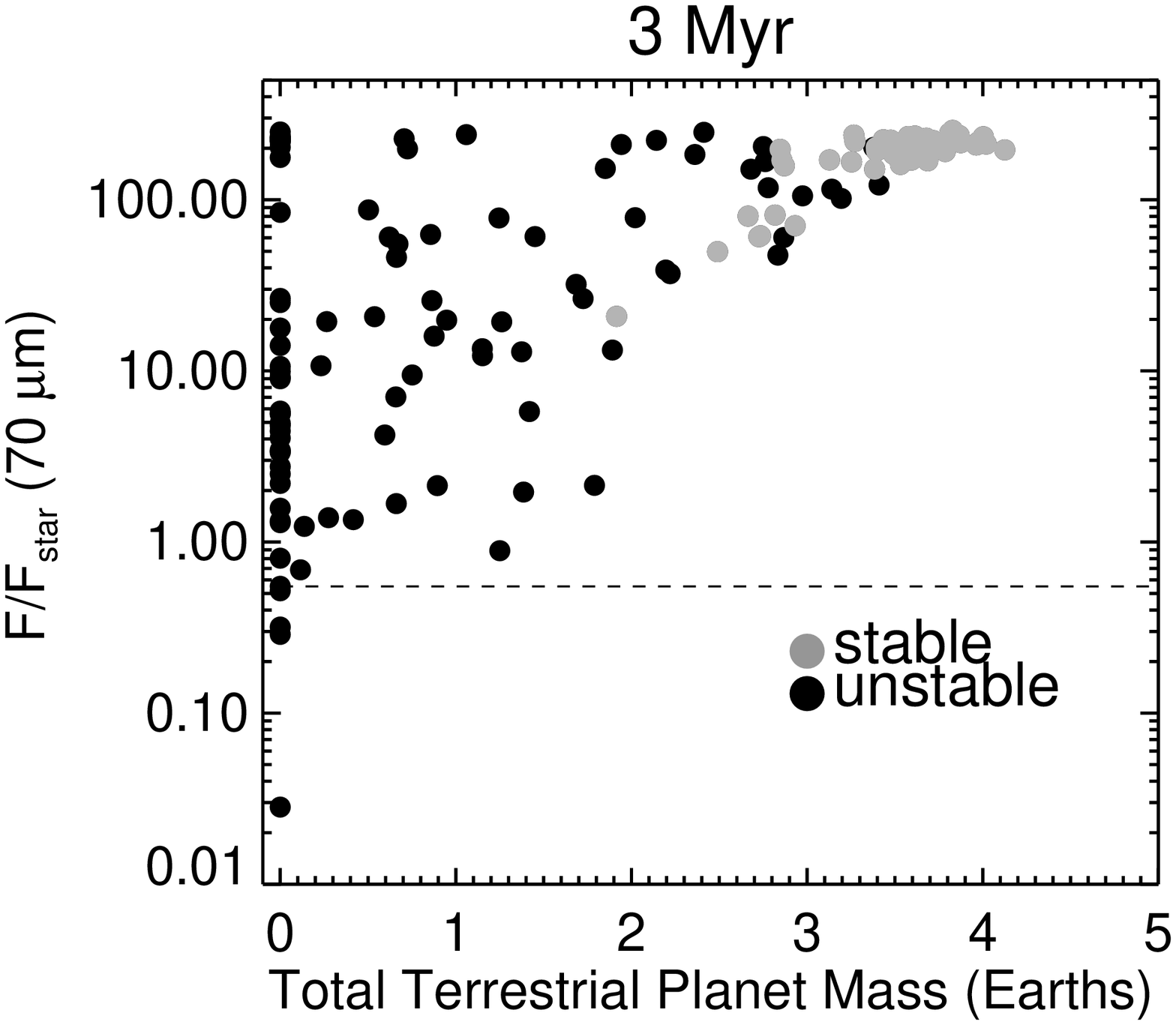}
\includegraphics[width=0.3\textwidth]{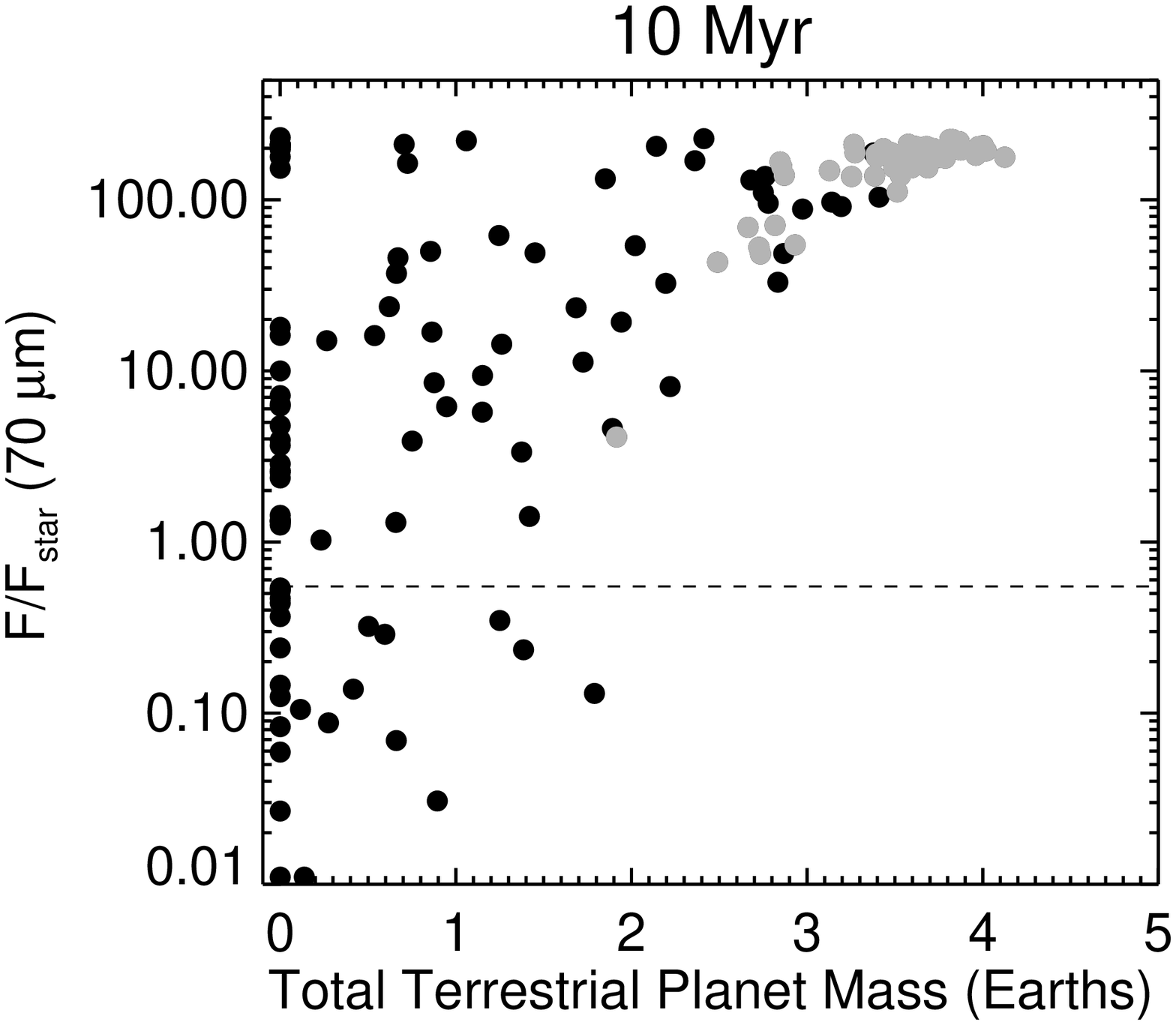}
%}
%\centerline{
\includegraphics[width=0.3\textwidth]{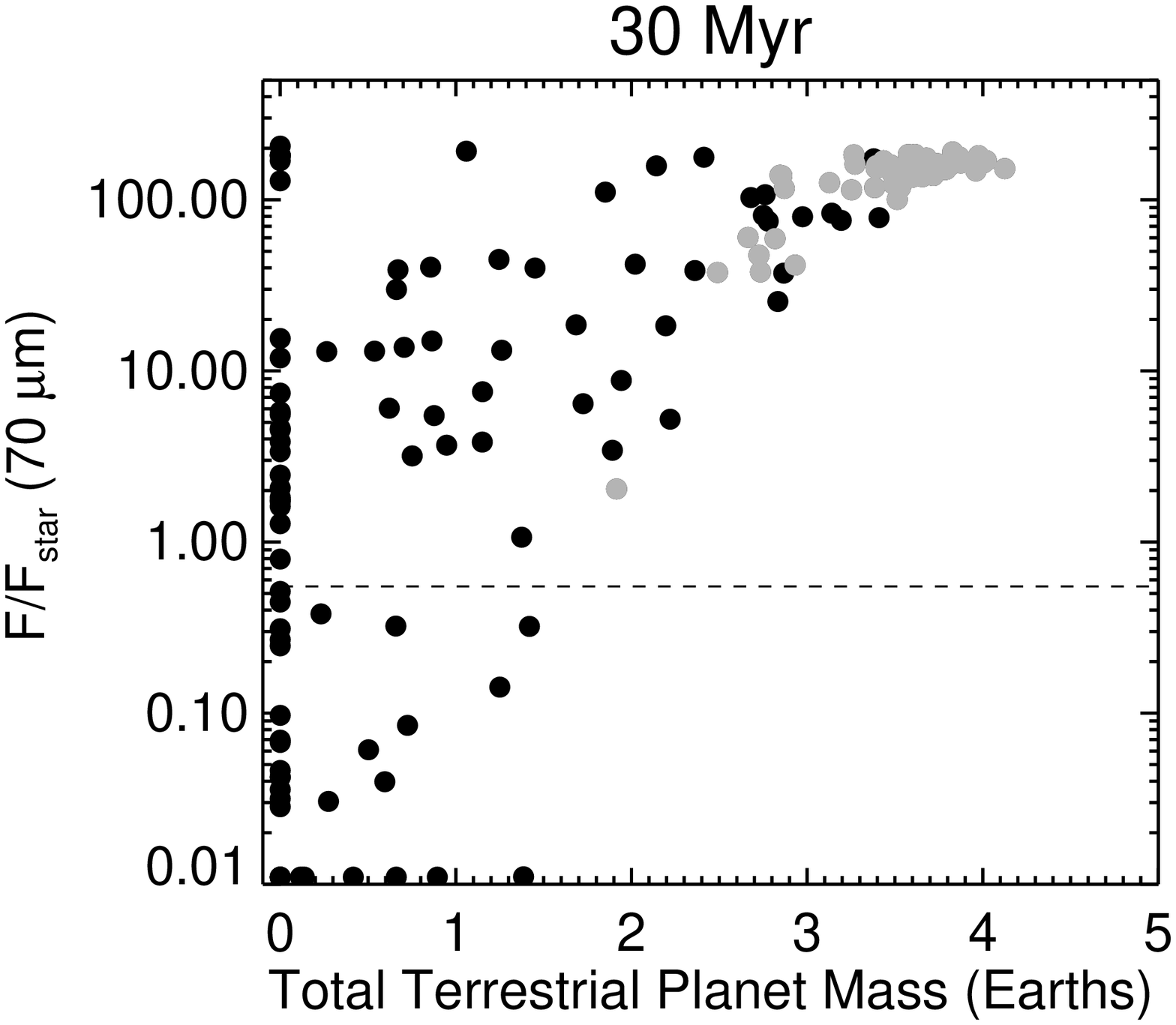}
\includegraphics[width=0.3\textwidth]{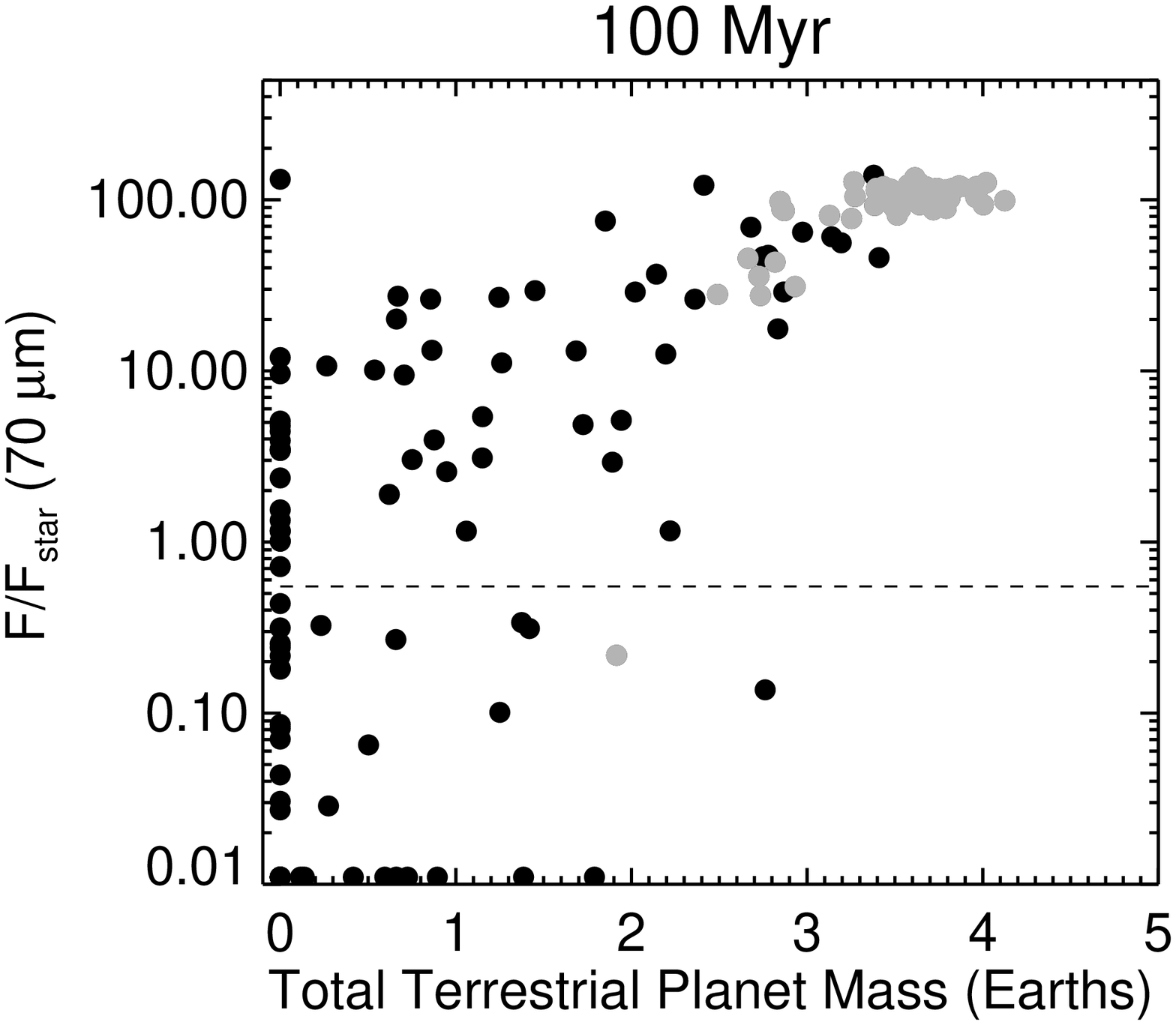}
\includegraphics[width=0.3\textwidth]{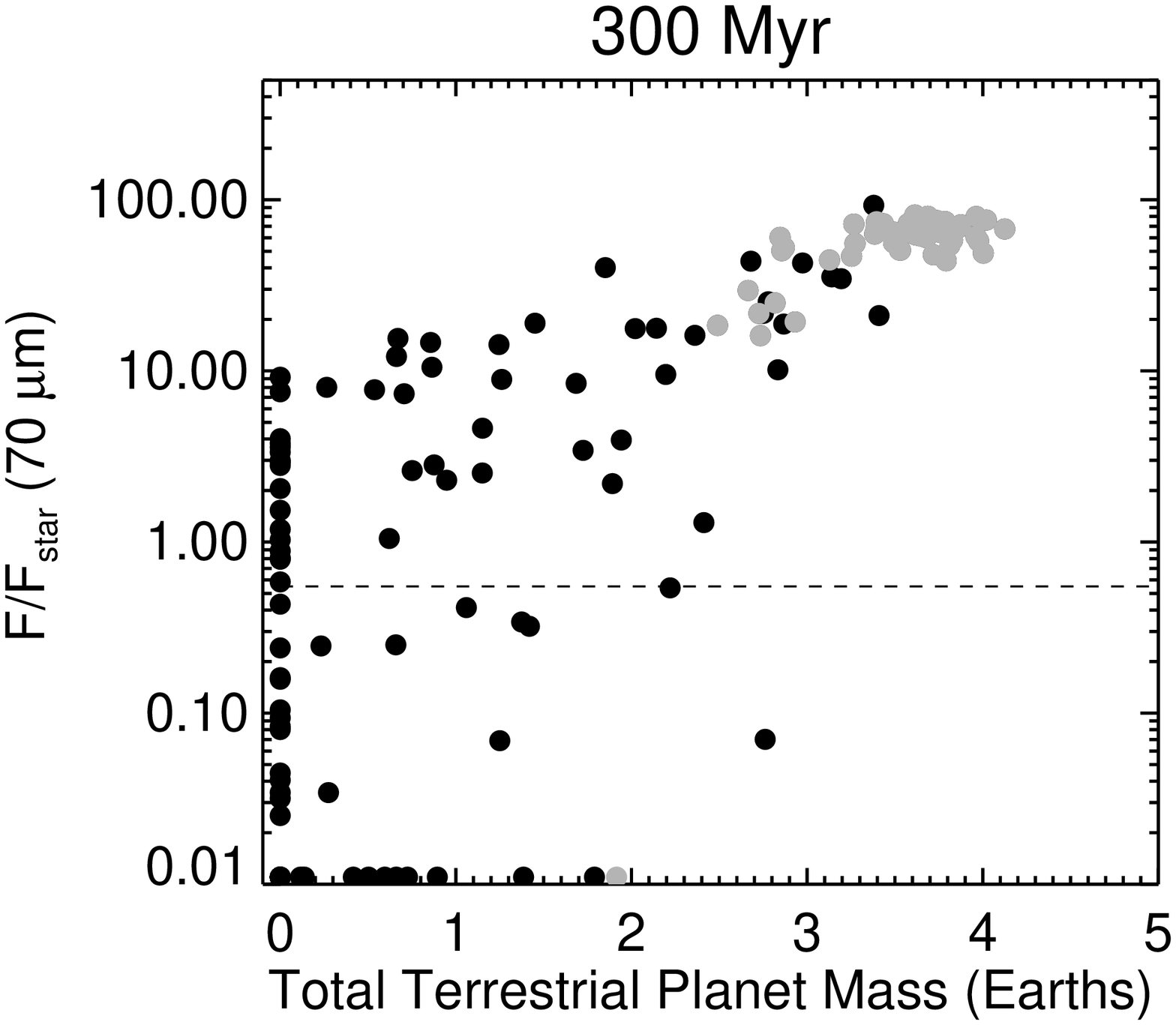}
%}
%\centerline{
\includegraphics[width=0.3\textwidth]{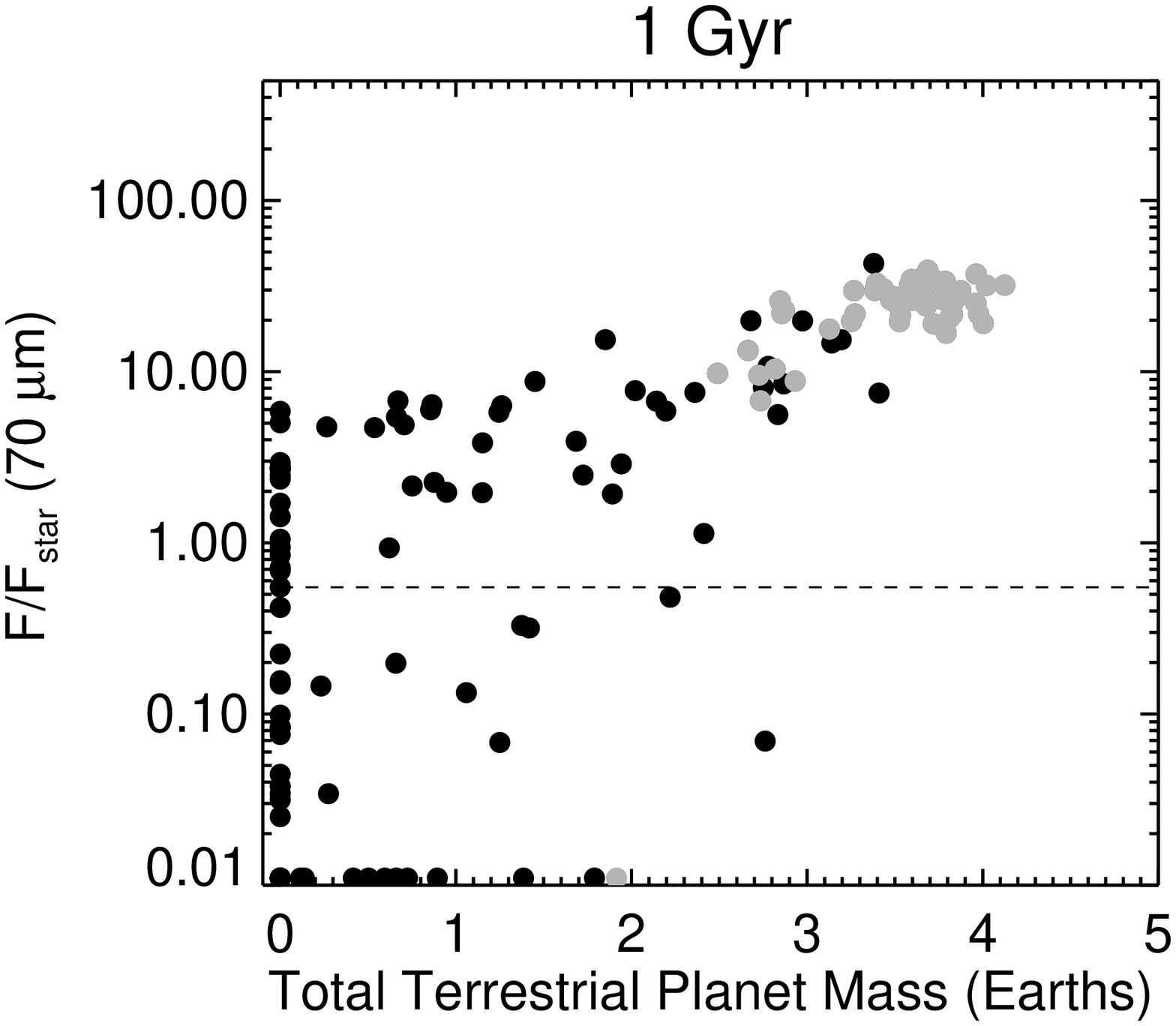}
\includegraphics[width=0.3\textwidth]{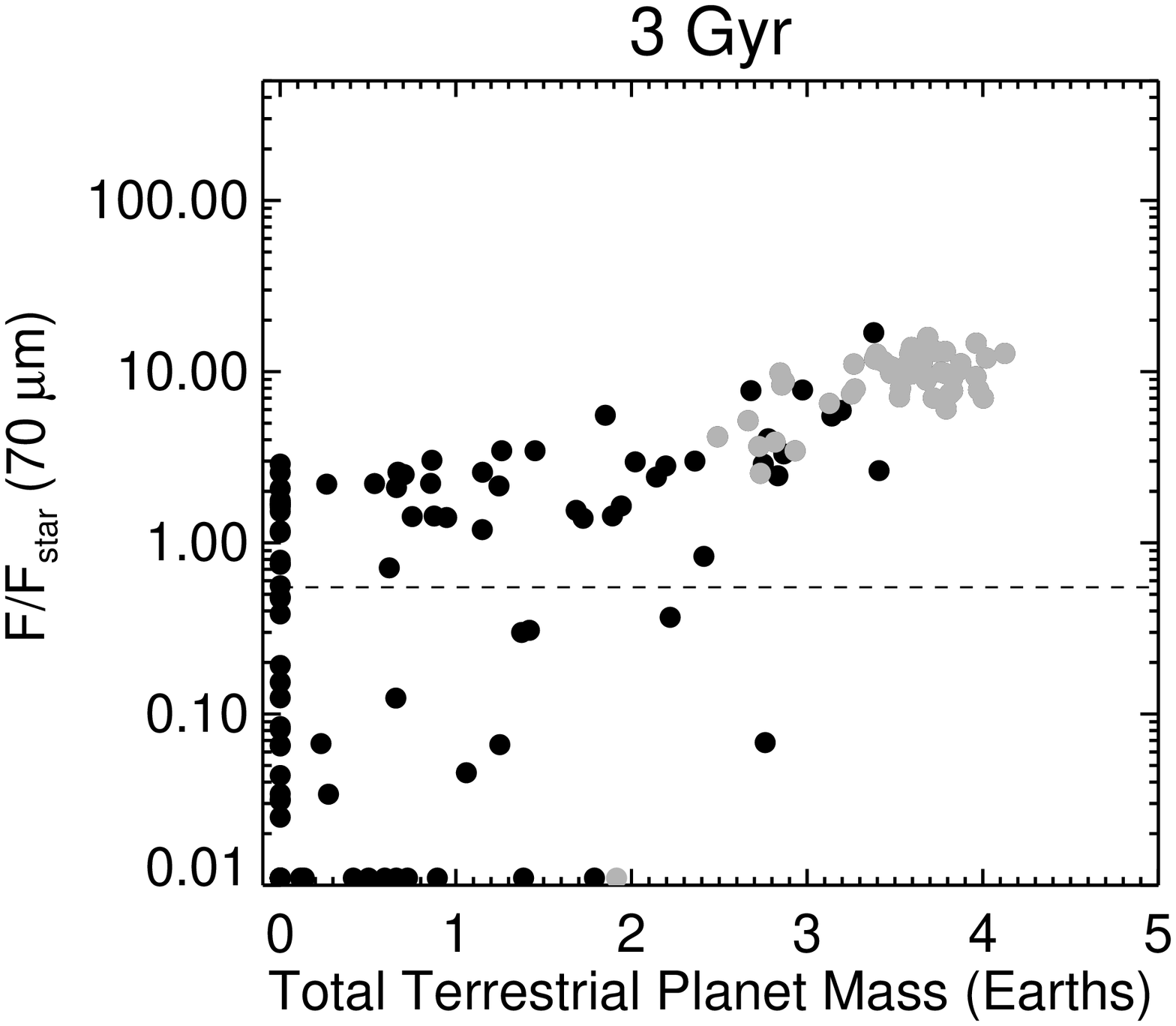}
%}
\caption{The dust-to-stellar flux ratio $F/F_{star}$ at $70 \micron$ as a function of the total mass in terrestrial planets for eight different times from 1 Myr to 3 Gyr.  The terrestrial planet mass refers to the {\em final} value such that simulations move vertically in time on the plot.  The {\it Spitzer} observational limit is shown with the dashed line~\citep{trilling08}.  Grey circles represent stable simulations and black circles represent unstable simulations.  }
\label{fig:ff-mterr-time}
\end{figure*}

The debris disk-terrestrial planet correlation is also a function of time.  Figure~\ref{fig:ff-mterr-time} shows $F/F_{star}$ at $70 \micron$ vs. the final terrestrial planet mass for all simulations at eight snapshots between 1 Myr and 3 Gyr.  At 1 Myr all systems are above the detection threshold, even those that have already undergone violent instabilities, as the timescale for clearing out planetesimals is generally closer to a few to 10 Myr from the time of the instability.  In a given snapshot, systems that have not yet become unstable but that eventually will are those for which the flux remains as high as the cluster of stable systems but for which the total terrestrial planet mass is low, indicating a strong future depletion of rocky material.  After an instability occurs the flux drops (though differently at different wavelengths; Fig.~\ref{fig:flux-t}) but the total terrestrial planet mass, measured at the end of the simulation, does not change, such that a given system moves vertically between snapshots.  In time, instabilities remove systems at low terrestrial planet mass and high $F/F_{star}$; the last two instabilities occurred at 168 Myr (in which one $2.4 \mearth$ terrestrial planet survived on a highly eccentric orbit) and 180 Myr (in which four fully-grown terrestrial planets were destroyed including a $1.3 \mearth$ planet in the habitable zone).  Beyond the end of our simulations one can imagine that a fraction of stable systems could actually become unstable and quickly lose a large fraction of their flux (and perhaps their terrestrial planets as well).  

Stable systems remain clustered at high fluxes at all times, but decreasing in time due to collisional grinding.\footnote{Note that one stable system ejected all of its planetesimals.  In that system the interactions between the giant planets excited eccentricities of $\sim 0.1$, causing an eventual complete depletion in the outer planetesimal disk and a decreased terrestrial planet formation efficiency compared with other stable simulations. Although the planets were all roughly one Jupiter mass, the two inner planets were actually also driven slowly across their 2:1 mutual mean motion resonance after 130 Myr (which actually decreased the eccentricities).}  The same collisional grinding affects the planetesimals that survive in unstable systems.  In some cases the outer planetesimal disk is only moderately perturbed by the instability, although in all cases it is somewhat disturbed as shown by the lack of unstable systems with fluxes as high as the stable systems.  After an instability, the mass in planetesimals decreases, although the planetesimals' eccentricities and inclinations both tend to increase (not always in a simple correlated fashion).  Mass loss causes the planetesimal population's collisional evolution to slow down and eventually stop, as is thought to have occurred in the Solar System's asteroid belt~\citep{petit01,bottke05}.  Thus, the dust flux decreases to a roughly asymptotic value.  Once the collisional timescale for the small particles becomes longer than the timescale for Poynting-Robertson our calculation breaks down although this occurs at a low enough flux that it should not affect our results~\citep{wyatt07a}.

Figure~\ref{fig:ff-mterr-time} shows that the debris disk-terrestrial planet correlation holds in time, especially after 10-100 Myr.  However, if the figure were plotted including the total terrestrial mass at a given time the correlation would hold at even earlier times because individual systems would not be restricted to move between panels on vertical lines.  

\section{Discussion}
In this section we first scale our simulations (section 5.1) to match the exoplanet semimajor axis distribution and infer the orbital distribution of terrestrial planets in the current sample (mainly drawn from the radial velocity sample).  In section 5.2 we compare our results with the statistics of known debris disks (including cases with known giant planets).  In section 5.3 we discuss the Solar System in the context of our results. In section 5.4 we discuss the limitations of our simulations.  

\subsection{Scaling to the observed $a-e$ exoplanet distribution}

The surviving giant planets in our unstable simulations provide a match to the observed eccentricity distribution (Fig.~\ref{fig:hist_etpgp}). However, in constructing a sample of simulations that represents the observed exoplanet systems we do not think that is realistic not to include any stable systems for several reasons.  First, current attempts to de-bias the observed eccentricity distribution infer a substantial fraction of systems -- up to $\sim 30\%$ -- with circular or near-circular orbits~\citep{shen08,zakamska10}.  Second, there exist individual systems that show no obvious signs of instability, for example with planets in resonance~\citep[e.g., GJ 876;][]{rivera10} or with many relatively closely-packed giant planets~\citep[e.g., 55 Cancri;][]{fischer08}.  

\begin{figure}%[p]
\includegraphics[width=0.4\textwidth]{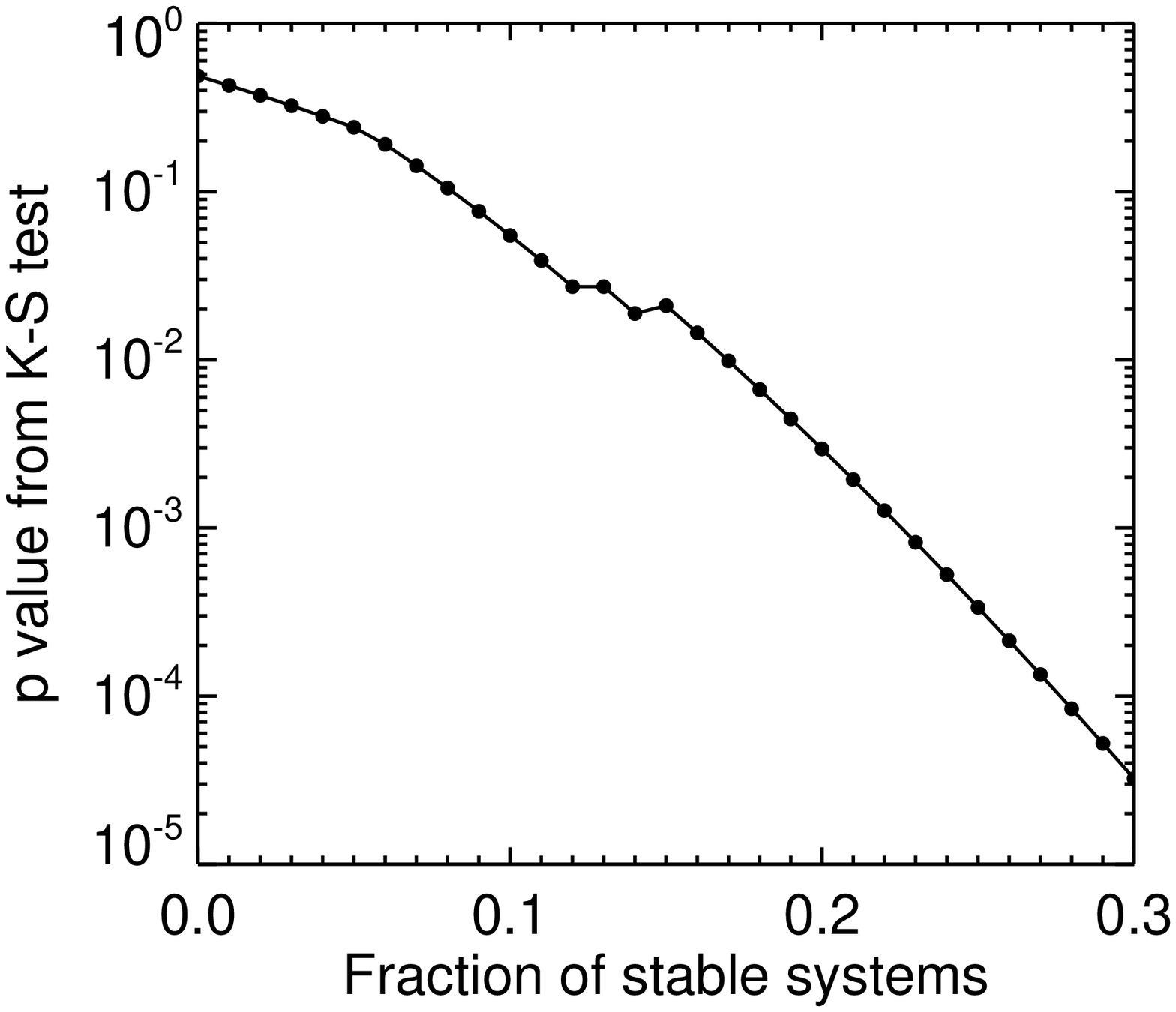}
\caption{The $p$ value from a Kolmogorov-Smirnov test comparing the eccentricities of our simulations vs. the observed exoplanet distribution as a function of the fraction of stable systems included in the sample.  The $p$ value drops below an acceptable level of 0.01 for more than a 17\% contribution from stable systems.  
}
\label{fig:fracstab}
\end{figure}

Figure~\ref{fig:fracstab} (left panel) shows the effect of the contribution from stable systems on the goodness of fit to the exoplanet eccentricity distribution.  The distribution is best matched when we do not include any stable systems in the sample, and drops below a nominal statistically acceptable limit of $p = 0.01$ for a contribution larger than 17\%.  We therefore construct our sample with a 10\% contribution of stable systems, which we think is a reasonable compromise between matching the exoplanet eccentricity distribution and allowing for stable systems.

The giant planets in our simulations are generally at larger orbital distances than the observed giant planets.  The observed sample of giant planets displays a rapid rise between 0.5-1 AU, a plateau to 2 AU, and then a decrease at larger orbital distances~\citep{butler06,udry07b}.  The rise and plateau are real but the decrease beyond 2 AU is an observational bias due to the long orbital periods of these planets.  The actual distribution of giant planets at Jupiter-Saturn distances is unknown, although observational constraints predict that at least 10\%, and perhaps more than 50\%, of stars have giant planets at these separations~\citep{cumming08,gould10}.

Our simulations can be scaled to match the combined semimajor axis-eccentricity distribution of giant planets beyond 0.5-1 AU.  At a smaller orbital distance, a giant planet's scattering power decreases.  This can be quantified by the quantity $\Theta^2$, the ratio of the escape speed from the planet's surface to the escape speed from the planetary system at that location~\citep{safronov69,goldreich04}:
\begin{equation}
\Theta^2 = \frac{M_p}{M_\star} \frac{R_p}{a},
\end{equation}
\noindent where $M_p$ and $M_\star$ are the planetary and stellar mass, respectively, $R_p$ is the planetary radius and $a$ is the orbital semimajor axis.  As $\Theta^2$ is inversely proportional to the $a$, close in giant planets scatter less strongly than at larger distances.  When $\Theta^2$ drops below 1, collisions become more important than scattering.  For our simulations, only the lowest-mass giant planets would drop to $\Theta^2 < 1$ by scaling them to 0.5-1 AU.  In addition, the planet-planet scattering mechanism has been proven at this range of distances~\citep{marzari02,juric08}.  Thus, we conclude that it is dynamically appropriate to scale the giant planets inward to 0.5-1 AU.  We cannot, however, scale to closer distances because the dynamical regime is different and giant planet-planet collisions are more likely than scattering events.  

We assume that the underlying distribution that is being probed by current (mainly radial velocity) observations rises linearly from zero at 0.5 AU to 1 AU, where it flattens off and remains constant to 5 AU.  We draw randomly from this distribution and scale the innermost giant planets from ten randomly chosen simulations -- nine that were unstable and one that was stable according to relative contribution of stable vs. unstable systems in our sample -- to match this value.  We then re-scale the terrestrial planets in each system to the same size scale as the inner giant planet's orbit.  Our sample was created by repeating this 100,000 times with different random numbers.

\begin{figure}%[p]
\includegraphics[width=0.45\textwidth]{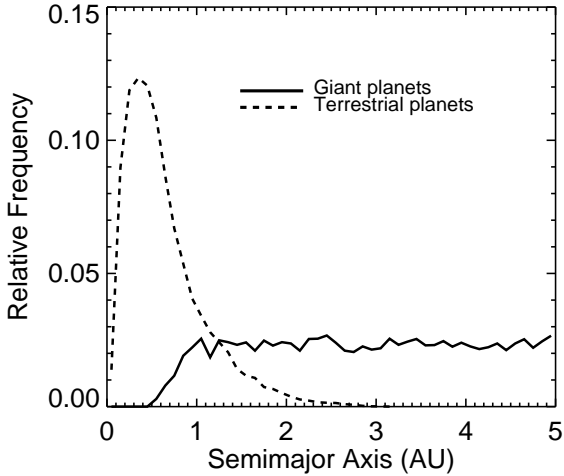}
\caption{Semimajor axis distribution of simulated terrestrial planets (dashed line) from our set of simulations, derived by scaling the innermost surviving giant planet in each simulation to match an assumed underlying distribution for relevant exoplanets that increases linearly from zero at 0.5 AU and is constant from 1-5 AU.}
\label{fig:atp-agp}
\end{figure}

Figure~\ref{fig:atp-agp} shows the radial distribution of terrestrial and giant planets after this scaling.  In essence, this figure shows the expected radial distribution of terrestrial planets in the known extra-solar {\em giant planet} systems.  In these systems, our simulations predict a factor of a few higher abundance of terrestrial planets at a few tenths of an AU than at 1 AU because, given the typical giant-terrestrial planet spacing, the formation of planets at 1 AU requires distant giant planets that are hard to detect by current methods.  Planets within $\sim 0.1$ AU are sparsely populated because of the assumed inner edge of the embryo disk at 0.5 AU.  The peak in the frequency of terrestrial planets at a few tenths of an AU is due to a combination of our inner disk edge and the giant planets' radial distribution.    

As noted above, the scaling we performed is dynamically permissible: it does not change the regime of accretion of the terrestrial planets nor the scattering regime of the giant planets.  However, our simulations each contained a constant mass in embryos and planetesimals in the inner disk.  By scaling the giant and terrestrial planets, we are effectively re-scaling the initial disk masses such that the same amount of mass would initially have been placed into an annulus whose position and width can vary.  In addition, if we had included initially closer-in material in our simulations, the peak in Fig~\ref{fig:atp-agp} would have shifted inward.  Despite these limitations, in the regime that we have considered we think that the shape of the curve is physical, and we predict that, at least within the known sample of extra-solar giant planets, terrestrial planets at $\sim 0.3$ AU should be several times more abundant than at 1 AU.  

\subsection{Comparison between our simulations and observed debris disks}

Observations (mainly with NASA's {\em Spitzer} Space Telescope) have shown that roughly 15\% of solar-type stars younger than 300 million years have measurable dust fluxes at $24 \micron$~\citep{gaspar09} but that this fraction decreases in time and flattens off at $\sim$3\%~\citep{carpenter09}. At $70 \micron$, 16\% of stars observable dust and there is no measured decrease in this fraction with age~\citep{trilling08,carpenter09}.  Considering the current exoplanet/debris disk sample, $9\% \pm 3\%$ of planet-hosting stars were detected at 24 or $70 \micron$ compared with $14\% \pm 3\%$ for stars without planets~\citep{bryden09}.  An update of that study found debris disks around the same fraction, $\sim$ 15\%, of stars with and without known giant planets~\citep{kospal09}. In addition, the strong correlation between stellar metallicity and the fraction of stars with planets~\citep{gonzalez97,santos01,fischer05} does not hold for the current sample of debris disks, whose presence is metallicity-independent~\citep{beichman06,greaves06,bryden06,kospal09}.  

Figure~\ref{fig:ff70-eg} shows that, in our simulations, the dust flux is anti-correlated with the giant planet eccentricity.  Almost all lower-eccentricity ($e < 0.1-0.2$) giant planets are in systems with debris disks, but at higher eccentricities the fraction of dusty systems decreases as does the dust brightness itself.  Figure~\ref{fig:slice-eg} shows that the fraction of systems that are detectable at all wavelengths from $5 \micron$ to $500 \micron$ decreases with increasing eccentricity of the innermost surviving giant planet.  In addition, the fraction of planets with $e_g > 0.1$ decreases with $F/F_{star} (70 \micron)$, including a dramatic drop for $F/F_{star} > 10$.  

We therefore expect to see a correlation between the orbital properties of exoplanets and the detectability of cold dust.  The dataset of~\cite{bryden09} detected debris disks at $70 \micron$ around 13 planet-hosting stars out of 146 for a detection frequency of $8.9 \% ^{+2.9\%}_{-1.8\%}$.  We arbitrarily divide the sample in two based on the eccentricity of the innermost giant planet in each system at a value of 0.2.  Debris disks are detected in 8 of 76 systems ($10.5\%^{+4.7\%}_{-2.6\%}$)in the low-eccentricity subsample and 5 of 70 systems ($7.1\%^{+4.4\%}_{-1.9\%}$) in the high eccentricity subsample.  Although the detection rate is somewhat higher in the lower-eccentricity subsample, the difference is not statistically significant and we must wait for better statistics from larger surveys.  

Our simulations do produce some systems with high-eccentricity giant planets and bright dust emission (Fig.~\ref{fig:ff70-eg}), in agreement with the detected debris disks in known exoplanet systems~\citep{moromartin10}.  In these cases the dynamical instability tends to be asymmetric and confined to the inner planetary system and these are therefore not generally good candidates for terrestrial planets, which also agrees with the observed systems~\citep{moromartin10}.  The outcome of a given system depends critically on the details of the instability, which is determined by the giant planet masses~\citep{raymond10}.

Our approach is not unique; other approaches based on dust production by collisional cascades can also reproduce the debris disk observations~\citep{krivov05,krivov06,wyatt07b,wyatt08,kenyon08,kenyon10,krivov10,kennedy10}.  In addition, there is a qualitative difference between models that consider planetesimal disks to be ``self-stirred''~\citep[i.e., eccentricities are excited by accreting bodies within planetesimal disks;][]{kenyon08,kenyon10,kennedy10} or those that consider external sources for planetesimal stirring~\citep[e.g.][]{wyatt10}.  \cite{mustill09} showed that self-stirred planetesimal disks tend to be fainter than disks stirred by giant planets.  Debris disks may be explained by some combination of these ideas. 

Nonetheless, we conclude that our simulations are consistent with the known sample of debris disks in exoplanet systems.  However, our initial conditions are biased in that all systems that could potentially produce debris disks also contain giant planets, which is not consistent with the observation that the frequency of debris disk + giant exoplanet systems is about the same as debris disks with no detected giant exoplanets~\citep{bryden09,kospal09}.  In paper 2 we use multiple sets of simulations to construct a sample that adequately matches the entire debris disk and exoplanet samples, and use that sample to infer the properties of as-yet-undetected terrestrial exoplanets.

\subsection{Our Solar System in context}
Our results suggest that the Solar System is unusual at the $\sim 15-25\%$ level.  This corresponds to the fraction of simulations that form three or four terrestrial planets (Fig.~\ref{fig:hist-nterr}; including a 10\% contribution from stable systems. By the same weighting 38\% of simulations destroy all their terrestrial material.).  

The Solar System lies at the very edge of the debris disk correlations in Figs.~\ref{fig:ff70-eg} and~\ref{fig:ff70-mterr} because of its combination of a rich terrestrial planet system, a low-eccentricity innermost giant planet, and a low dust flux.  To a distant observer, the Solar System's faint debris disk would suggest a dust-clearing instability in the system's past.  However, Jupiter's low-eccentricity orbit would imply that the instability was weak and that the system may in fact be suitable for terrestrial planets.\footnote{Of course, it would take about a decade of precise observations for these aliens to pin down Jupiter's orbital eccentricity.}  This naive argument is remarkably consistent with our current picture of the LHB instability as an asymmetric, outward-directed instability that included a scattering event between Jupiter and an ice giant but not between Jupiter and Saturn~\citep{morby10}.  

The Earth provides an interesting test case.  On long timescales Earth's eccentricity oscillates between 0.0002 and 0.058 and its inclination between zero and 4.3$\deg$~\citep{quinn91,laskar93}.  When compared with the stable simulations, the Earth's time-averaged eccentricity is significantly smaller than the median and its inclination is also smaller.  However, Earth's oscillation amplitudes are more than 50\% {\em larger} than the median values for the stable systems.  This situation is the same for Venus' orbit.  This presents a confusing situation; perhaps Earth and Venus' $e$ and $i$ are lower than these simulations because we are limited in the numerical resolution needed to adequately model dissipative processes.  But if that were the case, we would expect Earth and Venus' oscillation amplitudes to also be smaller than the simulated planets'.  One explanation for the Earth and Venus' orbits is that they formed in a dissipative environment but that were later dynamically perturbed during the instability that caused the late heavy bombardment~\citep{gomes05,brasser09,morby10}.  The perturbation was not large enough to disrupt the system's stability but sufficient to increase the amplitude of orbital oscillations of Earth and Venus.  

The instability that caused the LHB is not captured in our simulations nor in current exoplanet observations.  The degree to which our simulations interpret that the Solar System is unusual depends on how well the simulations characterize the instabilities in extra-solar planetary systems.  As our simulations reproduce the observed exoplanet eccentricity distribution with no free parameters, it appears that our simulations do in fact capture the essence of instabilities among the known exoplanets.  However, an instability such as the one proposed by the Nice model leaves little to no trace because the giant planets' eccentricities remain very small ($e_J$ and $e_S$ are only $\sim 0.05$).  It is plausible that Nice model-type instabilities are common in outer planetary systems, although if they systematically destroy their outer planetesimal disks then debris disk statistics constrains the fraction of stars that undergo such instabilities more than about 10 Myr after stellar formation to be less than about 10\%~\citep{booth09}.  In Paper 2 we present an additional set of simulations with larger mass ratios between the giant planets in which Nice model-type instabilities can occur.  Such instabilities do not change our conclusions; Nice model instabilities can effectively be lumped in with the stable systems as their impact on inner planetary systems is small.\footnote{\cite{brasser09} showed that some Nice model instabilities can destabilize the orbits of the terrestrial planets due to sweeping secular resonances and potentially cause collisions between the terrestrial planets.  However, this process does not remove all the terrestrial planets and so, in the context of our results, nothing has changed.}

\subsection{Limitations of our approach}

As with any numerical study, our simulations are limited in several ways.  

Our most important assumption is that the inner and outer regions of protoplanetary disks are connected such that observations of debris disks in the outer parts of planetary systems can tell us something about terrestrial planets in the inner parts of these systems.  However, there exist substantial uncertainties.  For example, several relevant processes -- notably the formation of planetesimals as well as giant planets -- are only modestly-well understood.  In addition, it is unknown whether the efficiencies and timescales of these processes vary with distance from the star.  If there exists a systematic bias to create an imbalance between the inner and outer disk mass, it could qualitatively affect our results.  For example, if planetesimal formation is much more efficient in the outskirts of planetary systems then the typical system may contain an outer planetesimal belt but no inner planetesimals or embryos from which to form terrestrial planets.  Alternately, one can imagine that outer planetesimal disks might be systematically destroyed in systems that do form terrestrial planets, as was the case for the Solar System.  If one of these scenarios is true, then the initial conditions for inner and outer planetary systems may not be coupled as strongly as we have assumed, and their outcomes may not correlate as strongly as in our simulations.  

Despite these uncertainties, we think that our approach, in particular the assumption that inner and outer disks are connected, is the simplest interpretation of current observations and theory.  The frequency of close-in planets is 12\% in the 3-10 $\mearth$ range and, by extrapolation, 23\% in the 0.5-2 $\mearth$ range~\citep{howard10}.  The observed frequency of debris disks around FGK stars of 16\%~\citep{trilling08} represents a lower limit for the frequency of outer planetesimal disks.  Preliminary results from the {\it Herschel} DUNES survey~\citep{eiroa10} suggest that roughly 1/3 of stars have debris disks (C. Eiroa, personal communication), which means that that the frequency of inner planets and outer planetesimal belts are probably within a factor of a few or less.  Although this certainly does not prove that our initial conditions are correct, it does provide circumstantial support for our basic assumption of a connection between the inner and outer disk although we note that there is as yet no observed correlation between the two.  

Our initial conditions, though chosen to match models of earlier phases of planetary growth, are ad hoc.  All planetary systems in our sample contain the same mass in terrestrial embryos and planetesimals ($9 \mearth$), form their innermost giant planet at 5.2 AU, and contain the same mass in an outer disk of leftover icy planetesimals ($50 \mearth$).  In reality, there is a spread of several orders of magnitude in the disk mass~\citep[e.g., ][]{andrews07a} that affects both the types of planets that form and the amount of leftover planetesimals~\citep[e.g.,][]{greaves07,thommes08}.  In addition, recent observations suggest that low-mass disks are more compact~\citep{andrews10} so there may be a correlation between the disk mass and the location of planet formation as well~\citep[see also][]{kennedy08}.  The disks that we modeled are comparable in mass to the minimum-mass solar nebula model and are probably more massive than the typical disk~\citep{eisner03,andrews07a}.  

Our simulations are confined to Solar-mass stars, which are a small minority of all stars~\citep{bochanski10}.  To expand on this work we should consider additional stellar types.  Debris disks are currently very difficult to detect around low-mass stars~\citep{gautier07} but are extremely interesting as planet hosts because they dominate the stellar population of the Galaxy and are very long-lived.  In contrast, debris disks are much easier to detect around A stars, but these are relatively few in number and their lifetimes are much shorter.  There may be interesting differences in the evolution of planetary systems around other stellar types.  

There are several physical processes not included in our simulations.  In particular, we did not include the effects of giant planet migration because population synthesis models are currently unable to reproduce the observed exoplanet mass and orbital distribution~\citep{howard10}.  Including migration would have the benefit of providing a natural trigger for giant planet instabilities~\citep{adams03,moorhead05}, although this depends on the details of the depletion of the gaseous disk~\citep{crida08}.  However, given that the giant planet observations can be matched with little to no giant planet migration, we chose not to include it.

\section{Conclusions}

Our main results are as follows:

\begin{itemize}

\item {\bf Giant planet instabilities are destructive to terrestrial planet formation.}  The survival of terrestrial embryos and planetesimals depends on the strength of the instability as measured by the minimum giant planet perihelion distance~\citep[Fig~\ref{fig:mterr-minperi}, see also][]{veras06}.  In about 40\% of our unstable simulations all rocky material was removed from the system, in large part by being thrown into the host star.  About 1/5 of our unstable simulations produced a system containing a single terrestrial planet (Fig.~\ref{fig:hist-nterr}). 

\item {\bf Terrestrial exoplanets on excited orbits should be common.}  The median eccentricity of surviving terrestrial planets in our simulations was about 0.1, but the distribution extends above 0.5 (Fig.~\ref{fig:hist_etpgp}).  The most excited orbits belong to single-terrestrial planet systems.  Compared with systems with many terrestrial planets, single-planet systems have only slightly higher eccentricities and inclinations but their oscillations in $e$ and $i$ are far larger (Fig.~\ref{fig:eiosc}).   

\item {\bf Debris disks are anti-correlated with strongly-scattered giant planets.}  Strong scattering events produce eccentric giant planets with large radial excursions that dynamically deplete the outer planetesimal disk by exciting their orbits until they cross a giant planet's at which point they are quickly ejected from the system.  Thus, we expect continued observations to show an anti-correlation between the fraction of systems with debris disks and the giant planet eccentricity.

\item {\bf Debris disks correlate with a high efficiency of terrestrial planet formation.}  Strong scattering events yield large giant planet eccentricities, and these eccentric giant planets tend to disturb both the inner and outer planetary system.  Thus, in a strongly perturbed system the giant planets tend to destroy both terrestrial planetary embryos -- aborting the growth of terrestrial planets -- and outer planetesimals -- preventing the creation of debris disks by long-term collisional evolution.  In contrast, in a calm system the giant planets will not strongly impede on the inner or outer planetary system, allowing for the formation of terrestrial planets and long-lasting cold dust.  The debris disk-terrestrial planet correlation holds for all wavelengths we tested but is clearer for $\lambda \gtrsim 15 \micron$ (Figs.~\ref{fig:ff-mterr-wav} and~\ref{fig:slice-mterr}).  The correlation also holds for all times later than about 10-30 Myr (Fig.~\ref{fig:ff-mterr-time}) and probably even earlier.  

\item {\bf Within the known sample of extra-solar giant planets, terrestrial planets at a few tenths of an AU should be several times more abundant than either terrestrial or giant planets at 1 AU.}  In section 5.1 we scaled the outcomes of our simulations to match the observed semimajor axis and eccentricity distributions of giant exoplanets.  This scaling produced a radial distribution of terrestrial exoplanets that peaks at a few tenths of an AU and drops below the giant planet frequency at 1.3 AU (Fig.~\ref{fig:atp-agp}).  However, we note that the distribution is incomplete in its inner regions due to our initial conditions, in particular the inner edge in our embryo distribution at 0.5 AU.

\item {\bf The Solar System lies at the outskirts of several correlations, probably because of the instability that caused the Late Heavy Bombardment.}  The Solar System has a rare combination of multiple terrestrial planets, low-eccentricity giant planets, and very low dust content~\citep[currently, $F/F_{star} \approx 2 \times 10^{-2}$;][] {booth09}.  In addition, Earth and Venus' orbits are more circular and coplanar than terrestrial planets in typical stable planetary systems but they have significantly larger-amplitude oscillations in those quantities.  This can be explained if the Solar System's formation was quiescent but it underwent a later punctual event that was strong enough to remove most of the outer planetesimal disk and give the inner Solar System a small kick but did not destabilize the inner Solar System or impart a large eccentricity to Jupiter.  This is in agreement with the current picture of the instability that caused the LHB~\citep{morby10}.  This type of instability is poorly constrained by observations, is much weaker than the instabilities inferred from the exoplanet eccentricity distribution, and is not captured in the simulations presented here.

\end{itemize}

One implication of our results is that exo-zodiacal dust clouds around stars with terrestrial planets may often be brighter than the Solar System's zodiacal cloud.  A major obstacle to the direct imaging of terrestrial exoplanets is the amount of bright dust close to those planets, i.e., their exo-zodiacal dust clouds~\citep{cash06,defrere10,noecker10}.  The Solar System's zodiacal dust has been shown to derive almost entirely from Kuiper Belt comets that were scattered by the giant planets into the inner Solar System, where they partially sublimated to produce warm dust before eventually being ejected~\citep{nesvorny10}.  Around other stars, cold debris disks should trace the same population of comets that produces exo-zodiacal dust: debris disks represent planetesimals on stable orbits in the outer system and exo-zodiacal dust is generated by the small fraction of bodies that has been destabilized and is in the process of being removed from the system.  Our results suggest that systems with bright debris disks are excellent targets in the search for terrestrial exoplanets.  These systems contain at least a few $M_\oplus$ (and often more than 20$M_\oplus$) in surviving cometary material, 1-2 orders of magnitude more than the current Kuiper Belt~\citep{gladman01}.  If the comet flux scales with the number of outer planetesimals then systems with bright debris disks should also harbor bright exo-zodiacal clouds close to the terrestrial planet zone.  However, the dynamics of the outer planetary systems -- in particular the architecture and masses of the giant planets -- are key in determining the fluxes of new comets in these systems as well as their residence lifetimes in the inner planetary systems.  In addition, there is almost certainly a significant population of systems with terrestrial planets without bright debris disks, i.e. what we have called ``false negatives''.  Those systems may be harder to find because they lack debris disks to signpost the presence of terrestrial planets, but they could prove easier to image because they may contain much fainter zodiacal clouds.  We plan to test these ideas in future work.

In a companion paper (Raymond et al 2011; referred to in the text as paper 2) we explore the effect of several other parameters on the results obtained here.  In particular, we present results of several other sets of simulations that vary the giant planets' mass distribution and total masses, the width of the outer planetesimal disk, the existence of icy embryos within the outer planetesimal disk, and the presence of disk gas at the time of giant planet instabilities.  In that paper we confirm the large-scale results presented here but with several important clarifications and dependencies.  We also carefully match our simulations to the observed statistics of giant exoplanets and debris disks to obtain an estimate for the fraction of stars that host terrestrial planets, $\eta_{Earth}$ in the famous Drake equation.  

In a second companion paper (Raymond et al 2011b), we explore the fate of bodies -- planetesimals, planetary embryos, and giant planets -- that are ejected from unstable planetary systems and pollute the galaxy.  By matching giant exoplanet and debris disk statistics we estimate the abundance of this population of free-floating bodies and the chances for their unambiguous detection either in interstellar space or entering the Solar System.

\section{Appendix A: Numerical Tests}

\begin{figure}%[p]
\center\includegraphics[width=0.4\textwidth]{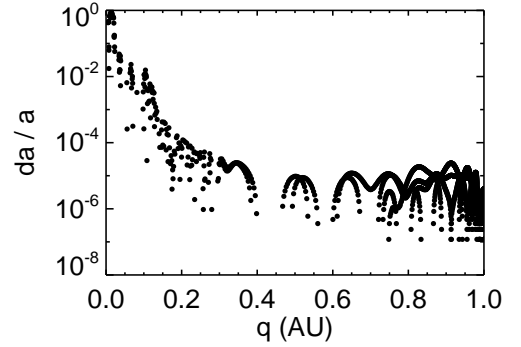}
\caption{Fractional error in the semimajor axis $a$ of a Mars-mass particle with initial $a = 1$ AU as a function of perihelion distance $q$.  The particle was initially placed on a circular, nearly polar orbit (initial inclination of 89.9$^\circ$) in the presence of an outer giant planet.  The Kozai effect forced the particle to fall into the star, and we tracked the integration error in time.}
\label{fig:a-q}
\end{figure}

We performed simple numerical tests to choose an inner boundary appropriate for this timestep by placing a particle at 1~AU on an orbit that was highly inclined with respect to a distant, Jupiter-mass planet\footnote{We thank Hal Levison for suggesting this numerical test.}. In this configuration, the Kozai effect~\citep{kozai62} drives a dramatic increase in the particle's eccentricity.  By tracking the integrator error at ever-smaller perihelion distance, $q = a(1-e)$, we saw that the fractional error in the semimajor axis $da/a$ increased to greater than $10^{-4}$ inside roughly 0.2 AU, as shown in Figure~\ref{fig:a-q}.  We therefore chose 0.2 AU as our inner particle boundary, inside which objects are removed from the simulation and assumed to have collided with the star.  We chose 100 AU as our corresponding outer boundary, beyond which bodies are assumed to have been dynamically ejected from the system.

\begin{figure} %[p]
\center\includegraphics[width=0.4\textwidth]{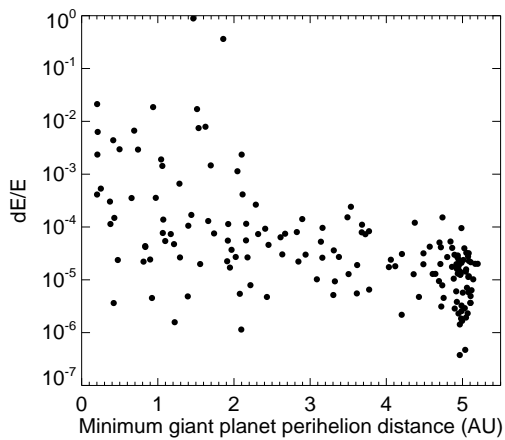}
\caption{Fractional error in the system's energy budget ${\rm d}E/E$ as a function of the smallest perihelion distance of a giant planet for all simulations.}
\label{fig:dee-q}
\end{figure}

With our choice of timestep and inner boundary, the majority of our simulations maintained 
excellent (${\rm d}E/E < 10^{-4}$) energy and angular momentum conservation properties\footnote{We are aware that removing particles from the simulation, at any radius larger than the physical 
one (approximately the size of the star), can in principle result in unphysical behavior, 
{\em even when energy and angular momentum are well-conserved}. Unfortunately, it is currently 
infeasible to run long-duration terrestrial planet formation simulations with dramatically 
shorter timesteps and smaller inner boundary radii.}. However, as shown in Figure~\ref{fig:dee-q}, 
a fraction of the simulations in which the minimum giant planet perihelion distance was 
small ($q < 2$~AU) exhibited substantially worse conservation properties. Given this 
behavior, we adopted an empirical energy conservation threshold of ${\rm d}E/E < 1 \times 10^{-2}$, 
and discarded from the final sample any runs that failed to meet this limit.

Does our cutoff in energy at ${\rm d}E/E < 1 \times 10^{-2}$ bias our results?  The true outcome of our numerical experiment depends on the details of the giant planets' orbital evolution, and it is only relatively extreme cases with $q \leq 2$ AU for which significant integration error occurs.  It is for these close perihelion passages that all terrestrial material is destroyed, called mode 2 accretion in the main text.  In the case of the five simulations with ${\rm d}E/E > 1 \times 10^{-2}$, two contained a single surviving terrestrial planet and three had destroyed all of their terrestrial planets.  By removing these cases we are therefore slightly biasing our sample at the 5\% level away from accretion modes 2 and 3, and so we include this small extra contribution in our estimates in the paper of the fraction of systems for which the different accretion modes occur. 

\begin{figure*}%[p]
\centerline{
\includegraphics[width=0.4\textwidth]{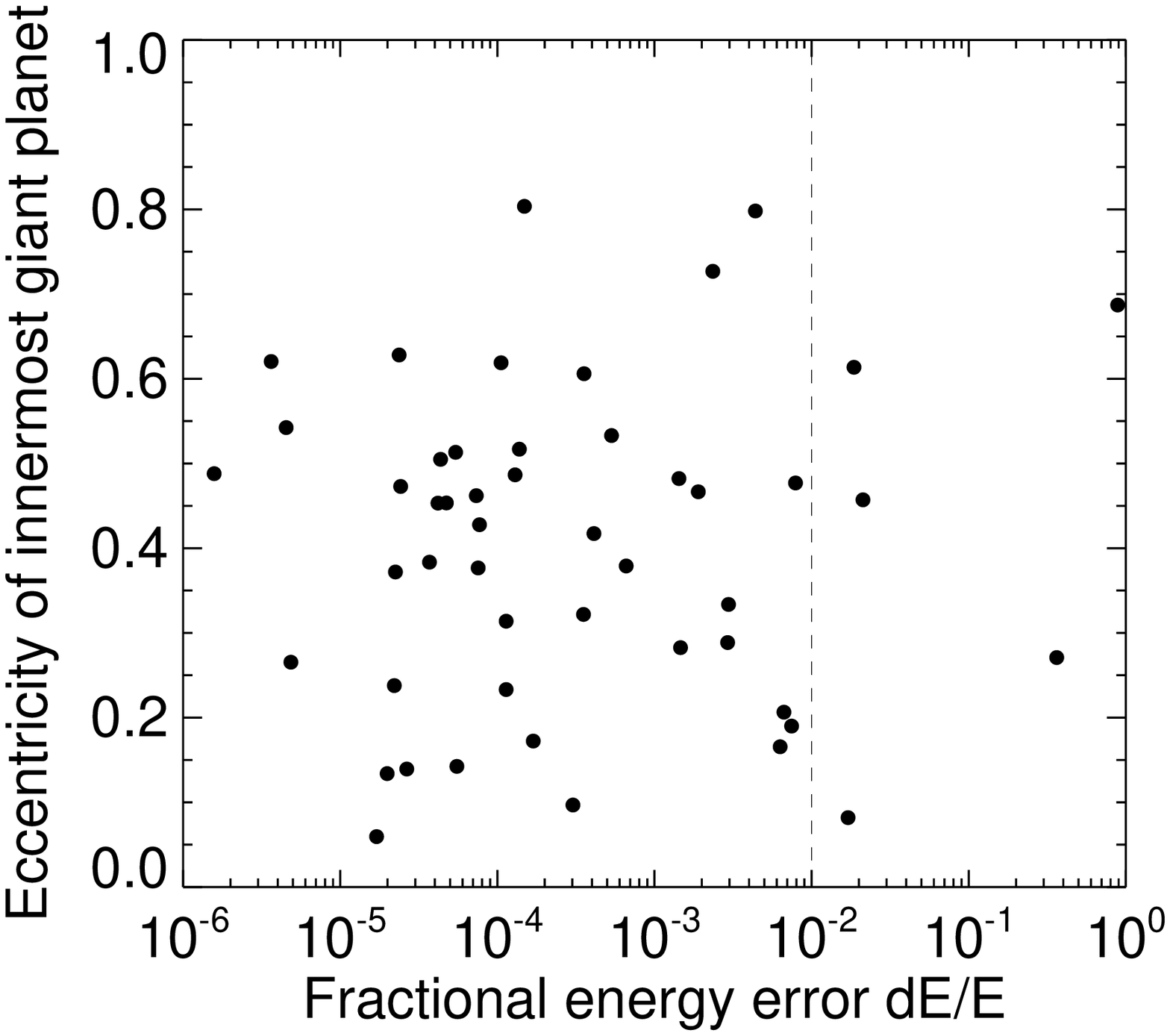}
\includegraphics[width=0.4\textwidth]{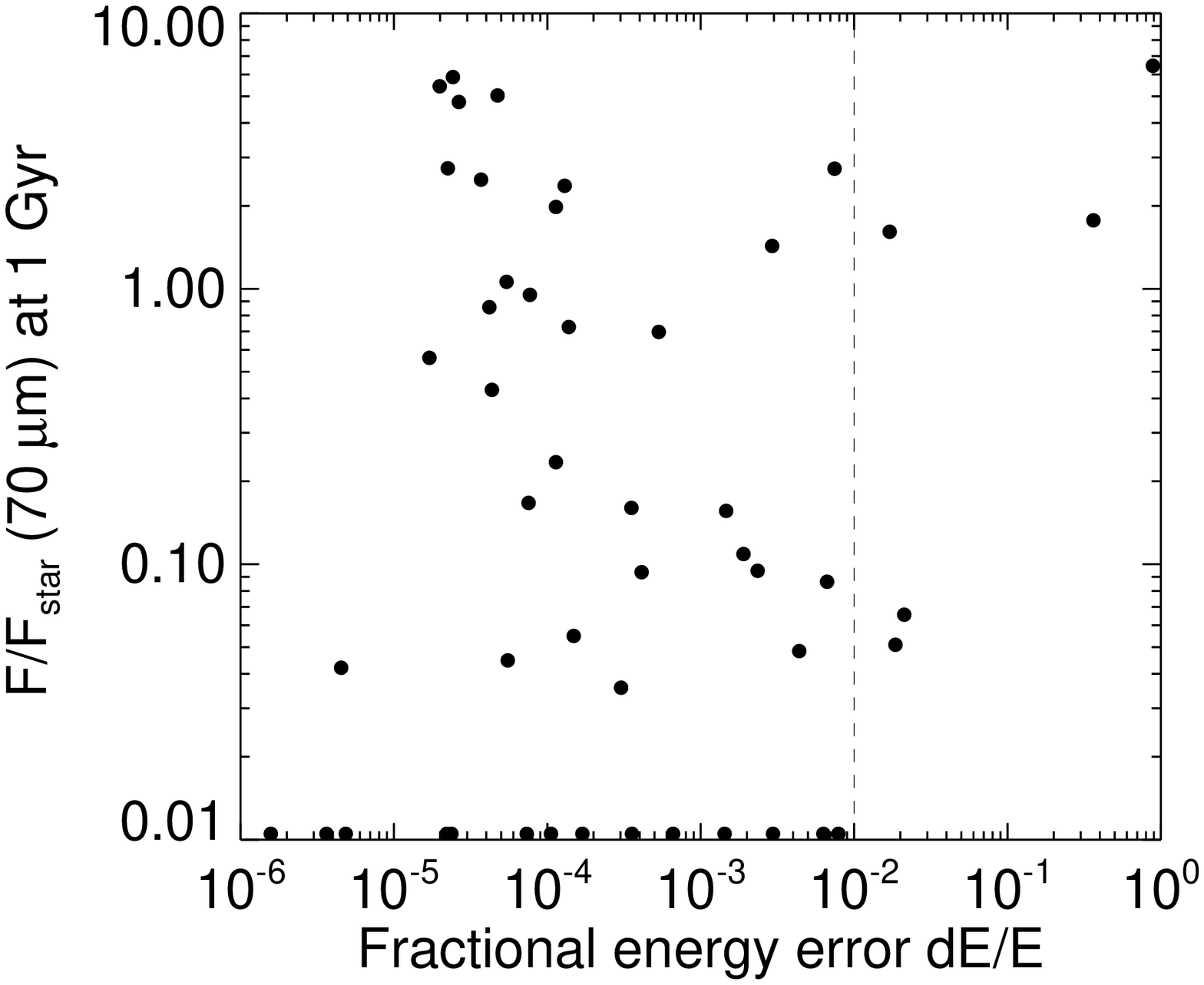}}
\caption{Eccentricity of the innermost giant planet (left panel) and dust-to-stellar flux ratio at 70$\mu$m after 1 Gyr of evolution (right panel) as a function of the fractional error in the system's energy budget ${\rm d}E/E$.  This plot is only for systems with minimum giant planet perihelia of less than 2 AU (see Fig.~\ref{fig:dee-q}).  Our energy error cutoff at 0.01 is shown with the dashed line.  There is no evidence for contamination of our sample.}
\label{fig:dee}
\end{figure*}

We do not see any clear signature of other bias introduced in our analysis from our energy cutoff.  To assess this possibility, we restrict ourselves to systems for which the minimum giant planet perihelion was less than 2 AU because this is where large errors occur.  Figure~\ref{fig:dee} shows the giant planet eccentricity and the dust-to-stellar flux ratio at 70$\mu$m after 1 Gyr as a function of ${\rm d}E/E$ for this subsample.  Both panels are scatter plots, with no clear trend or any indication that the computed ${\rm d}E/E$ value changes the outcome in any way.  We therefore conclude that our chosen cutoff in integrator energy, while less stringent than some other studies, has no measurable impact on our results.

\begin{acknowledgements}
We thank the referee, Scott Kenyon, for a very thorough review that helped us improve the paper.  Simulations were run at Weber State University and at Purdue University (supported in part by the NSF through TeraGrid resources).  Some of the collaborations vital to this paper (in particular, between S.N.R, P.J.A., A.M.-M., M.B. and M.C.W.) started during the Isaac Newton InstituteÕs Dynamics of Disks and Planets program in Cambridge, UK. S.N.R. acknowledges funding from the CNRS's PNP and EPOV programs and NASA Astrobiology Institute's Virtual Planetary Laboratory lead team. P.J.A. acknowledges funding from NASAÕs Origins of Solar Systems program (NNX09AB90G), NASAÕs Astrophysics Theory and Fundamental Physics program (NNX07AH08G), and the NSFÕs Division of Astronomical Sciences (0807471).

This paper is dedicated to office B329 in the University of Washington's Physics and Astronomy building, which from 2000-2003 housed S.N.R., J.C.A., and A.A.W. and was the site of many ridiculous pursuits.  
\end{acknowledgements}

\bibliographystyle{aa}
\bibliography{refs.bib}

\end{document}